King Saud University

College of Computer and Information Sciences

Computer Science Department

# Dual-Language General-Purpose Self-Hosted Visual Language and new Textual Programming Language for Applications

# لغة مرئية ذاتية الإستضافة ثنائية اللغة للأغراض العامة ولغة برمجة نصية جديدة للتطبيقات

DISSERTATION

by

MAHMOUD SAMIR FAYED

439106337

Supervisor

Dr. Yousef Ahmed Alohali

Submitted in partial fulfilment of the requirements for the Degree of Doctor of Philosophy in the Department of Computer Science at the College of Computer and Information Sciences.

# Examination Committee Page

The committee for

**[Mahmoud Samir Fayed]**

certifies that this is the approved version of the following dissertation and is acceptable in quality and form for publication in paper and in digital formats:

## Dual-Language General-Purpose Self-Hosted Visual Language and new Textual Programming Language for Applications

Dissertation Committee Members:

Committee Supervisor: Dr. Yousef Ahmed Alohali
Signature: ____________
Date: ١/١٢/١٤٤٧هـ

Committee First Member: Prof. Abdulmalik Al-Salman
Signature: ____________
Date: ١/١٢/١٤٤٧هـ

Committee Second Member: Prof. Abdel Monim Artoli
Signature: ____________
Date: ١/١٢/١٤٤٧هـ

Committee Third Member: Prof. Khalil El Hindi
Signature: ____________
Date: ١/١٢/١٤٤٧هـ

Committee Forth Member: Dr. Mohamed Tounsi
Signature: ____________
Date: 28/05/2025

King Saud University

2025

# Declaration

I, Mahmoud Samir Fayed, hereby declare that the work presented in this thesis has not been submitted for any other degree or professional qualification, and that it is the result of my own independent work.

This Ph.D. thesis continues the path I began during my master's study at King Saud University from 2012 to 2017, where the first generation of PWCT was introduced as a general-purpose visual programming language for MS-Windows, primarily focused on developing desktop database applications. Besides working on my master's thesis at that time, I made an early attempt to create an initial prototype for the Ring programming language compiler and virtual machine. However, after finishing my master's study, it became clear that additional years of research and development would be needed to create a production-ready textual programming language capable of supporting advanced projects, such as the development of a modern self-hosting visual programming language. It was evident that I needed to redesign and reimplement most of the work done in the prototype to add more features, achieve a faster and safer implementation of the language, and work on adding extensions, libraries, and tools.

Through my Ph.D. study from 2018 to 2025, the PWCT2 visual programming language was developed, along with extensive work on the design and implementation of the Ring programming language. The focus was on scientific contributions and achieving a level of efficiency that enables the use of the Ring language in developing the PWCT2 visual programming language. Additionally, the work involves testing, fixing many bugs during development and adding support for more platforms, such as WebAssembly and 32-bit microcontrollers like Raspberry Pi Pico. The contributions of the Ring textual programming language were introduced in a 2024 research paper, and the contributions of the PWCT2 visual programming language were introduced in a 2025 research paper.

I authorize King Saud University to lend this report to other institutions or individuals for the purpose of scholarly research.

# Abstract

Most visual programming languages (VPLs) are domain-specific, with few general-purpose VPLs like Programming Without Coding Technology (PWCT). These general-purpose VPLs are developed using textual programming languages and improving them requires textual programming. In this thesis, we designed and developed PWCT2, a dual-language (Arabic/English), general-purpose, self-hosting visual programming language. Before doing so, we specifically designed a textual programming language called Ring for its development. Ring is a dynamically typed language with a lightweight implementation, offering syntax customization features. It permits the creation of domain-specific languages through new features that extend object-oriented programming, allowing for specialized languages resembling Cascading Style Sheets (CSS) or Supernova language. The same Ring implementation allows us to create projects for desktops, WebAssembly, and the Raspberry Pi Pico microcontroller. The Ring Compiler and Virtual Machine are designed using the PWCT visual programming language where the visual implementation is composed of 18,945 components that generate 24,743 lines of code (written in ANSI C language), which increases the abstraction level and hides unnecessary details. Using PWCT to develop Ring allowed us to realize several issues in PWCT, which led to the development of the PWCT2 visual programming language using the Ring textual programming language. PWCT2 provides approximately 36 times faster code generation and requires 20 times less storage for visual source files. It also allows for the conversion of Ring code into visual code, enabling the creation of a self-hosting VPL that can be developed using itself. PWCT2 consists of approximately 92,000 lines of Ring code and comes with 394 visual components. Moreover, using Ring in this project demonstrates the feasibility of utilizing the language for large-scale projects. PWCT2 is distributed to many users through the Steam platform and has received positive feedback, On Steam, 1772 users have launched the software, and the total recorded usage time exceeds 17,000 hours, encouraging further research and development in the field of general-purpose VPLs.

# الخلاصة

معظم لغات البرمجة المرئية تم تصميمها لمجالات محددة، مع عدد قليل من لغات البرمجة المرئية التي صممت للأغراض العامة مثل تقنية البرمجة بدون كتابة الكود النصي. ايضا يتم تطوير هذه اللغات المرئية باستخدام لغات البرمجة النصية مما يعني ان تحسينها وتطويرها يتطلب استخدام البرمجة النصية. في هذه الأطروحة، قمنا بتصميم وتطوير الجيل الثاني من تقنية البرمجة بدون كتابة الكود النصي، وهي لغة برمجة مرئية ثنائية اللغة (العربية/الانجليزية)، ومتعددة الأغراض، ومستضافة ذاتيًا (اي يمكن تطويرها من داخلها باستخدام البرمجة المرئية). قبل القيام بذلك، قمنا بتصميم لغة برمجة نصية تسمى رينق صممت خصيصًا لتطويرها. رينق هي لغة برمجة ديناميكية لها بناء داخلي صغير الحجم، وتوفر ميزات تخصيص الكلمات المفتاحية. إنها تسمح بإنشاء لغات مخصصة لمجال محدد من خلال ميزات جديدة تعمل على توسيع البرمجة الموجهة للكائنات، مما يسمح ببناء لغات تشبه لغة أوراق الأنماط المتتالية (سي اس اس) أو لغة سوبرنوفا. يسمح لنا نفس الإصدار من رينق بإنشاء مشاريع لأجهزة سطح المكتب، والويب اسمبلي، والمتحكم الدقيق راسبيري باي بيكو. تم تصميم كل من المترجم والالة الافتراضية بإستخدام الجيل الأول من تقنية البرمجة بدون كتابة الكود النصي، حيث ان البناء المرئي مكون من 18,945 مكون مرئي بينما الكود النصي الناتج عبارة عن 24,743 من اسطر الكود بلغة السي والذي يعني زيادة مستوى التجريد وإخفاء التفاصيل غير الضرورية. سمح لنا إستخدام الجيل الاول من التقنية لتطوير لغة البرمجة رينق بإدراك العديد من المشكلات، مما أدى إلى تطوير الجيل الثاني والجديد من تقنية البرمجة بدون كتابة الكود النصي بإستخدام لغة البرمجة النصية رينق. يوفر الجيل الثاني عملية توليد الكود النصي بشكل أسرع بنحو 36 مرة ويتطلب مساحة تخزين أقل بنحو 20 مرة لملفات المصدر المرئية مقارنة بالجيل الاول. كما يسمح بتحويل كود رينق النصي إلى كود مرئي، مما يوفر لنا لغة برمجة مرئية ذاتية الإستضافة يمكن ان يتم تطويرها عبر إستخدامها. يتكون الجيل الجديد من حوالي 92000 سطر كود بلغة الرينق ويأتي مع 394 مكونًا مرئيًا. علاوة على ذلك، يوضح إستخدام رينق في هذا المشروع إمكانية إستخدام اللغة لتطوير مشاريع بهذا الحجم. تم توزيع الجيل الجديد على العديد من المستخدمين من خلال منصة إستيم وقد تلقى ردود فعل إيجابية، على إستيم قام 1772 مستخدم بتشغيل المشروع وتجاوز إجمالي وقت الإستخدام 17000 ساعة مما يشجع على المزيد من البحث والتطوير فى مجال لغات البرمجة المرئية المصممة للأغراض العامة.

# Associated Publications

- Fayed, Mahmoud Samir, and Yousef A. Alohali. "PWCT2: A Self-Hosting Visual Programming Language Based on Ring with Interactive Textual-to-Visual Code Conversion." Applied Sciences 15, no. 3 (2025): 1521.
- Fayed, Mahmoud Samir, and Yousef A. Alohali. "Ring: A Lightweight and Versatile Cross-Platform Dynamic Programming Language Developed Using Visual Programming." Electronics 13, no. 23 (2024): 4627.
- Fayed, Mahmoud Samir, Muhammad Al-Qurishi, Atif Alamri, M. Anwar Hossain, and Ahmad A. Al-Daraiseh. "PWCT: a novel general-purpose visual programming language in support of pervasive application development." CCF Transactions on Pervasive Computing and Interaction 2 (2020): 164-177.

# Acknowledgements

*In the name of Allah, Most Gracious, Most Merciful*

Firstly, I would like to express my gratitude to my supervisor, Dr. Yousef Alohali, for his exceptional support and guidance throughout my PhD journey. He began by teaching me multiple PhD courses and the fundamentals of conducting research studies. His unwavering direction continued during my PhD research, helping me connect my work with recent research studies.

Also, I would like to say thanks to the members of the Ring project team for their contributions in testing the software and adding more samples.

- Bert Mariani
- Ahmed Hassouna
- Mansour Ayouni
- Majdi Sobain
- Youssef Saeed
- Azzeddine Remmal
- Mounir Idrassi
- Ilir Liburn
- Mohannad Al-Ayash
- Ahmed Zakaria
- Khalid Abid
- Gal Zsolt
- Jose Rosado
- Marino Esteban
- Magdy Ragab

Mahmoud Fayed
May 2025

This thesis is dedicated to my father, my wonderful mother, and my beloved family.

# Table of Contents

# List of Figures

# List of Tables

# List of Abbreviations

| | |
|---|---|
| ANSI | American National Standards Institute |
| CDF | Cognitive Dimensions Framework |
| CGT | Code Generation Time |
| CSS | Cascading Style Sheets |
| DSL | Domain Specific Language |
| GCR | Graphical Code Replacement |
| GIL | Global Interpreter Lock |
| GUI | Graphical User Interface |
| IDE | Integrated Development Environment |
| IDSL | Internal Domain-Specific Language |
| IoT | Internet of Things |
| JSON | JavaScript Object Notation |
| KLOC | Thousands of lines of code |
| LLM | Large Language Model |
| MVC | Model View Controller |
| OOP | Object Oriented Programming |
| PWCT | Programming Without Coding Technology |
| REPL | Read-Eval-Print-Loop |
| SQL | Structured Query Language |
| VFP | Visual FoxPro |
| VM | Virtual Machine |
| VPL | Visual Programming Language |

# Chapter 1: Introduction

## 1.1 Overview

The demands for software applications are increasing because computers are now a very important part of our daily lives. Today, software runs on a variety of devices, including high performance clusters, personal computers, embedded devices, and distributed systems. Applications are developed in different areas and the cost can vary greatly as free and open-source software competes with proprietary software. Reducing costs, improving reliability, and increasing scalability are among the requirements that software developers face. In the age of information technology, software development plays a vital role in responding to the needs of companies and organizations for high-quality information systems. This leads to the need for more programmers and more productive software development tools to be able to respond quickly to the needs of companies. As a result of this complexity in software requirements, many aspects of software development have evolved, and many tools have been developed to help programmers [1-2].

Integrated development environments (IDEs) such as Microsoft Visual Studio, Qt Creator, and Eclipse are very important for large projects. Unfortunately, these tools do not eliminate the need for programmers to know the strict syntax of each programming language they use, where understanding of general programming paradigms is essential but not sufficient. Representation of the software in the textual source code files is limited to text as we cannot include images and graphics to make them part of the source code. Also, the more expressive a programming language is, the more complex the syntax becomes, making programs difficult to understand or write. This challenge opens the door to the uses of visual programming languages and tools that attract more users to programming and increase software development productivity [3-7].

Visual programming languages allow the development of applications and computer programs using more than one dimension and provide a programming system that is based on interaction with graphical elements that combine text, shapes, colors, and time instead of writing source code based on text. There are many visual programming languages, but most of the successful and widely used visual languages are used in education, such as Alice and Scratch (shown in Figure 1.1) or in a specific field such as Blueprints (Unreal game engine) and LabView (Industrial automation).

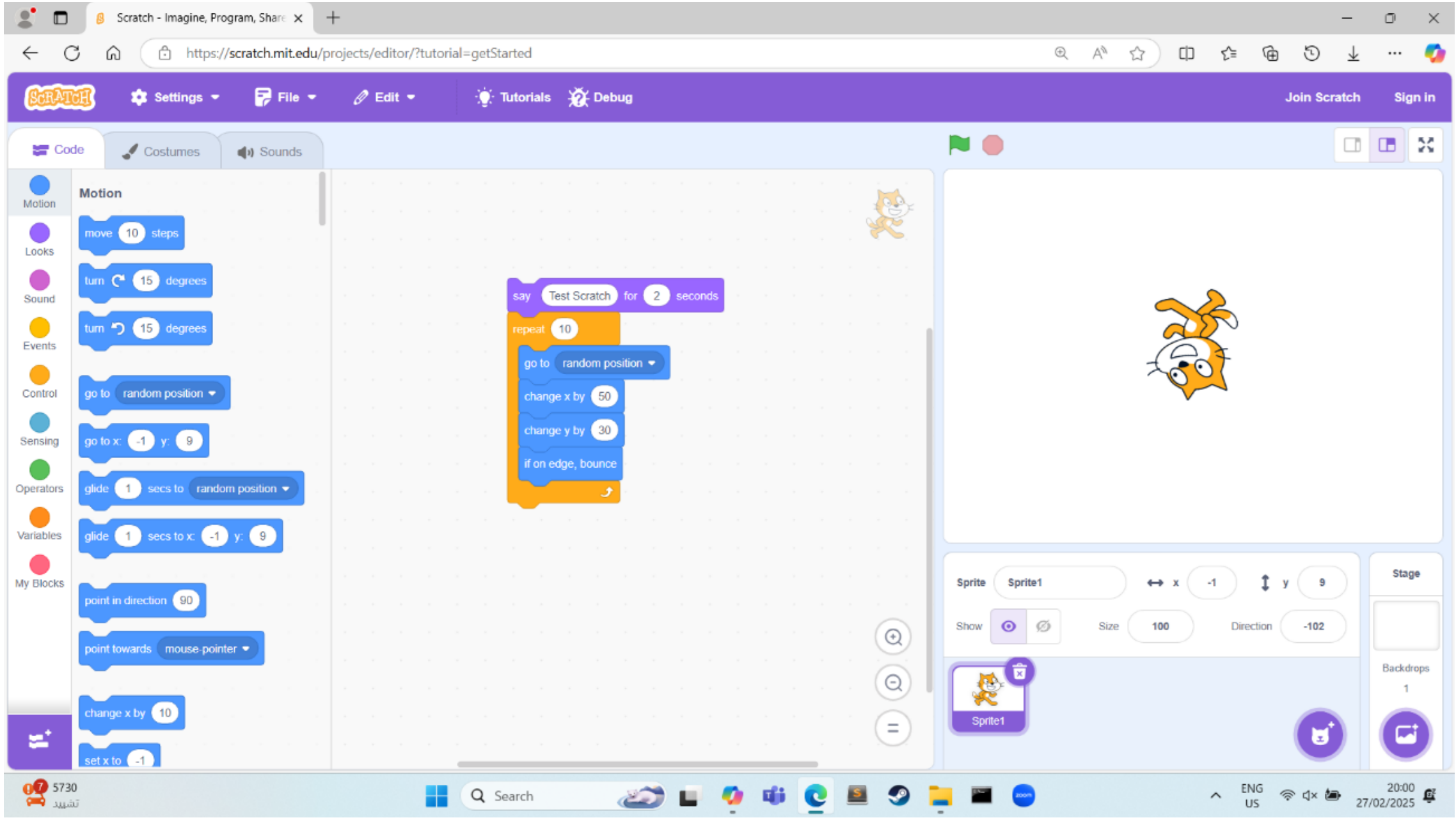

*__Figure 1.1__ The Scratch Visual Programming Language*

There are general-purpose visual programming languages and systems like Lava, Tersus, Limnor, Envision, and Unit. However, these VPLs and systems are not widely used, according to the TIOBE Index, which measures the popularity of programming languages. Additionally, there are few studies that evaluate such systems through the development of large-scale applications and systems [8-10].

Programming languages like Visual Basic and Visual C# are not Visual Programming Languages (VPLs). All these languages are textual programming languages where the programmer must write the textual code using the language's syntax to create useful applications. Environments such as Microsoft Visual Studio and Qt Creator are not considered VPLs. These environments enable the software developers to create parts of the application using visual components, but the textual code is necessary to complete useful and real applications. On the other hand, VPLs use only visual components instead of writing textual code [11-12]. In textual programming languages like C++ and Java, the code is text-based. It is one dimensional. The compiler reads the source code token by token. In VPLs, the graphical representation uses more than one dimension. Each graphic object has its place in 2D or 3D worlds. Each object can have its own shape, color, and image. There are many relationships that can appear between objects, such as: Inside, outside, touching, next to, etc. Some visual languages also use the Time dimension (before/after) as another dimension in the graphic code [13].

A VPL can have one of the following five visual representations [14-15]:

**1. Diagrammatic:** uses components of shapes and text and uses links to connect between shapes and represent the control or data flow.

**2. Iconic:** uses icons from the domain of the problem.

**3. Form-based:** uses forms such as spreadsheet or data-entry forms.

**4. Block-based:** uses blocks that are pieced together to create programs.

**5. Hybrid:** uses a mixture of any of the above four.

Visual programming systems can use different interaction methods. For example, Scratch uses drag-and-drop because it's designed for children, while the Envision visual structured editor (shown in Figure 1.2) uses command-based interactions because it's designed for programmers. Some VPLs use a syntax-directed editor that recognizes the syntax and prevents the programmer from making syntax errors; other VPLs provide a free editor in which the programmer can make mistakes during the development process and the VPL compiler can detect errors during the compilation process. The syntax-directed editor is more suitable for novice programmers, while the free editor offers more flexibility for advanced programmers [16].

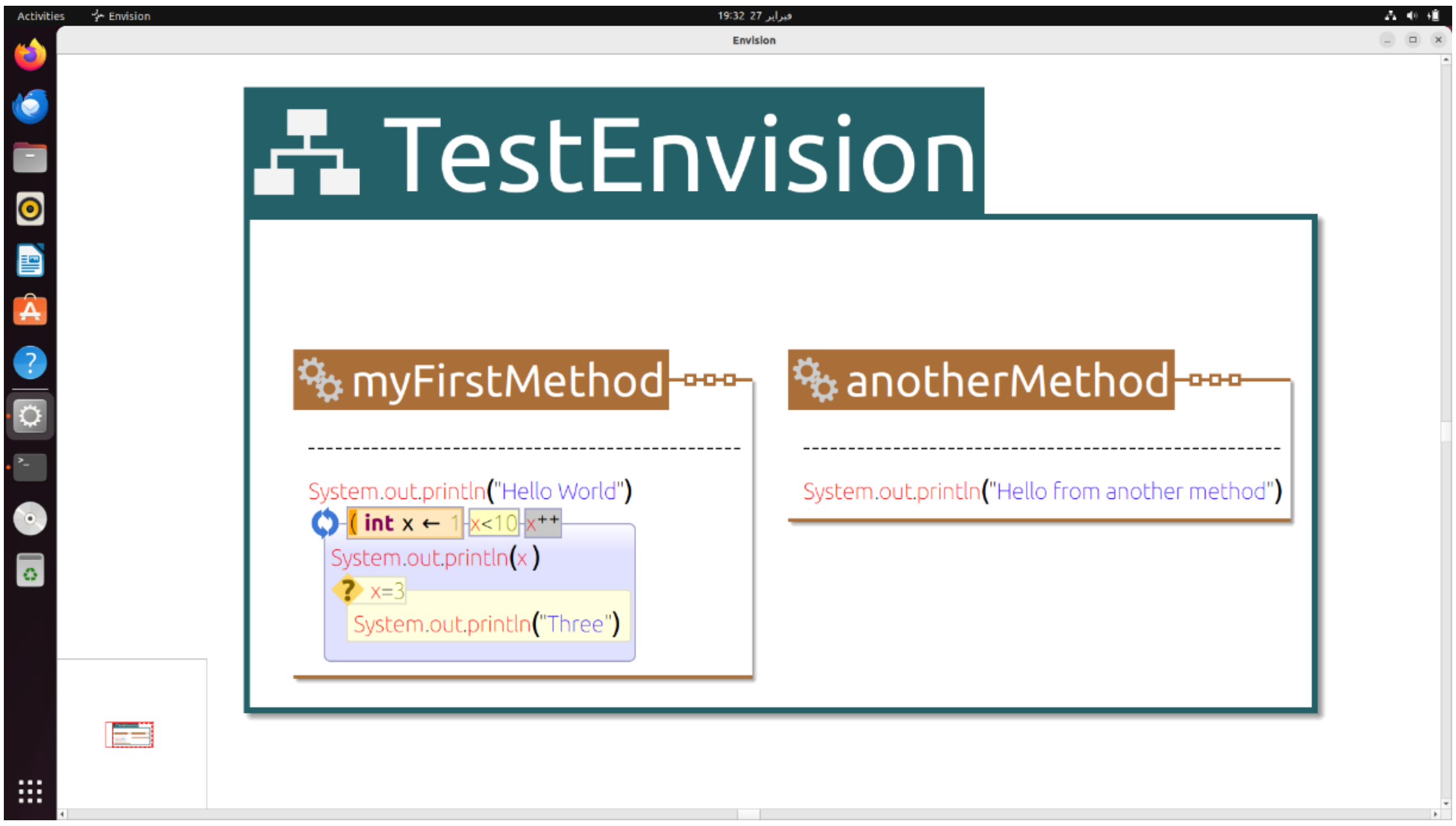

***Figure 1.2*** *The Envision Visual Programming System.*

A VPL's framework is a collection of tools that enable the developers to create VPLs in less time with less effort and better quality by utilizing ready-to-use and well-tested tools. Programming Without Coding Technology (PWCT) is a general-purpose visual programming system designed with the concept of enabling the creation of visual programming languages (VPLs) in mind. PWCT includes tools that facilitate this process and come with multiple groups of visual components that generate code in various textual programming languages. [12, 17].

There are many issues related to visual programming languages. Almost all visual representations are physically larger than the text they generate, so the space used to show a program in a VPL is greater when compared with a text-based program. Many large programs created by VPLs look like a maze of wires that are hard to understand. Many VPLs don't provide a place for writing comments.

The most successful VPLs are designed for specific applications (not for general purposes). There are few VPL's frameworks, and most of them are designed for a specific category of visual programming languages. Most VPLs use a drag-and-drop visual programming approach and are not designed for fast interactions using the keyboard. Also, a lot of visual programming languages don't support advanced dimensions like the Time dimension. Another issue is that the VPLs users can't improve the visual language itself using visual programming because it's based on textual programming language (TPL) code, and it's common to use advanced languages like C++ to develop VPLs [18-21]. The PWCT visual programming language (illustrated in Figure 1.3) addresses some of these issues through the following design decisions [12]:

- Utilizing a visual representation based on the TreeView control, which solves the visual representation size issue and avoids the maze of wires problem. In PWCT, the program is represented as a group of steps called the Steps Tree.
- Including hundreds of visual components that provide a general-purpose VPL.
- Using a visual programming approach suitable for keyboard-based interaction.

However, many issues remain unsolved. PWCT is not a self-hosting VPL, and developing or maintaining it requires writing textual code. Additionally, PWCT is limited to the Windows environment. Furthermore, PWCT has not been evaluated through the development of a large or complex project over many years, which could help in discovering more practical issues based on serious usage and analysis.

This thesis focuses on the development of PWCT2, a self-hosting visual programming language for application development. This research identifies the limitations in the previous PWCT implementation, defines the requirements for PWCT2, and includes the implementation and evaluation of the proposed visual programming language. The thesis also includes the design and implementation of a textual programming language developed to achieve two goals. The first goal is to evaluate PWCT, and the second is to be used in the development of the proposed visual programming language to ensure its quality and support future research projects. This textual language is designed for developing applications and tools. The proposed visual language could represent a step forward in the field of visual programming languages, aiming to create a more powerful general-purpose visual programming language.

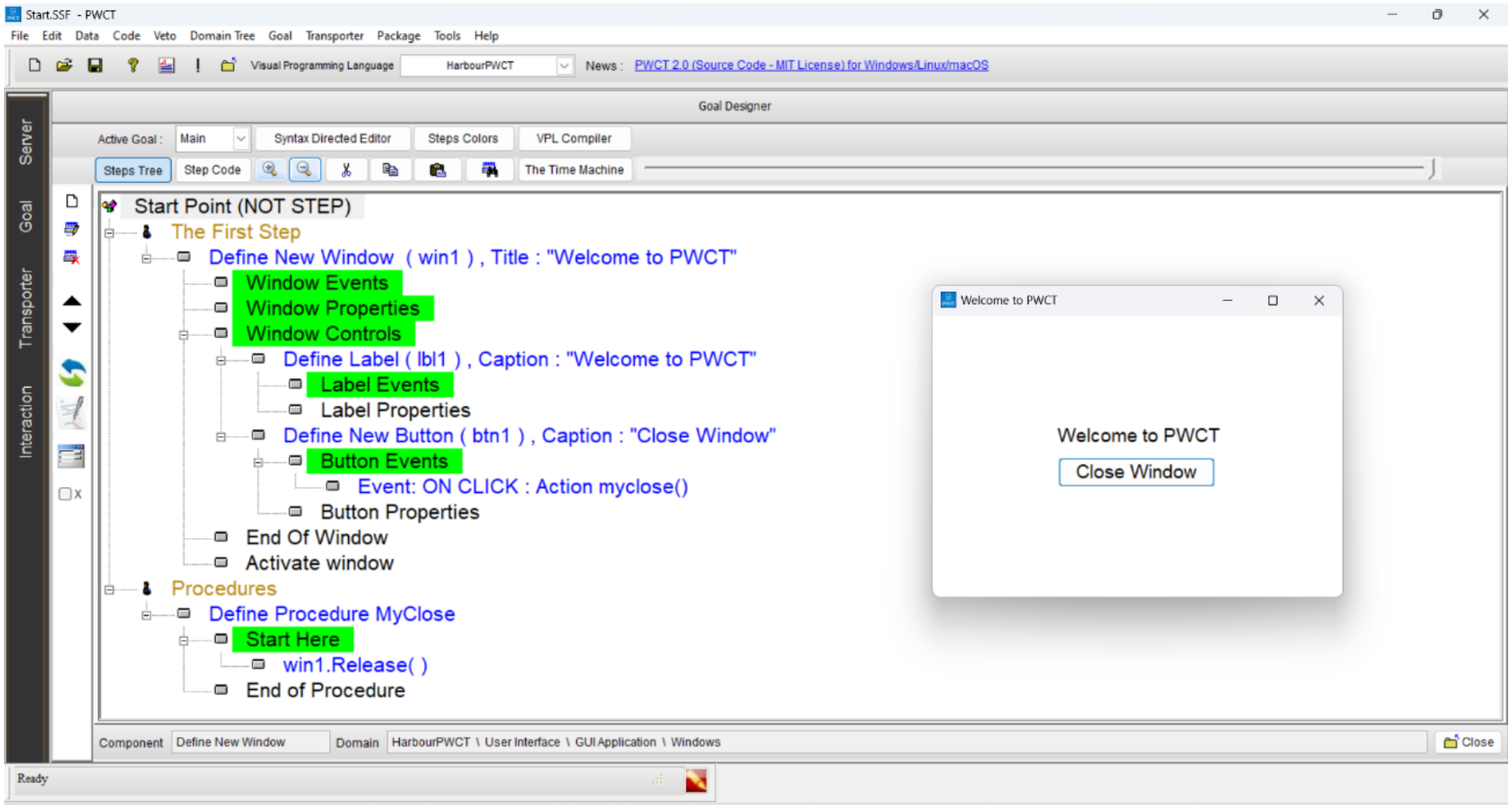

***Figure 1.3*** *The PWCT visual programming language.*

## 1.2 Motivation

After the success of many domain-specific visual programming languages like Scratch, Alice and Blueprints and reaching millions of users worldwide, it's expected to find more interest in creating new visual programming languages. These languages help novice programmers to learn programming and help mainstream programmers to create high-quality programs faster, but these languages must be designed carefully to solve the problem without adding other critical problems and this is an important factor for new visual programming languages to gain popularity.

Additionally, the adoption of these new visual programming languages will not increase unless they can integrate with other development tools, highlighting the necessity of being able to import and export textual source code.

Many software projects require different programming skills and different programming languages. This causes a problem for many companies and researchers that need to hire many programmers to develop a complete solution. In this research, we expect to provide a general-purpose visual programming language that can be used in developing complex and large software projects using visual programming.

Also, designing a new textual programming language to be used in tools development like a general-purpose visual programming language will help many similar projects in their mission. In computer science, developments in programming languages and development tools are considered very helpful in practical and scientific projects. Through this research, we are going to help in this active and interesting area.

## 1.3 Problem Statement

Most large and complex software projects are developed using textual programming languages. The adoption of general-purpose visual programming languages by mainstream programmers is still in an early stage. A lot of general-purpose visual programming languages and systems are no longer under continuous development (Lava, Envision, etc.).

These languages have features that can be merged, and a lot of features that can be improved. Also, most of these projects (PWCT, Limnor, etc.) don't provide good and high-quality support for modern technologies that appeared after their initial development. There is an open space for innovation and producing useful ideas through research and development.

To our knowledge, there is no dual-language, self-hosting, general-purpose visual programming language that can be used for developing large and complex software projects and provides support for various platforms, including Desktop, Web, WebAssembly, and 32-bit Microcontroller platforms, and be widely adopted by many software developers in real-world projects.

The research questions that we aim to answer during this research are:

**RQ1:** What is the design of a modern textual programming language suitable for the development of the proposed dual-language, self-hosting, general-purpose visual programming language?

**RQ2:** What are the advantages and disadvantages of using a general-purpose visual programming language like PWCT to develop and maintain a textual programming language compiler and virtual machine over many years?

**RQ3:** What design decisions could be employed to create a lightweight, multi-paradigm dynamic programming language suitable for Desktop, Web, WebAssembly, and 32-bit Microcontrollers, using the same implementation?

**RQ4:** What novel features could be used to extend the object-oriented programming paradigm to enable the development of internal domain-specific languages that resemble external domain-specific languages like CSS and Supernova?

**RQ5:** What is the design of a modern general-purpose visual programming language that leverages advancements in visual programming research and considers current technology trends, such as the use of Large Language Models (LLMs) for generating textual source code?

**RQ6:** Can we maintain and continue developing the new visual programming language using itself and have a self-hosting general-purpose visual programming language?

## 1.4 Research Goal and Objectives

The main goal of this thesis is the design and implementation of a new dual-language self-hosting general-purpose visual programming language powered by a new textual programming language for applications and tools development. The new visual language must come with modern features and advantages that encourage and enable usage in advanced projects and applications. Visual language design must support improving the current projects or creating new projects from scratch based on the proposed textual programming language. The objectives of this research subject are:

- The design and implementation of a textual programming language (called Ring) for applications and tools development to use it in developing the new self-hosting visual programming language.

- The design and implementation of a dual-language general-purpose self-hosting visual programming language (called PWCT2) for applications development. This visual language should support importing and exporting the textual source code written in the proposed dynamic programming language.

## 1.5 Thesis Contributions

This thesis provides the next contributions:

• The first study to use visual programming to develop and maintain a compiler and virtual machine for a dynamic programming language for over eight years. We used the PWCT visual programming language to develop and maintain the Ring programming language. We have provided multiple releases each year to improve the design and respond to community feedback.

• Novel features that can extend the object-oriented programming paradigm, enabling the development of domain-specific languages that resemble CSS and Supernova. Additionally, Ring's customization features, such as syntax modification, could support multiple languages (e.g., Arabic, English).

• The design and implementation of a dynamic programming language with broad cross-platform compatibility, featuring a lightweight implementation that still provides rich features.

• The design and implementation of the research prototype PWCT2, which offers enhanced features, lower storage requirements for visual source files, and better code generation performance compared to the first generation.

• The design and implementation of the first VPL that supports code generation in the Ring language (RingPWCT), containing 394 visual components.

• The design and implementation of a textual-to-visual code conversion tool called Ring2PWCT that can import Ring programming language code. Using this tool enables a self-hosting VPL based on Ring.

• Testing the feasibility of using the Ring programming language compiler and virtual machine in the development of projects on a scale similar to PWCT2.

• Arabic Translation for the PWCT2 Environment and the RingPWCT Visual Components.

## 1.6 Thesis Outline

The remainder of this thesis is divided into the following chapters:

- **Chapter 2:** This chapter offers an overview of dynamic programming languages and visual programming languages. It begins by introducing dynamic programming languages and their intriguing features. Following that, it highlights the main characteristics of visual programming languages.
- **Chapter 3:** This chapter presents the existing research and developments relevant to the design of dynamic and visual programming languages. We begin by exploring the literature related to dynamic language design. Our review classifies the existing and related dynamic languages into various categories, providing an organized framework for comparison. We then select key dynamic languages from these categories that are most relevant to our work and conduct a comparative analysis with our proposed dynamic language, Ring. This analysis aims to identify the unique features and advancements of Ring, as well as highlight the gaps and areas for improvement within the existing literature. Subsequently, we shift our focus to visual programming languages (VPLs). We classify related VPLs into different categories and pinpoint those that have significantly influenced our design. By comparing these selected VPLs with our proposed visual programming language, PWCT2, we aim to underscore the distinctive aspects of PWCT2 and address the research gaps that our language fills. Through this literature review, we establish the foundation for our proposed languages and provide a thorough understanding of the existing landscape, setting the stage for the features introduced by Ring and PWCT2.
- **Chapter 4:** This chapter presents the proposed dynamic programming language and its important features, such as syntax customization and novel features that extend object-oriented programming and enable the development of internal domain-specific languages resembling CSS and Supernova. Additionally, this chapter introduces the critical details about the visual implementation and the significant design decisions made during development.
- **Chapter 5:** This chapter presents the design of the proposed dual-language self-hosting general-purpose visual programming language (PWCT2) and highlights the important features of the visual programming environment compared to the first generation of PWCT.

- **Chapter 6:** This chapter presents the experiments and results of the evaluation of the proposed textual programming language (Ring) and the proposed self-hosting visual programming language (PWCT2). The evaluation includes various measurements related to abstraction level and performance. Additionally, we provide different use cases.
- **Chapter 7**: This chapter presents the discussion and highlights the discovered advantages and limitations of various experiments. We will analyse the findings in detail, discussing their implications and how they support the objectives of this thesis. Additionally, we will address the limitations encountered during our research, providing a critical evaluation of the potential challenges and areas for improvement. By examining these aspects, we aim to offer a thorough understanding of the strengths and weaknesses of the Ring dynamic programming language and the PWCT2 visual programming language.
- **Chapter 8:** This chapter presents the conclusion, future work, and various research directions that become available after developing the Ring textual programming language and the PWCT2 visual programming language.

# Chapter 2: Background

## 2.1 Introduction

New programming languages are often designed to keep up with technological advancements and project requirements while also learning from previous attempts and introducing more powerful expression mechanisms. However, most existing dynamic programming languages rely on English keywords and lack features that facilitate easy translation of language syntax. Additionally, maintaining multiple implementations of the same language for different platforms, such as desktops and microcontrollers, can lead to inconsistencies and fragmented features. Furthermore, they usually do not use visual programming to fully implement the compiler and virtual machine. In this chapter, we will introduce dynamic programming languages and their interesting features. Then, we will introduce visual programming languages and their main characteristics.

## 2.2 Dynamic Programming Languages

Programming languages play a crucial role in producing systems and applications. They serve as the means of communication between us and the computer, enabling control and the creation of software and applications. Initially, there was machine language, which allowed us to program by directly controlling the operations provided by the hardware. Soon, many programming languages evolved, each with different goals—such as ease of learning, specific domain usability, using new programming paradigms, performance improvement, security, portability, or achieving flexibility [22–25]. During the evolution of programming languages, a category known as dynamic programming languages emerged. Examples of such languages include Lisp, Smalltalk, Erlang, Python, Lua, and Julia, as demonstrated in Figure 2.1. (The vertical lines are designed to improve the figure's readability).

These languages exhibit several features that defer determination and execution to runtime rather than compiling time. Notable characteristics include dynamic typing, flexible data structures, reflection, metaprogramming, and the ability to evaluate code from strings using functions like eval(). Additionally, dynamic languages often provide a Read-Eval-Print-Loop (REPL) for interactive development. The overarching goal of these languages is to achieve simplicity, flexibility and reduced compile time. Ultimately, this speeds up the development cycle and facilitates the creation of project prototypes in less time [26–33].

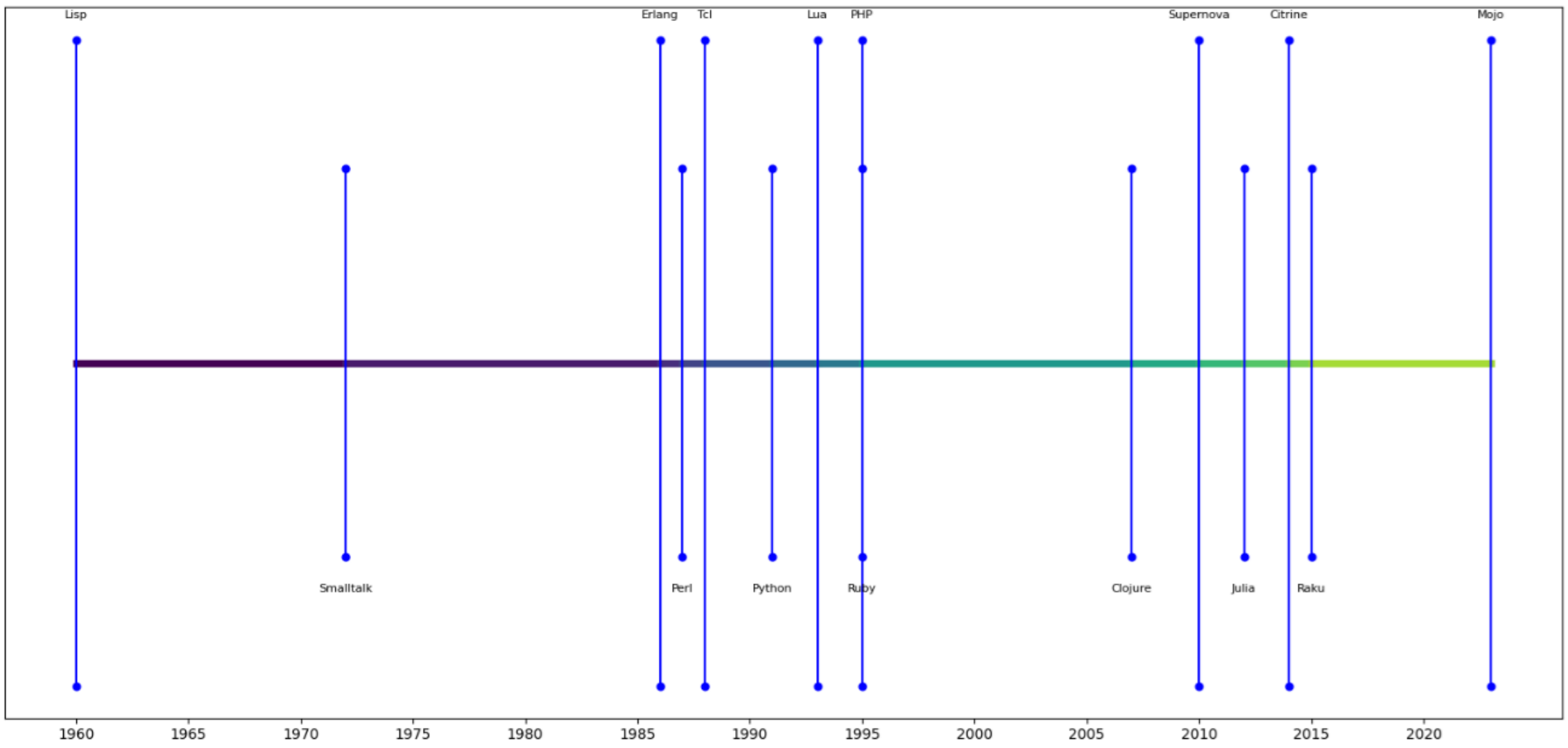

***Figure 2.1*** *Some of the dynamic programming languages, starting in 1960.*

Another category of programming languages is visual programming languages (VPLs). These languages use more than one dimension to create computer programs graphically through text, shapes, colors, etc. They have achieved notable usage in education through projects like Scratch [34–36]. Unlike the Scratch visual programming language, which enables children to create multimedia applications using a user interface in their native language, most dynamic programming languages rely on English keywords. Unfortunately, these dynamic programming languages lack features that facilitate easy translation of language syntax and libraries into other human languages [37–42].

While most visual programming languages are domain-specific, there are projects classified as general-purpose and applicable to a wide range of programming tasks. One such project is the Programming Without Coding Technology (PWCT) software, a visual programming language that supports code generation in multiple textual programming languages, including the C programming language [34]. Most of the popular dynamic programming language implementations are based on using textual programming languages like C, C++, etc. We assume that using visual programming to create the dynamic programming language compiler and virtual machine is possible and provides a more user-friendly implementation by avoiding syntax errors and increasing the abstraction level.

Dynamic programming languages as software products differ from one another in terms of design, syntax, semantics, paradigms, features, implementation, execution methods, libraries, tools, and supported platforms, resulting in variations in the domains where they are most suitable for use [43–52]. Some dynamic programming languages are specifically tailored for domains like R, MATLAB, and dBase [53–55]. On the other hand, some dynamic programming languages serve as general-purpose tools suitable for a wide range of tasks like Python [56–61]. Domain-specific languages are designed for a specific domain, and they can be classified into two main types. The first type is Internal/Embedded DSLs, which are embedded inside general-purpose languages and use their constructs, while the second type is External DSLs (Like CSS, SQL, Supernova, etc.), which use its syntax and semantics. Dynamic programming languages like Ruby could be used to create internal DSLs. However, these internal DSLs will not resemble external DSLs [62–65].

The Supernova dynamic programming language is a domain-specific language distributed with the PWCT Visual Programming language [66]. This language was developed by the author to explore creating simple GUI applications using command-based syntax that looks natural, as demonstrated in Figure 2.2.

After developing Supernova, we considered whether we could develop a new programming language that supports object-oriented programming and extend it with novel features to enable the development of embedded domain-specific languages resembling CSS and Supernova [34,66].

```
I want window and the window title   is Hello.
I want label  and the label  caption is Welcome.
I want button and the button caption is Close.

The window back color is white.
```

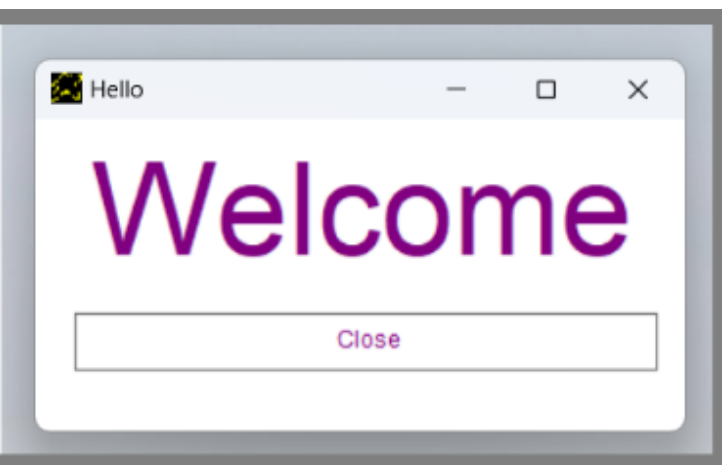

***Figure 2.2*** *Using commands in the Supernova programming language.*

With the rise of popularity of the Internet of Things (IoT) [67–69], numerous projects—such as MicroPython and mRuby—have endeavored to leverage popular dynamic programming languages for embedded systems and microcontroller development. This requires developing a lightweight implementation and has led to the challenge of maintaining different implementations for the same programming language, where one implementation could miss features that exist in another implementation [70–73].

Also, while we find many dynamic languages used for web application development on the server side, JavaScript has dominated the scene as the language used at the front-end inside web browsers. With the emergence of WebAssembly (binary instruction format that can be executed by modern web browsers), it has become more practical to use several other languages within the browser. However, this led to the development of different language implementations to allow dynamic languages to fully benefit from this leap. We assume that creating a new dynamic programming language in this era may necessitate considering this evolution to maximize its advantages [74–76].

## 2.3 Visual Programming Languages

Programming languages are essential for the development of systems and applications. They act as a bridge between humans and computers, facilitating control and the creation of software. Over time, a variety of programming languages have emerged, most of which use textual source code to create and represent computer programs [23,25]. During the evolution of programming languages, a category known as visual programming languages (VPLs) emerged. These languages create and represent computer programs graphically, using more than one dimension and incorporating a mix of text, colors, and shapes in their visual representations [77,78].

In Figure 2.3, we present some of the VPLs and systems developed from 1966 to 2024, showcasing a clear trend of increasing innovation and development in this field. The vertical lines are designed to improve the figure's readability. Early pioneering efforts are illustrated by the development of GPE (Graphical Program Editor) in 1966 and Pygmalion in 1975 [79,80], followed by subsequent languages such as Prograph, Simulink, and Lab-VIEW in the 1980s [81–83]. The 2000s saw the emergence of educational and accessible VPLs, like Scratch and Alice, which have become instrumental in teaching programming to younger audiences. For example, Scratch enables children to create stories, multimedia applications, and computer games using a user interface in their native language (Arabic, English, etc.) [84]. More recent advancements introduced in the literature include Envision, Node-RED, Blueprints, and FlowPilot, reflecting the continuous expansion. [85–88].

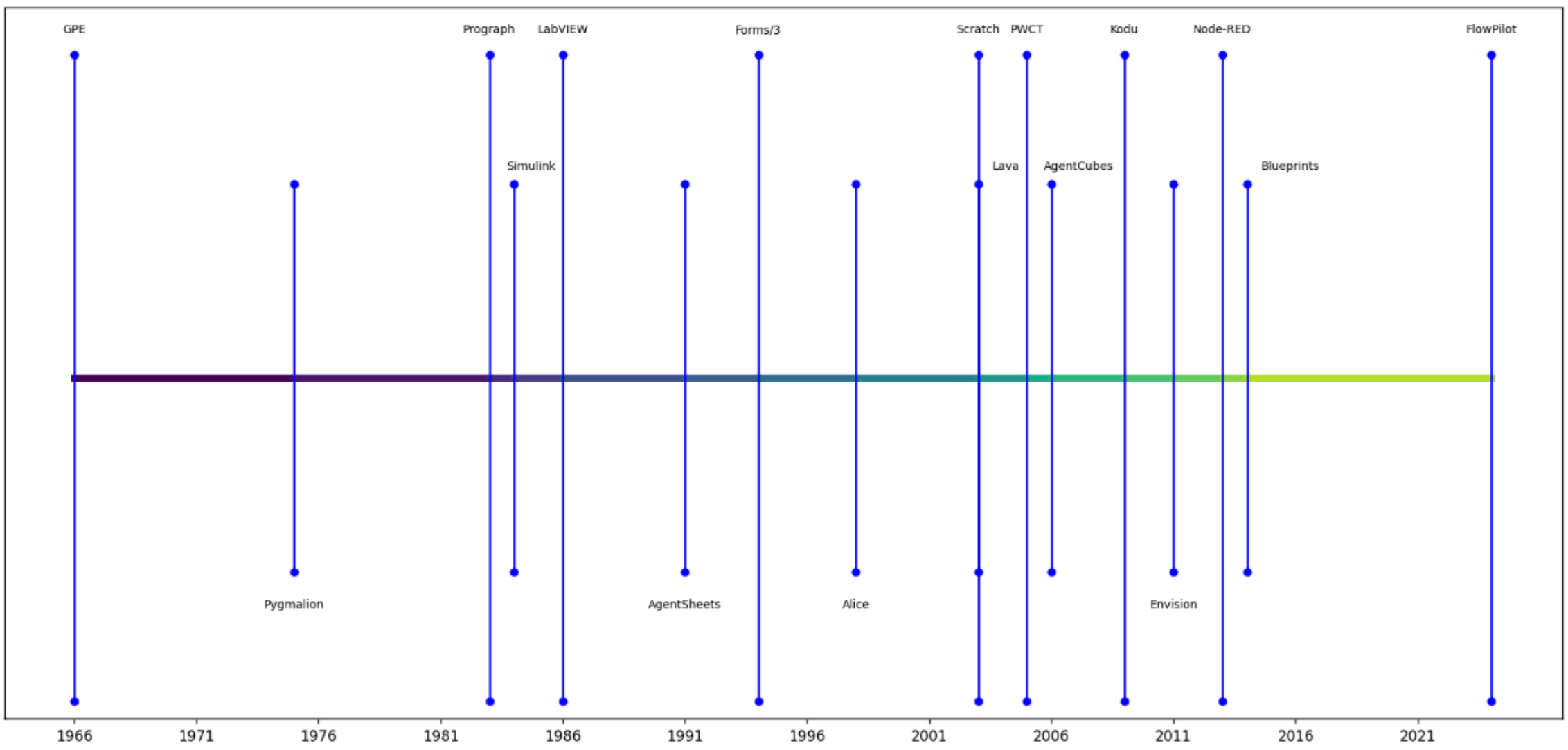

***Figure 2.3*** *Some of the visual programming languages starting from 1966.*

Most visual programming languages are either used in education or specific fields. Only a few projects have been designed to be general-purpose and versatile. One of these projects is Programming Without Coding Technology (PWCT), which supports code generation in multiple textual programming languages, such as C, Harbour, Python, and Supernova. PWCT introduces a visual programming approach called the Graphical Code Replacement (GCR) method, which is an alternative to the traditional Drag-and-Drop approach. GCR is based on Automatic Steps Tree Generation and Update in response to interaction with components that provide users with simple data entry forms. This method combines programming using a Diagrammatic approach and programming using a Form-based approach, seamlessly integrating the two through an Automatic Visual Representation Generation/Update process. GCR enables the design and implementation of advanced visual components that could include optional features that change the structure of the generated visual representation. Also, PWCT is designed to support fast interactions through the computer keyboard where using the Mouse is optional. Additionally, PWCT incorporates the Time Dimension at the program design level and supports a feature called play programs as movies that enables step-by-step implementation visualization [34,66].

PWCT does not support importing textual source code and is designed to operate exclusively on Microsoft Windows. Furthermore, the implementation of PWCT is based on Microsoft Visual FoxPro, which is no longer under active development. These issues need to be addressed when developing a new generation of PWCT. We assume that using PWCT to develop and maintain the Ring programming language compiler and virtual machine will enable discovering more issues. The proposed PWCT2 design could be influenced by advancements in other VPLs where a literature review could be done to learn about useful features introduced in the literature. By integrating these proven features from other successful VPLs, PWCT2 can provide a more flexible and user-friendly environment. Most VPLs and systems are developed using textual programming languages; for example, the first generation of PWCT was developed using Visual FoxPro and Envision was developed using C++. Self-hosting PWCT2 is crucial, as it allows the development and modification of the PWCT2 environment using the same visual programming tools that it provides to users. This means that developers can update and enhance PWCT2 through visual programming rather than writing textual code, making the process more intuitive and accessible [89].

We assume that using the proposed Ring programming language to develop PWCT2 could yield better results. The choice to transition from Visual FoxPro to the Ring programming language for the second generation of PWCT was driven by several key factors. Both Visual FoxPro and the proposed dynamic languages are designed to support object-oriented programming (OOP), and each comes with an integrated development environment (IDE), a Graphical User Interface (GUI) framework, and a form designer, making them both suitable for many similar programming tasks. However, the proposed dynamic programming language could distinguish itself with the advantage of compatibility across multiple modern operating systems, which ensures that the new generation of PWCT can operate efficiently on various systems. The proposed dynamic language could include features and libraries that are specifically designed to be used in developing PWCT2. This decision not only leverages Ring's strengths but also provides an excellent test of its features and capabilities, particularly since PWCT2 is an advanced project. [90,91].

## 2.4 Chapter Summary

In this chapter, we introduced dynamic programming languages, highlighting their unique and interesting features. Following this, we explored visual programming languages and discussed their main characteristics. These two types of programming languages offer distinct approaches and benefits to developers. In the next chapter, we will delve into the literature review, examining relevant studies, theories, and works that underpin the concepts discussed throughout this thesis.

# Chapter 3: Literature Review

## 3.1 Introduction

In this thesis, we develop and introduce two programming languages: the Ring dynamic programming language and the PWCT2 visual programming language. This chapter presents the existing research and developments relevant to the design of dynamic and visual programming languages.

We begin by exploring the literature related to dynamic language design. Our review classifies the existing and related dynamic languages into various categories, providing an organized framework for comparison. We then select key dynamic languages from these categories that are most relevant to our work and conduct a comparative analysis with our proposed dynamic language, Ring. This analysis aims to identify the unique features and advancements of Ring, as well as highlight the gaps and areas for improvement within the existing literature.

Subsequently, we shift our focus to visual programming languages (VPLs). We classify related VPLs into different categories and pinpoint those that have significantly influenced our design. By comparing these selected VPLs with our proposed visual programming language, PWCT2, we aim to underscore the distinctive aspects of PWCT2 and address the research gaps that our language fills.

Through this literature review, we establish the foundation for our proposed languages and provide a thorough understanding of the existing landscape, setting the stage for the features introduced by Ring and PWCT2.

## 3.2 Related Dynamic Programming Languages

The design of the proposed programming language is associated with various categories of dynamic programming languages. In Table 3.1, we present some of the different categories along with examples of dynamic programming languages that could fit within them. It is worth noting that some programming languages can be classified in more than one category. For instance, Ruby could be classified in both categories two and three, while a language like Tcl could fall into the first three categories.

*Table 3.1* *Some categories of dynamic programming languages and examples.*

| Refs. | Category | Examples |
|---|---|---|
| [43–46] | Lightweight and Embeddable | Lua, Squirrel, Wren, etc. |
| [50,57] | Comes with Ready-to-Use Libraries | Tcl, Perl, Python, etc. |
| [33,64] | Support creating Embedded DSLs | Lisp, Ruby, etc. |
| [43,55] | Comes with Powerful IDEs | Smalltalk, Visual FoxPro, etc. |
| [40,41] | Supporting Non-English Syntax | Supernova, Citrine, etc. |
| [53–55] | Domain-specific dynamic languages | R, xBase, etc. |
| [30,92,93] | Concurrency-oriented design | Erlang, Elixir, etc. |
| [27,94,95] | Comes with a focus on Performance | Julia, Mojo, etc. |
| [70–73] | Other implementations | MicroPython, mRuby, etc. |

The first category is Lightweight and Embeddable languages [43–46], designed for ease of integration and minimal resource consumption, making them suitable for various environments. The second category includes Languages that Come with Ready-to-Use Libraries, offering a rich set of pre-built functionalities to accelerate development [50,57]. Support for Creating Embedded DSLs is the third category, providing flexibility for niche applications [33,64]. The fourth category is Languages that Come with Powerful IDEs, which enhance the development experience through robust tools and features [32,55]. The fifth category encompasses Languages Supporting Non-English Syntax, broadening accessibility for developers worldwide [40,41]. Domain-Specific Dynamic Languages form the sixth category, tailored for fields to optimize efficiency and effectiveness [53–55]. The seventh category focuses on Concurrency-Oriented Design, managing simultaneous tasks crucial for high-performance applications [30,92,93]. Languages with a Focus on Performance make up the eighth category, ensuring rapid execution and responsiveness [27,94,95]. The ninth category, Other Implementations, includes different implementations of popular dynamic languages that focus on supporting microcontrollers or embedded systems. These implementations could be developed by the same team that created the original language (like mRuby) or by another team of developers who take the original language implementation and modify it by adding or removing features or changing the implementation (like MicroPython) [70–73]. In Table 3.2, we present the key features of our proposed language and its connections to some other dynamic programming languages.

*Table 3.2* *The main features of the proposed dynamic programming language.*

| Criteria | Lua [45] | Python [96] | Ruby [97] | VFP [55] | Supernova [41] | Proposed Language (Ring) |
|---|---|---|---|---|---|---|
| Open Source | √ | √ | √ | X | √ | √ |
| Portable | √ | √ | √ | * | * | √ |
| Lightweight | √ | * | * | X | √ | √ |
| Embeddable | √ | √ | √ | X | X | √ |
| Dynamic Typing | √ | √ | √ | √ | √ | √ |
| Function like Eval() | √ | √ | √ | √ | X | √ |
| Classes Concept | * | √ | √ | √ | X | √ |
| Inheritance Concept | * | √ | √ | √ | X | √ |
| Private Attributes | * | * | √ | √ | X | √ |
| Batteries Included | * | √ | √ | √ | * | √ |
| IDE | * | √ | * | √ | * | √ |
| Form Designer | * | * | * | √ | * | √ |
| Non-English Syntax | * | * | * | * | √ | √ |
| Case insensitive | X | X | X | √ | √ | √ |
| 1-based indexing | √ | X | X | √ | √ | √ |
| Change Keywords | X | X | X | X | X | √ |
| Internal DSL | √ | √ | √ | √ | X | √ |
| IDSL (Custom Syntax) | X | X | X | X | X | √ |
| Visual Implementation | X | X | X | X | √ | √ |
| VI Based on CPWCT | X | X | X | X | X | √ |
| Desktop | √ | √ | √ | √ | √ | √ |
| Web | √ | √ | √ | √ | X | √ |
| WebAssembly | * | √ | √ | X | X | √ |
| Microcontroller | * | * | * | X | X | √ |
| No-GIL | √ | * | * | X | X | √ |
| Register based VM | √ | X | X | X | X | X |
| Off-side rule | X | √ | X | X | X | X |
| xBase (Database DSL) | X | X | X | √ | X | X |

Deliberately, we chose at least one programming language from each relevant category (the first five categories) to provide a broader context for our language design. As we compare these languages, if a feature is absent in the basic distribution but available through external libraries, tools, or ongoing projects, we denote it with a (star) in the corresponding cell.

The Lua programming language stands out due to its compact language features and efficient implementation, which are written in ANSI C. It serves as an embeddable language, making it suitable for integration into projects that require scripting capabilities through a relatively fast scripting language. Lua is commonly used for scripting in game development. Notably, Lua does not include the concept of classes but instead emulates object-oriented concepts through its small and extensible language features. Additionally, Lua lacks a rich standard library [43–46]. Some programming languages, such as Squirrel and Wren, are designed to compete with Lua for game scripting. They use a different syntax based on braces and provide direct support for classes. Similar to Lua, they also lack a rich standard library. Although Wren seems no longer under active development, a smaller version of the language called Lox is used to introduce how interpreters are developed, allowing us to observe the usage of dynamic languages in introducing compiler and virtual machine concepts [47]. On the other hand, we have another lightweight scripting language called Tcl, which was introduced five years before Lua. While not as lightweight as Lua, Tcl comes with a rich standard library. Tcl is known for its command-based syntax—where everything is treated as a string—and its popular GUI library (Tcl/Tk), which is used in other programming languages like Perl, Python, and Ruby for GUI tasks [48–52].

Microsoft Visual FoxPro (VFP) is a fast-commercial dBase dialect that natively supports object-oriented programming. It includes a powerful IDE with auto-complete features and a GUI builder (Form Designer) like Visual Basic. However, the latest release of the language (Visual FoxPro 9.0 SP2) is a 32-bit Windows product and is no longer actively developed [55].

Python is immensely popular. Although not as lightweight as Lua or Tcl, Python boasts a rich standard library and supports various programming paradigms. It has found widespread use in scientific computing and machine learning. Python comes with an integrated development environment (IDE) called IDLE, although it lacks a Form Designer/Builder, which exists in Visual FoxPro. While such tools exist for Python through external libraries and tools, having such features in the standard tools of the language could increase its usage in GUI development, especially since most standard IDEs for desktop platforms come with these features [56–58].

The Ruby programming language is an example of a dynamic language that is used to create DSLs. This reduces the development cost, and the learning curve required to create a DSL. Unfortunately, these DSLs will not look like Supernova or SQL because they are influenced by the Ruby syntax [64,65].

With respect to syntax translation, numerous projects have attempted to address this gap. For example, several Arabic programming languages (Like Supernova) have been developed with a focus on using Arabic syntax. However, most of these projects remain unused in production due to limited features and are no longer actively developed [37–42]. Another approach involves creating packages that introduce translation as a feature. For instance, zhpy is a Python package that enables the writing of source code using traditional Chinese keywords, which is then converted to Python. However, this approach can suffer from multiple issues, including additional development and testing efforts, as well as lower compile-time performance due to the extra layer of translation before invoking the Python interpreter. Some programming language designers have recognized this challenge and intentionally introduced syntax localization. For instance, the Supernova programming language supports both Arabic and English syntax simultaneously, allowing an easy way to share libraries written in different languages. However, adding translation support for additional human languages without modifying the language implementation remains a complex task. In contrast, the Citrine programming language provides multiple versions that support over 100 human languages. Unfortunately, it does not offer an easy mechanism for sharing code across these language versions within the same project, as sharing code requires translation [39–41].

The implementation of dynamic programming language virtual machines could use a Global Interpreter Lock (GIL) to ensure that only one thread can access the interpreter at a time. This provides safety and avoids race conditions but prevents better performance from using threads on multi-core systems for CPU-bound tasks [59–61]. For a programming language like Python, there is ongoing work towards removing the GIL in recent versions.

The proposed dynamic programming language is designed to incorporate and improve features related to the first five categories: Lightweight, Embeddable, Scripting, and Batteries-Included, providing powerful support for embedded DSLs, and it comes with an IDE suitable for GUI development. Additionally, it provides syntax flexibility and supports non-English syntax. Importantly, it does not belong to other categories, such as domain-specific languages, concurrency-oriented languages, or those focused solely on performance. While the proposed language is not a domain-specific language itself, it could be used to create domain-specific languages. Additionally, while it is not specifically designed around concurrency or performance, its proposed implementation—using a VM without a GIL—allows the use of threads to improve the performance of CPU-bound applications. The proposed programming language is designed to have a small implementation and provide direct support for multiple programming paradigms in the first place, then be fast enough and provide better runtime performance.

## 3.3 Related Visual Programming Languages

Languages like Visual Basic and Visual C# fall into the category of textual programming languages, not visual programming languages (VPLs). Programmers need to write text-based code using the specific syntax of these languages to develop large and complex real-world applications. Tools like Microsoft Visual Studio 2022 are known as integrated development environments (IDEs) rather than VPLs. These environments allow software developers to create portions of applications using visual elements, but textual code is essential to achieve full control over the application's functionality. In contrast, VPLs rely exclusively on visual components without the need to write textual code directly. In text-based programming languages like C++ and Java, the code is linear and one-dimensional, with the compiler processing it token by token. In contrast, VPLs utilize graphical representations that span multiple dimensions. Each graphical element occupies a specific position within a 2D or 3D space and can have unique shapes, colors, and images. Various relationships can be depicted among these objects, such as being inside, outside, touching, or adjacent to one another. Additionally, some visual languages incorporate the time dimension (before/after) to further enhance the graphical code representation [11,78].

A visual programming language can have several different representations. One option is diagrammatic, which uses shapes and text components with links to connect shapes and illustrate control or data flow. Another representation is iconic, which utilizes icons derived from the problem's domain. There are also form-based representations that incorporate forms like spreadsheets or data-entry forms and block-based representations that feature blocks assembled to form programs. Additionally, a VPL can be hybrid, combining elements of any of these four representations [14,15].

The design of the proposed visual programming language is associated with various categories of VPLs. In Table 3.3, we present some of the different categories based on their visual representations or usage scope, along with examples of visual programming languages that fit within them. It is worth noting that some VPLs can be classified into more than one category. For instance, Scratch could fall into both the first and fifth categories, as it is a visual programming language that uses block-based programming and is designed for use in education and teaching children about programming.

***Table 3.3*** *Some categories of visual programming languages/systems and examples.*

| Ref. | Category | Examples |
|---|---|---|
| [98–100] | Block-based | Scratch, Snap!, etc. |
| [101–104] | Diagrammatic | Tersus, RAPTOR, etc. |
| [105–108] | Iconic | Kodu, Limnor, etc. |
| [109–111] | Form-based and spreadsheet-based | Forms/3, FAR, etc. |
| [112,113] | Domain-specific | Blueprints, Pure Data, etc. |
| [114,115] | General-purpose | PWCT, Envision, etc. |

In Table 3.4, we present the key features of our proposed visual programming language and its connections to some other visual programming languages from different categories. As we compare these visual languages, if a feature is absent but available through external ongoing projects, we denote it with a (star) in the corresponding cell.

We selected Scratch because it is a popular VPL for education that uses blocks-based programming. Forms/3 is chosen as an example of form-based and spreadsheet-like programming, with support for the time dimension. Lava is included because it is a VPL that uses the TreeView control and supports object-oriented programming (OOP). Envision is selected as a research prototype for a general-purpose visual programming system that features interactive visualizations. Finally, PWCT is chosen because it is a general-purpose VPL used in advanced projects, such as developing the Ring programming language.

***Table 3.4*** *The main features of the proposed visual programming language.*

| Criteria | Scratch [116] | Forms/3 [117] | Envision [118] | Lava [119] | PWCT [120] | Proposed VPL (PWCT2) |
|---|---|---|---|---|---|---|
| Open Source | √ | X | √ | √ | √ | √ |
| Portable | √ | X | √ | √ | X | √ |
| Rich Colors | √ | X | √ | X | X | √ |
| Time Dimension | X | √ | X | X | √ | √ |
| Auto-Run | √ | √ | X | X | X | √ |
| Rich-Comments | X | X | √ | X | X | √ |
| Interactive Visualization | X | X | √ | X | X | √ |
| Self-hosting | X | X | * | X | X | √ |
| Form Designer | X | √ | X | √ | √ | √ |
| Steps Tree/Blocks | √ | X | X | √ | √ | √ |
| Steps Tree/Blocks (DAD) | √ | X | X | X | X | √ |
| Play programs as movie | X | X | X | X | √ | √ |
| Supports OOP | X | X | √ | √ | √ | √ |
| Children Oriented | √ | X | X | X | X | X |
| Research Oriented | X | √ | √ | √ | X | X |

In the efforts made by the researchers to make Envision a self-hosting visual programming environment, significant strides were achieved. A code generation framework was designed and implemented to represent macros in Envision when importing code from C++. Additionally, an extra stage in the existing C++ import system was developed to facilitate macro import by reconstructing them from expanded code using preprocessor information from Clang. Although the authors reported that time constraints and issues in existing components prevented the complete achievement of the goal, it is important to note that using C++ for Envision implementation introduces some challenges in implementing a self-hosting VPL because of the numerous features and preprocessor usage [89].

Multiple research studies highlight the importance of ensuring the construction of correct programs through visual programming languages. This addresses key aspects of program correctness and reliability, which are critical for enhancing both usability and the practical effectiveness of visual programming environments. PWCT provides two modes of operation. The first mode follows the concept of a syntax-directed editor and prevents composition errors when connecting components. The second mode uses a free editor where mistakes can occur and are then detected by the compiler. In PWCT2, the visual editor prevents composition errors, and the user can use the customization window to allow or disallow errors when typing expressions in the interaction page [16,121].

The first generation of PWCT is influenced by Lava 0.7.2 (using a TreeView control to represent the program structure) and Forms/3 (using the Time Dimension). It introduces new features like the Graphical Code Replacement (GCR) method (instead of drag-and-drop) and playing programs as a movie using the Time Dimension. The proposed visual programming language (PWCT2) builds on PWCT by incorporating the GCR method, Steps Tree, Time Dimension, and playing programs as a movie. It also draws inspiration from Scratch, incorporating rich colors and block-level drag-and-drop support. Additionally, PWCT2 is influenced by the Envision visual programming system, enabling rich comments and supporting interactive visualization.

The proposed VPL is implemented using the Ring language, supports Ring code generation, and enables importing Ring code, making it a self-hosting VPL. These capabilities are particularly important in the age of large language models (LLMs) and code generation, as they enable the use of LLMs to generate Ring code that can be used in PWCT2 and updated using visual programming. Since Forms/3, Lava, and Envision are no longer under active development, and while Scratch is actively developed, it is domain-specific, and PWCT, though general-purpose, is designed for MS-Windows, we expect that the proposed VPL, with its support for multiple platforms and modern features, could be a valuable addition to the landscape of VPLs. It could be especially useful for Ring programmers or novice programmers who want to learn about the Ring language through visual programming, as this proposed VPL is the first to support the Ring programming language.

While visual tools such as Blockly 2022 and Node-RED 4.0.2 are very popular and serve as the foundation for many visual programming languages, we believe that the PWCT2 approach and its interactive textual-to-visual code conversion offer notable flexibility. This could attract more users with coding backgrounds to try the PWCT2 visual programming approach. Additionally, such features could influence future updates to Blockly and Node-RED if more users find them useful and necessary [85,86,122,123].

No-code platforms provide highly intuitive drag-and-drop interfaces and prioritize rapid development, allowing non-technical users to quickly build entire applications without any coding knowledge. These tools simplify the development process for specific application types, making them accessible to a wider audience. While visual programming languages (VPLs) provide a visual approach to traditional programming [124,125].

Both PWCT2 and no-code development tools aim to simplify the creation and development of software and applications, but they serve different needs. For example, PWCT2 focuses on making traditional coding more accessible by providing flexibility and customization for a wide range of applications. Through further development, the PWCT2 visual programming language can act as an intermediate-level abstraction layer between traditional coding and no-code. General-purpose visual programming languages like PWCT2 can be used as the foundation for building no-code platforms, enabling higher levels of abstraction and ease of use while maintaining full control through visual programming.

## 3.4 Chapter Summary

In this chapter, we conducted a thorough literature review, classifying and comparing existing dynamic and visual programming languages. Through this analysis, we have determined the key characteristics and unique features of both the proposed dynamic programming language and the proposed visual programming language, highlighting the research gaps they address.

- Most of the dynamic languages are developed using textual programming.
- There are few studies on developing a language with a lightweight implementation and rich features.
- Few programming languages offer built-in support for easy translation.
- Embedded DSLs doesn't resemble external DSLs, such as CSS or Supernova.
- There is limited research on the use of VPLs in large and complex system projects.
- Many GPVPLs are no longer under active development.
- Importing textual code is uncommon or incomplete in most VPLs.
- The Time Machine in PWCT does not support the Auto-Run feature.
- There are no self-hosting GPVPL.

In the next chapter, we will delve into the design and implementation of the Ring programming language. We will explore its most important features and contributions, demonstrating how it advances the current state of dynamic programming languages.

# Chapter 4: The Ring Programming Language

## 4.1 Introduction

The primary aim of developing the Ring programming language, as demonstrated in Figure 4.1, is to use visual programming to develop a lightweight and embeddable dynamic programming language and environment that facilitates easy and rapid translation of language syntax. Additionally, the language will empower developers to create embedded domain-specific languages (DSLs) resembling external DSLs like CSS and Supernova. The language is a multi-paradigm, providing direct support for object-oriented concepts such as classes, objects, encapsulation, and inheritance. Furthermore, our language offers cross-platform support for desktop, web, WebAssembly, and 32-bit microcontrollers—all using a unified implementation. This implementation is based on a visual programming design that generates ANSI C code for the bytecode compiler and the virtual machine. As a “batteries-included” language, it comes with rich libraries and tools, including an integrated IDE with a form designer. This chapter addresses the first research question (RQ1).

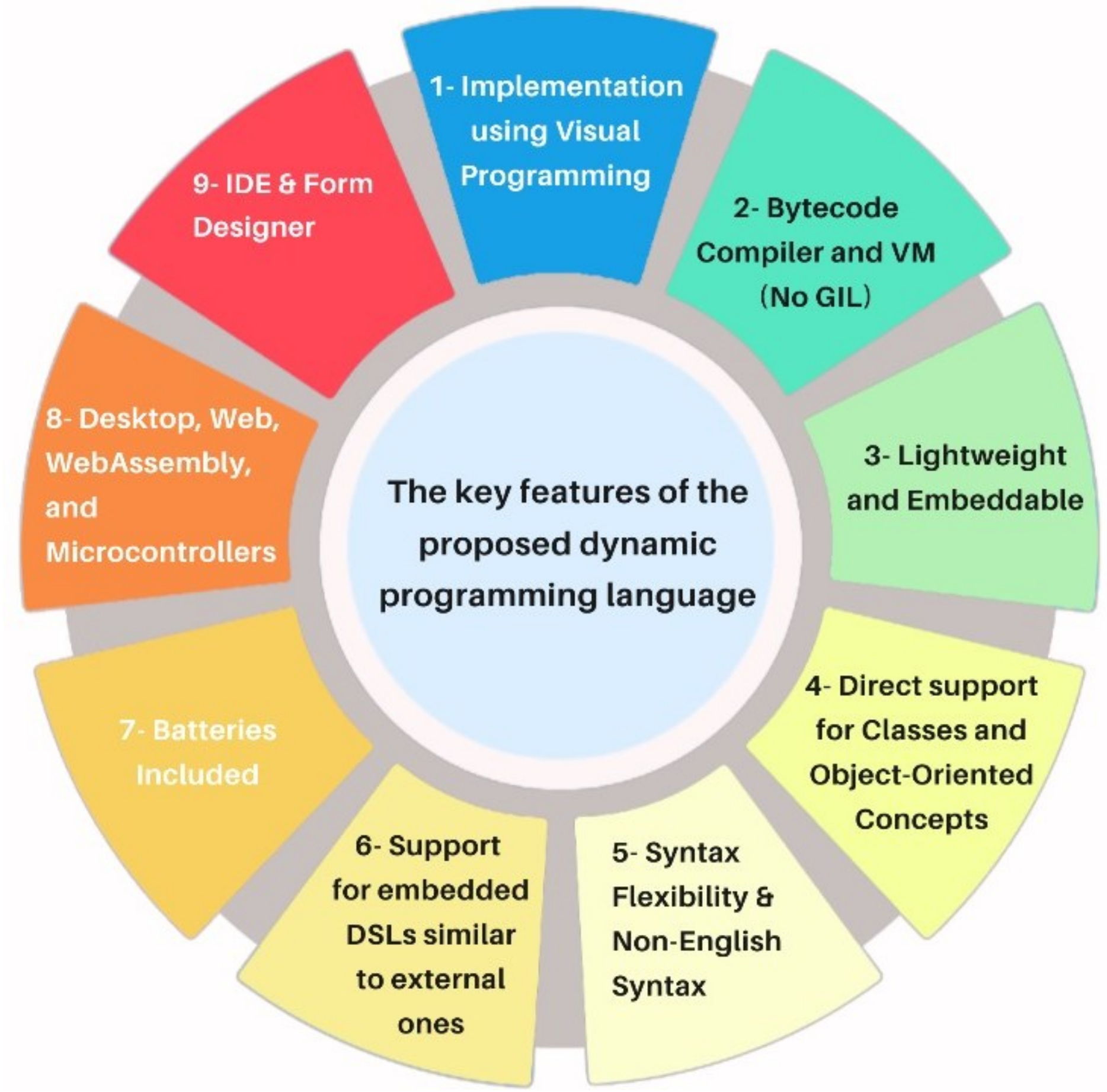

***Figure 4.1*** *The key features of the proposed dynamic language and environment.*

In this chapter, we delve into our system design and implementation. We highlight the essential features of the proposed dynamic programming language, Ring, and present the system architecture. Our focus lies on the language features that facilitate localization, syntax customization, and the development of domain-specific languages. The language has been meticulously designed to offer syntax flexibility and empower users to customize the language syntax according to their specific needs. This leads to the ability to create internal domain-specific languages (IDSLs) that look like external domain-specific languages without the need to create specific parsers for them where the language constructs will be enough to achieve this goal.

## 4.2 System Architecture

In Figure 4.2, we present the system architecture, which comprises three layers: the language layer, the batteries-included layer, and the tools layer. The language layer is closely tied to the compiler and the virtual machine implementation. It defines the core programming language features, syntax, and semantics. In the Batteries-Included Layer, we encounter various extensions and libraries that cater to different domains. These include support for GUI, databases, web development, game development, and even platforms like Raspberry Pi Pico. The tools layer encompasses both command-based utilities (such as the package manager and REPL) and graphical tools like the form designer.

In the language layer, we have visual implementation, generated code, build scripts, and automated tests. Visual Implementation is developed using Programming Without Coding Technology (PWCT) software (version 1.9) [34,66]. The generated code is written in the C programming language (specifically ANSI C) and necessitates a C compiler to build the Ring executable. Throughout development, we employed multiple compilers, as illustrated in Table 4.1. With respect to the build scripts and the automated tests, we have employed batch files and shell scripts to automate the build process. Additionally, we have a CMake file that can generate the C project for multiple compilers [126]. This file uses CMake version 3.5. After each update to the project's source code and before committing code using Git [127], we used to run a comprehensive suite of tests. This process is now automated through a Ring program that executes each test in a separate process and verifies the output against the expected results. We are using Git version 2.42.

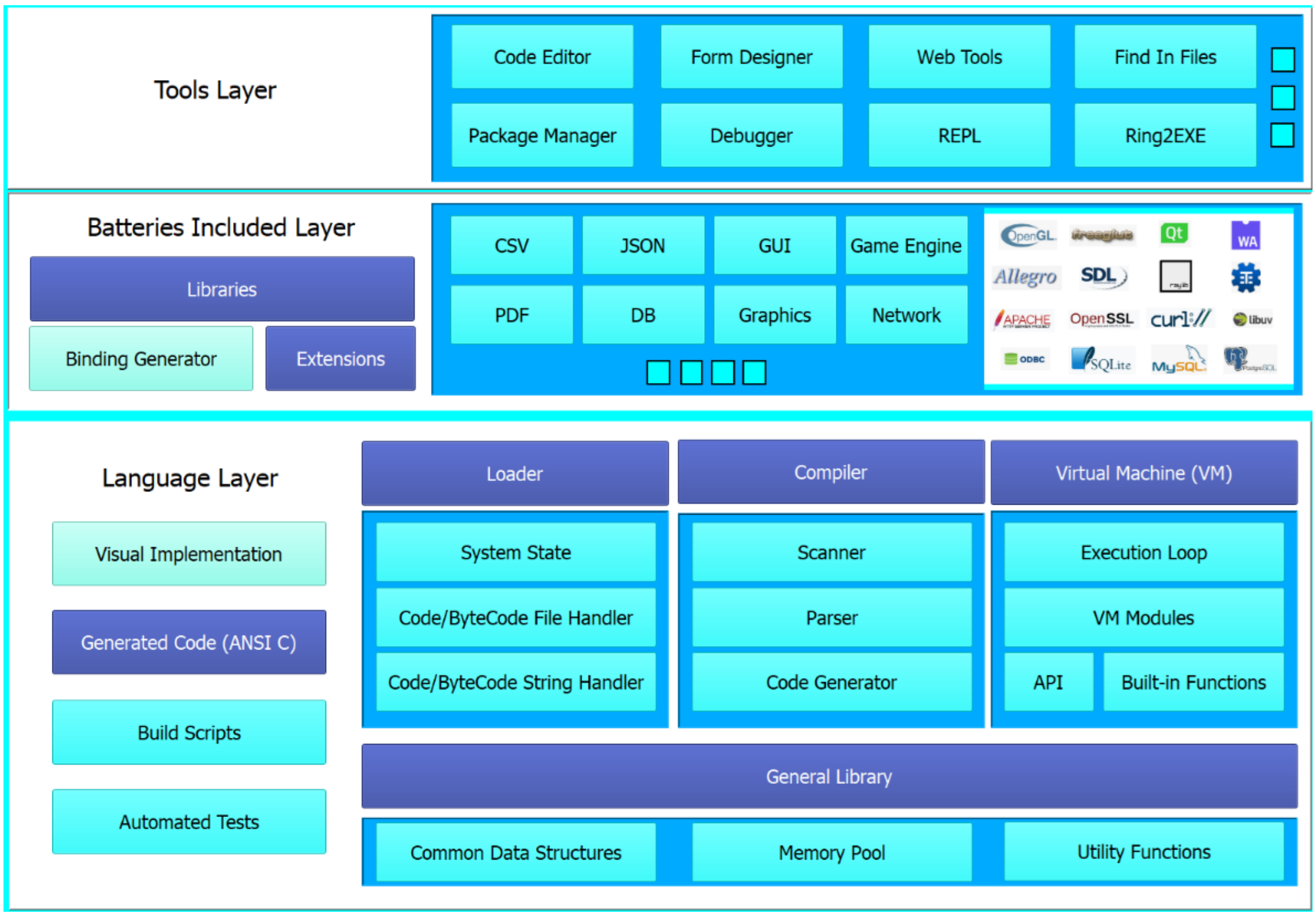

***Figure 4.2*** *The proposed system architecture.*

***Table 4.1*** *C Compilers used for building our Ring Compiler/VM.*

| C Compiler | Platform/OS (Target) |
|---|---|
| Watcom C/C++ | MS-DOS |
| Microsoft Visual C/C++ | Microsoft Windows |
| GNU C/C++ | Ubuntu Linux |
| Clang | macOS |
| Android-clang | Android |
| Emscripten | WebAssembly |
| GNU ARM embedded toolchain | Raspberry Pi Pico |

The General library provides common features used by other components, such as the loader, compiler, and VM. These features include functions for processing files and directories, especially when Ring is used on an operating system that provides a file system. Additionally, the library implements Strings, Lists, and Hash Tables. One of the crucial features offered by the library is the Memory Pool, which pre-allocates memory. The size of the pre-allocated memory depends on the environment: a few kilobytes are allocated when using Ring on microcontrollers like the Raspberry Pi Pico, while several megabytes of memory are pre-allocated when using Ring on desktop environments such as Windows, Linux, and macOS.

Since the proposed programming language can serve as both a scripting language and an embeddable language, it needs features that fulfill these dual roles. This is achieved through the loader, which plays a managerial role in our system design. The loader determines what actions will be taken and what will be avoided. It can print usage information, process source code or bytecode files, execute code from strings, halt operations at specific points (such as obtaining scanner tokens), or display applied grammar rules, among other tasks. To achieve these objectives, the loader calls the Compiler/VM components.

Since the implementation never uses C global variables, the loader also creates the system state, and the state pointer is passed to different functions that require access to common information about the processed files, the current stage, the memory pool, and so on.

In the Ring compiler, we have three main modules: the scanner, the parser, and the code generator. The scanner reads the textual source code and converts it into tokens (such as keywords, operators, identifiers, and constants). The parser processes these tokens, checking for correct adherence to the language grammar, and then invokes the code generator functions to produce bytecode. Ring employs a single-pass compiler [128], where parsing, code generation, and optimization are interleaved. However, the language performs only a few optimizations during code generation.

All these decisions are made in favor of maintaining a small implementation. The language implementation utilizes a stack-based virtual machine [24]. This virtual machine is specifically designed for the language and contains many instructions that directly map to its features. In total, there are 128 instructions within the virtual machine. The VM comes with 255 built-in functions and provides an API for extensions written in the C language.

In the Batteries Included Layer, we have a powerful tool called the Binding Generator (Like SWIG for Python [129]). Our tool is written in Ring itself and allows us to use straightforward configuration files that describe and customize the functions and classes available in C/C++ libraries. Once these configuration files are in place, the generator works by producing the extension code. This code enables us to seamlessly use those C/C++ functions and classes from within the Ring language programs. To build extensions, we can employ a C/C++ compiler, resulting in dynamic link libraries (DLLs), shared objects (SOs), or dynamic libraries (Dylibs) according to the platform (Windows/Linux/macOS) [130].

In Table 4.2, we find a list of external C/C++ libraries used by the standard Ring extensions provided by the language. The selection of these libraries is based on our experience of using them in previous projects. Most of these libraries enjoy popularity within the C/C++ community and cover various programming domains, including Database, Graphics, Multimedia, Games, Terminal, GUI, Network Programming, and Web Development.

These libraries exhibit different characteristics that impact the produced software. For instance, the Qt GUI Framework offers an extensive array of classes and features, but delving into it necessitates investing more time to study the framework [131]. Consequently, programs built with Qt may have larger runtimes. On the other hand, a lightweight GUI library like Libui has fewer features compared to Qt, but it excels in being compact. Notably, most of the GUI tools in the Tools layer are based on RingQt.

***Table 4.2*** *A list of C/C++ Libraries used by Ring Extensions.*

| Domain | C/C++ Libraries/Tools | Count |
|---|---|---|
| Terminal User Interface (TUI) | ConsoleColors and RogueUtil | 2 |
| Network and Security | LibCurl, Libuv, and OpenSSL | 3 |
| Web Servers | HTTPLib and Apache Web Server | 2 |
| Database | ODBC, SQLite, MySQL, and PostgreSQL | 4 |
| Games & multi-media | Allegro, LibSDL, RayLib and Tilengine | 4 |
| Graphics | OpenGL, FreeGLUT and StbImage | 3 |
| Graphical User Interface (GUI) | Qt, Libui, and NAppGUI | 3 |
| Common Files | MiniZip, PDFGen and CJSON | 3 |
| SDK for Specific Platforms | Android SDK and Raspberry Pi Pico SDK | 2 |

In the Tools layer, we have a group of command-based tools such as the Ring Package Manager, Ring2EXE, and the REPL. Additionally, we have GUI-based tools like the Ring Notepad, which serves as our code editor, and the Form Designer, which is used for designing application user interfaces and generating code following the MVC design pattern [132]. Furthermore, Ring includes an application for searching text in multiple files—a common feature required for large projects. All these tools are written using the Ring programming language itself. Software documentation helps users learn and use it.

The language is distributed with documentation of over 2000 pages in the English language that cover the different features and concepts. Also, there are chapters that cover the different extensions, libraries, and tools provided by the language. The documentation is created using Sphinx, a Python-based documentation tool. Specifically, we are using Sphinx version 6.2.1, HTML Help Workshop version 4.74.8702, and MiKTeX 23.4.

## 4.3 Non-English Syntax

The language scanner (The first phase in the compiler) supports specific commands (illustrated in Table 4.3) that allow users to change the language keywords and operators multiple times, facilitating easy translation of the language syntax.

***Table 4.3*** *Scanner commands provided by the Ring Compiler.*

| Command | Parameters | Usage |
|---|---|---|
| ChangeRingKeyword | OldKeyword NewKeyword | Change language keyword |
| ChangeRingOperator | OldOperator NewOperator | Change language operator |
| LoadSyntax | Syntax file name as literal | Load syntax file |
| EnableHashComments | None | Support using # for comments |
| DisableHashComments | None | Disable using # for comments |

In Figure 4.3, we present a WebAssembly application developed using Ring for online language experimentation. Additionally, we provide an example of how Scanner commands can be used to switch language keywords to Arabic syntax. The code begins by translating keywords (such as put, get, if, elseif, and endif) from English to Arabic. Subsequently, it employs this Arabic syntax to create a program that prompts the user for their age and delivers a message based on that input.

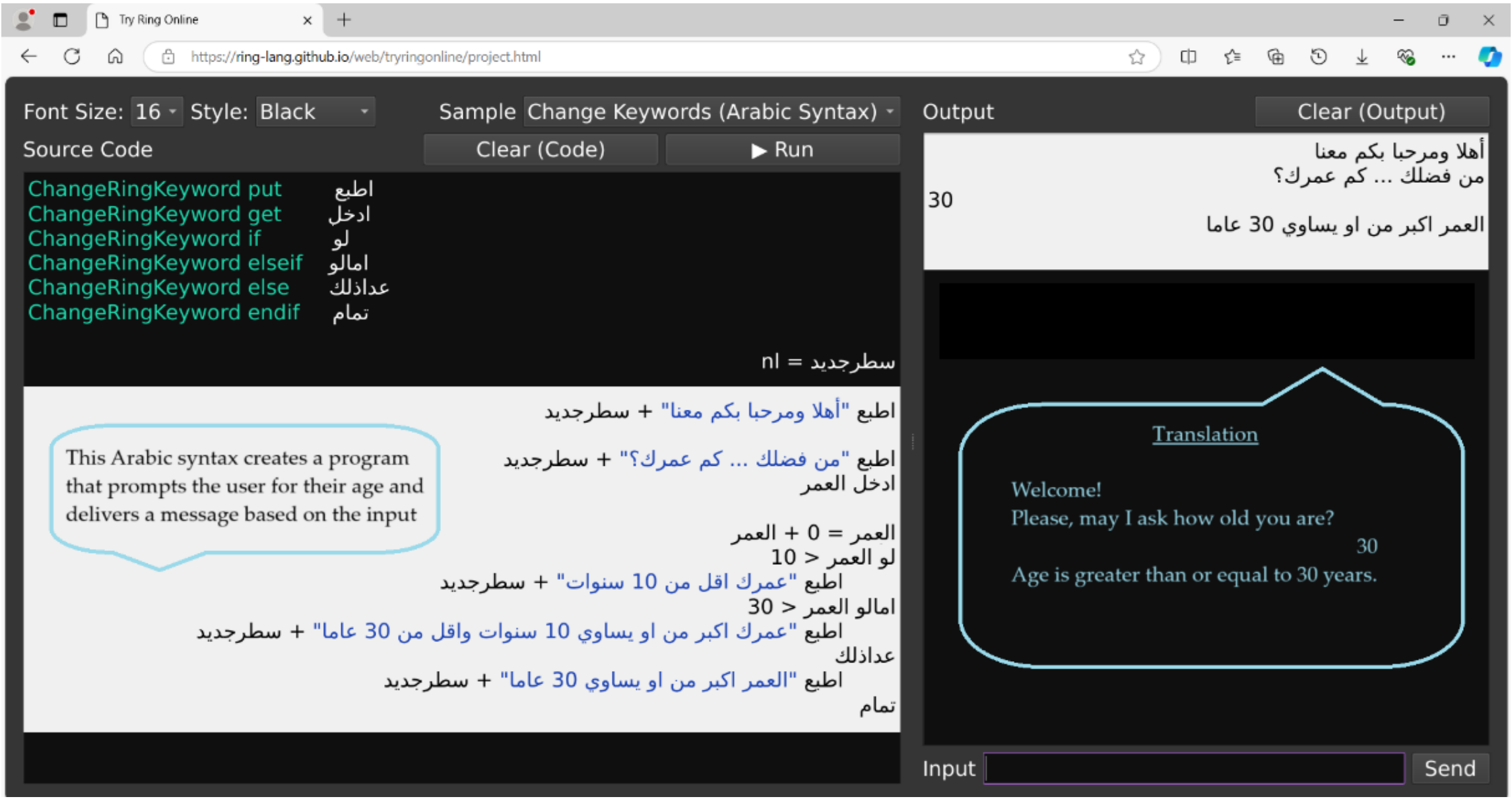

***Figure 4.3*** *Arabic syntax within a WebAssembly application developed using Ring.*

Rather than including these Scanner commands at the start of every Arabic source code file, we can use the LoadSyntax command. This command allows us to load syntax files containing groups of these commands. Additionally, instead of placing the LoadSyntax command at the beginning of each source code file, we can simply add a file named (ringsyntax.ring) to the Arabic project folder. The Scanner will automatically load this file whenever we use any Ring source code file in the same folder.

This approach draws inspiration from the use of (__init__.py) files in Python modules and the concept of (.htaccess) files in the Apache HTTP Server [133,134].

The application is developed using Ring through the following steps:

1. The user interface is designed using the Ring form designer. The form file (try.rform) generates the tryView.ring file, which contains the RingQt source code that defines the window controls and layouts and sets the default style;
2. In the controller class (tryController.ring), we determine the Ring code that will be executed based on user interaction with the application GUI;
3. The (style.ring) file contains the Style class, which changes colors based on the selected style. The default style is Black, and the user can change it to another predefined style (Black, White, Blue, Modern, or Windows);
4. The (samples.ring) file contains a Ring list that provides the predefined samples, where each sample is represented through a nested list containing the sample name and the sample code;

5. The (ringvm.ring) file contains a class that enables the sample's source code to be run in an isolated Ring virtual machine. If the sample produces a runtime error or terminates the program, we can still use the Try Ring Online application executed by the caller VM;
6. The (onlinering.ring) file contains functions that are automatically called before executing any sample code in the nested VM. These functions override the standard functions used for input/output operations. When the console application requires an input, these functions will pause the sub VM. Later, when the user enters the required data, the caller VM will update the state of the sub VM, set the variable value, and then resume the sub VM.

## 4.4 Domain-Specific Languages

One of the features provided by the Ring language is the ability to create domain-specific languages (DSLs) on top of classes. These DSLs can employ specific syntax, and we have the freedom to design this syntax. The underlying idea relies on using braces to access objects, granting us the ability to utilize the attributes and methods provided by those objects. This section addresses the fourth research question (RQ4).

Unlike some other programming languages that offer the "with" statement, in Ring, this feature is provided through an operator. This operator allows us to use this feature within expressions and in various places throughout the code. Notably, Ring does not require semicolons or new lines between statements. We can type different statements on the same line without any fuss. Additionally, in Ring, every valid expression is an acceptable statement, giving us the freedom to write various values, all of which will be accepted by the compiler.

Ring classes also support properties. Typing a property name can invoke the getter method and execute the associated code. Moreover, Ring goes a step further by allowing us to define methods like braceStart() and braceEnd(). These methods are automatically called when we access an object using braces. Furthermore, the language automatically invokes a method called braceExprEval() when we write an expression inside braces. With these features, coupled with the ability to customize language keywords and operators, we can construct domain-specific languages that resemble external DSLs such as CSS, QML, SQL, and Supernova.

As an example, we will implement a tiny DSL. This simple DSL accepts a group of numbers. While entering numbers, we can highlight some of these numbers as important. Additionally, we could stop the processing using the “stop” command. The results of the process include both the summation of the entered numbers and a list of the important numbers. When typing numbers, using new lines is optional. Also, we do not need to use () or [] to group these numbers.

Table 4.4 presents an example of how to use this tiny DSL and the expected output. In the example, we entered a group of numbers while asking for some of them to be highlighted using the (Important) word. Then, we decided to stop processing after the number 60 using the (Stop) word.

***Table 4.4*** *A simple domain-specific-language implemented using Ring.*

| Usage (Ring Code) | Output |
|---|---|
| new DSL {<br>200<br>400 Important<br>50<br>600 Important<br>60<br>10 20 30<br>40 50 60 Stop<br>70 80 90<br>800 Important<br>} | Sum: 1520<br>Important:<br>400<br>600 |

To implement this tiny DSL, we only need to write one Ring class containing seven lines of compact code. Figure 4.4 demonstrates the implementation and analysis of the Ring features used in this class code.

```
class DSL

 nSum = 0        lastvalue
 stop = False    important    aImportant = []

 func getStop()               stop        = True
 func getImportant()          if !stop    aImportant + lastvalue    end
 func braceExprEval(value)    if !stop    nSum += value             end     lastvalue = value
 func braceEnd()              ? "Sum: "+nSum ? "Important: " ? aImportant
```

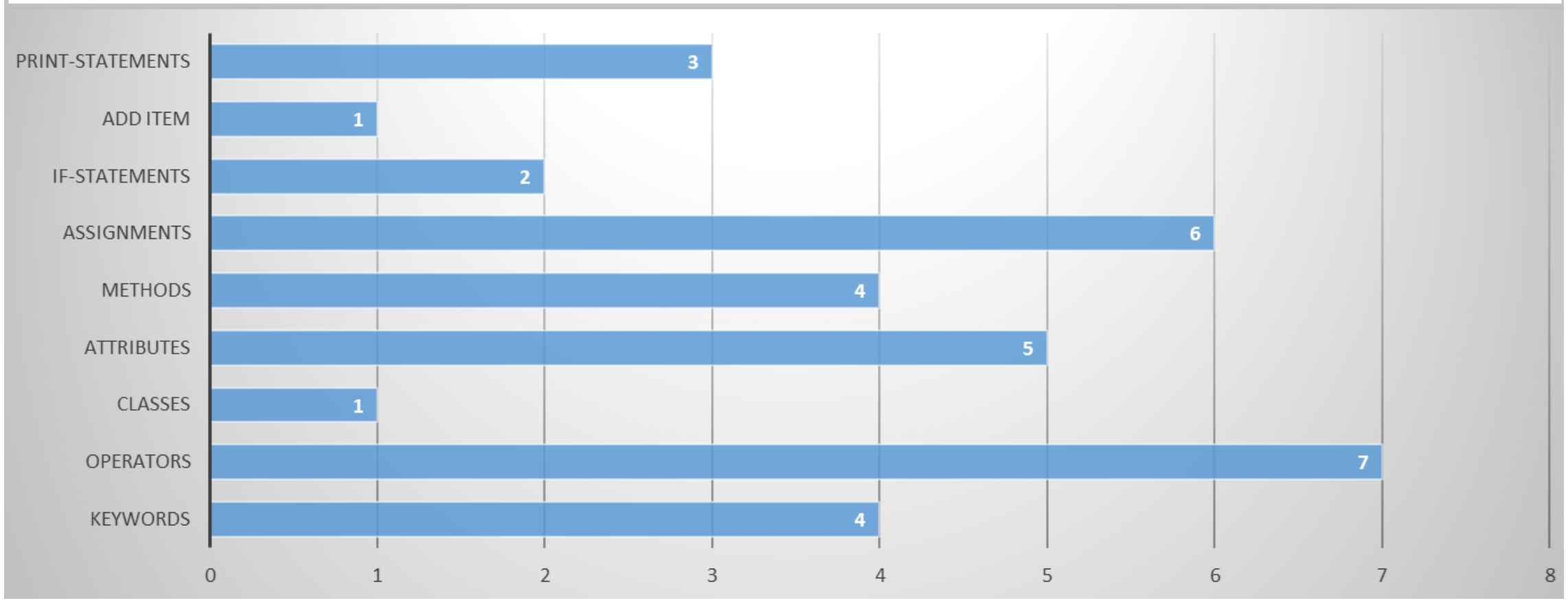

***Figure 4.4*** *Ring code to implement a simple domain-specific language.*

We begin by defining the class and setting its name (DSL). After that, we declare five attributes within our class: nSum, aImportant, important, stop, and lastvalue. Since our tiny DSL has two commands (Important and Stop), after declaring them as attributes, we define the methods getImportant() and getStop() to determine what happens when these words are used. The braceExprEval() method is called each time the language processes a number inside the braces that access the object. We sum up these numbers by adding them to the nSum attribute, and we store the last encountered number in the lastvalue attribute. Finally, when we finish using braces, the braceEnd() method is automatically called, printing the results.

## 4.5 Object Oriented Programming

The Ring programming language provides direct support for many features related to the object-oriented programming paradigm [135], such as classes, objects, encapsulation, composition, aggregation, inheritance, polymorphism, and operator overloading. Additionally, we have extended these features by introducing new capabilities, such as using braces to access objects (braceStart(), braceEnd(), braceExprEval(), etc.) to facilitate the implementation of domain-specific languages (DSLs) using classes.

This allows us to seamlessly blend our DSL implementation with the well-known features of object-oriented programming. For instance, consider Figure 4.5, where we introduce an update to our domain-specific language. This updated DSL supports retrieving groups of computer prices and highlighting acceptable prices while displaying the output through a Graphical User Interface (GUI). The figure contains four sections. The first section represents the data used as input for the object created from the PickPrice class. The second section presents the PickPrice class. The third section is related to the DSL class. The fourth section, on the right side of the figure, shows the program's output: a simple GUI window containing a list box.

We define a new class called PickPrice, which inherits from the existing DSL class (still present in our code, starting from line 25). Inside the PickPrice class, we introduce a new attribute called Acceptable (used in place of the word "Important"). We define the getAcceptable() method, which is called when the Ring language encounters the word "Acceptable" within code, which uses braces to access an object created from the PickPrice class. Essentially, it acts as a wrapper method, invoking the existing getImportant() method from the parent class (DSL) to reuse its functionality.

```
1  new PickPrice {
2      Name                        Price
3      Computer1                   7000                  Acceptable
4      Computer2                   4400                  Acceptable
5      Computer3                   9000
6  }
7
8  class PickPrice from DSL
9      Acceptable func getAcceptable() getImportant()
10     func braceEnd
11         load "guilib.ring"
12         import System.GUI
13         app = new App {
14             win = new Window() {
15                 setWindowTitle("Acceptable Prices")
16                 resize(400,200) setWinIcon(self,"bestprice.png")
17                 list = new ListWidget(win) { addList(Sort(this.aImportant)) }
18                 setLayout( new VBoxLayout() { addWidget(list) } )
19                 show()
20             }
21             exec()
22         }
23     func braceError
24
25 class DSL
```

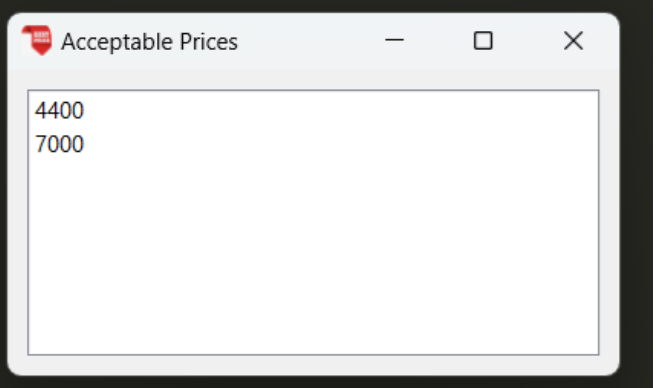

***Figure 4.5** Extending our DSL using inheritance and the GUI library.*

At line 10, we override the braceEnd() method implementation. Instead of the command-based user interface, we replace it with a GUI. The GUI code leverages the GUILib library provided by the Ring programming language, utilizing the popular Qt framework. In line 17, we sort the numbers in our list (aImportant) using the Sort() function and then add the items to a ListWidget. Line 23 introduces the braceError() method, designed to prevent errors when typing words like “Name”, “Price”, “Computer1”, etc.

While we are inside class methods, using braces to access objects changes the current object. Consequently, we need different variables to reference the opened object or the object instance created from the current class. That is why we have (self) and (this). In line 16, while we are inside the window object, we pass this object to the setWinIcon() function using the (self) variable. In line 17, to access the aImportant attribute (inherited from the DSL class), we use the (this) variable to refer to the object instance that will be created from the PickPrice class.

## 4.6 Batteries Included

The language ships with numerous extensions that wrap up popular C/C++ libraries such as Allegro, LibSDL, OpenGL, Qt, and SQLite. These extensions provide convenient bridges between Ring and these well-established libraries, making it easier for developers to harness their capabilities.

Ring libraries add another layer of abstraction on top of Ring extensions. One standout example is the Game Engine for two-dimensional (2D) Games. This engine consists of a set of classes built around the Allegro and LibSDL game programming libraries [136,137]. What is interesting is that the engine encourages a declarative coding style reminiscent of CSS or QML. So, when working with it, we can express the game logic in a way that feels quite intuitive and expressive.

The language comes with an extension called RingPico, which supports Raspberry Pi Pico SDK [138]. In Figure 4.6, we see an example of using this extension. We can employ procedural programming and directly call the functions provided by the extension, much like we would when coding in a language such as C. Additionally, we have the option to build classes around these functions or leverage the features of the Ring language for more declarative code as demonstrated in the Figure.

In this example, we load the file circuit.ring, which contains the classes Circuit, LED, and LEDSwitch. Inside this file, an object called Circuit is created from the Circuit class as a global variable. We can access this object directly using braces, as shown in Line 8.

Once we are inside the object, using words like LED or LEDSwitch is equivalent to accessing attributes defined in the Circuit class.

```
1   load "circuit.ring"
2
3   SWITCH_PIN = 14
4   LED_PIN    = 15
5
6   func main
7
8       Circuit {
9           LED {
10              Pin = PICO_DEFAULT_LED_PIN
11              Blink   = True
12              Delay   = 0.1
13          }
14          LEDSwitch {
15              Pin = SWITCH_PIN
16              LED {
17                  Pin = LED_PIN
18                  Blink   = True
19                  Delay   = 3
20              }
21          }
22      }
```

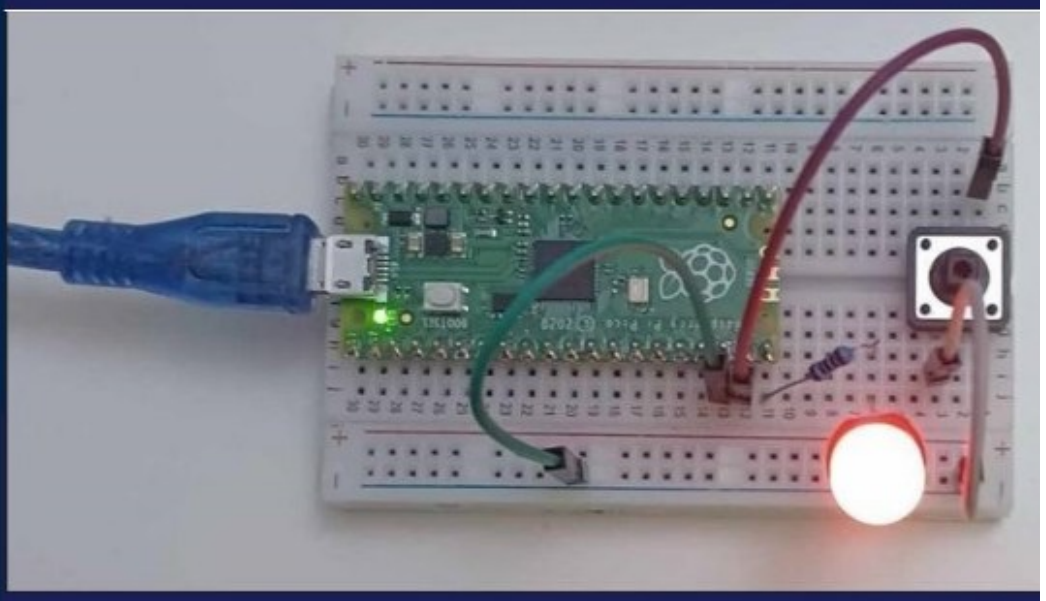

***Figure 4.6*** *Using Declarative Style in Ring for Raspberry Pi Pico programming.*

Using these attributes will call the getLED() and getLEDSwitch() methods. Invoking these methods creates new objects from the LED and LEDSwitch classes, which are then added to a list of objects defined inside the Circuit class. Additionally, the getLED() and getLEDSwitch() methods return these newly created objects. We can access these objects using braces, just as we did in lines 9, 14, and 16. Once we have access to an object, we can set its attributes. The additional magic related to the execution loop and responding to attribute values (such as Pin, Blink, and Delay) is handled through the braceEnd() method. This method initiates the execution loop and calls other methods that check the defined objects and their attributes.

## 4.7 The IDE and the Form Designer

We developed GUI-based tools (Demonstrated in Figure 4.7) like the Ring Notepad, which serves as our code editor, and the Form Designer, which is used for designing application user interfaces and generating code following the MVC design pattern [134]. Furthermore, Ring includes an application for searching text in multiple files—a common feature required for large projects. All these tools are written using the Ring programming language itself, totaling around 15,000 lines of Ring code. Developing these tools based on the Qt framework requires knowledge of GUI development, object-oriented programming, and how to organize large programs in the Ring language.

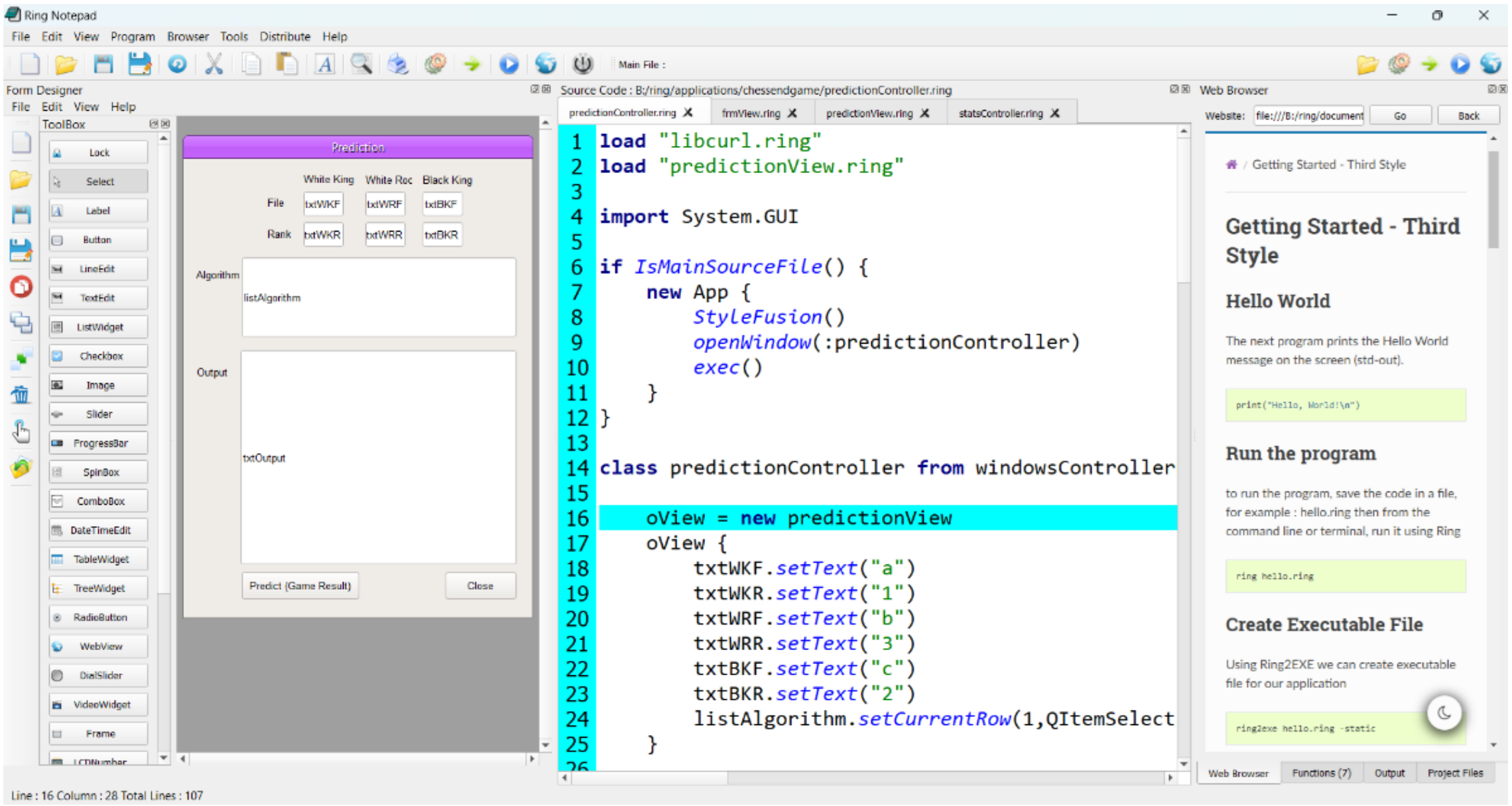

***Figure 4.7*** *Ring IDE (Code Editor, Form Designer, etc.) is developed using Ring itself.*

## 4.8 The Implementation (Using Visual Programming)

Ring Visual Implementation is developed using Programming Without Coding Technology (PWCT) software (version 1.9) [34,66]. This software offers various visual programming languages, including HarbourPWCT, PythonPWCT, SupernovaPWCT, C#PWCT, and CPWCT [12,120].

Each visual language corresponds to a specific textual programming language used in the code generation process. In our case, we utilized CPWCT to design different components, such as the General Library, Loader, Compiler, and Virtual Machine, through visual programming. Subsequently, we obtained the source code in the C programming language.

PWCT is designed to provide precise control, like what we experience with textual code editors, while also offering visual programming advantages such as reducing errors and the ability to work with multiple dimensions and a rich user interface. Notably, PWCT includes a powerful feature called the Time Dimension during visual programming. With this feature, each step or block generated in the program stores information about development time. Programmers can watch the program evolve step by step, revise the order of the construction process, and even run the program at specific points in time.

The General library plays a crucial role in successfully implementing the Ring language as a lightweight programming language. Other components extensively reuse the library functions to implement many language features using lists and hash tables instead of specific C structures. The list implementation (demonstrated in Figure 4.8) uses a Doubly Linked List, Deque (Double-Ended Queue), and Singleton Cache to store the pointer of the last visited item and the index of the next item. This allows for quick traversal of the list through a function that receives the item index as a parameter.

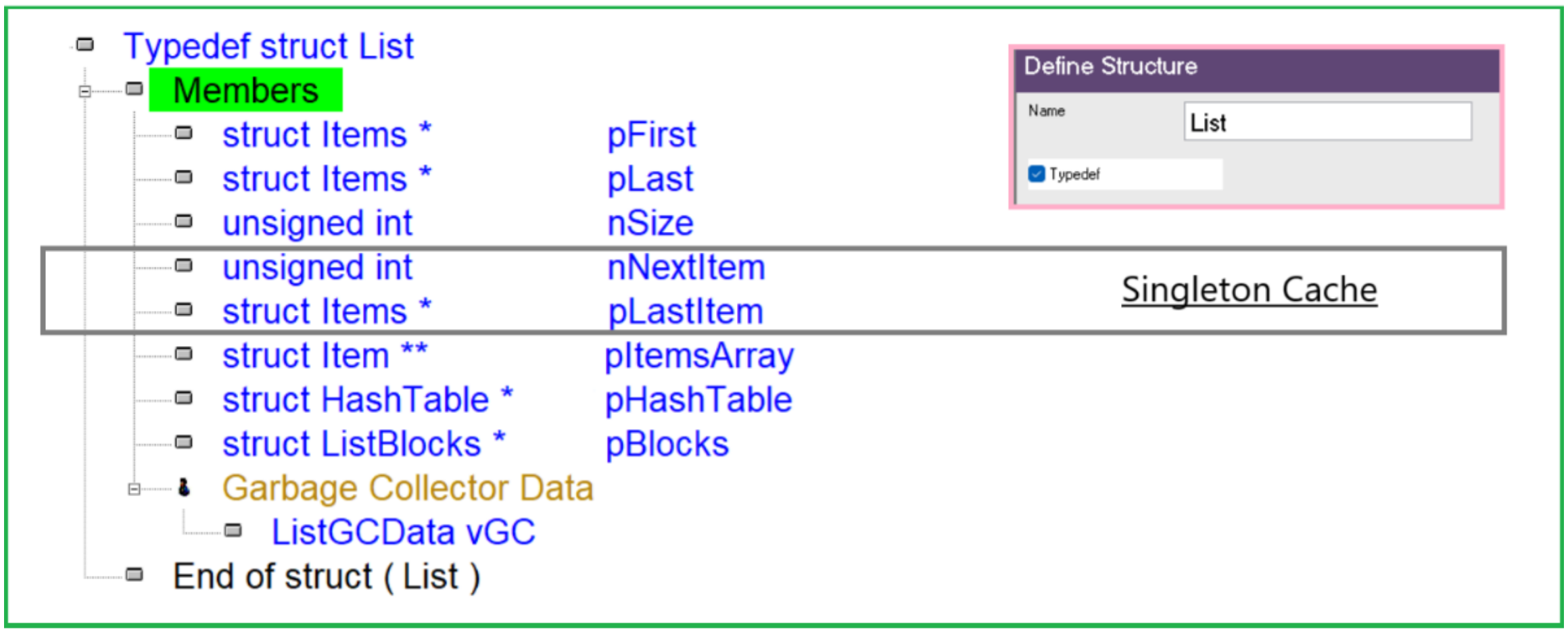

***Figure 4.8*** *Using PWCT to define the List structure which uses a singleton cache.*

Additionally, there is an optional array of pointers that can be used in specific situations to quickly find an item (through the item index) and use it without the need for the traverse process or using the singleton cache. Furthermore, if the size of the list items is known when creating new lists, memory could be allocated as a continuous block to be cache-friendly and minimize cache misses. Using this data structure to implement language features enables us to create a lightweight language with numerous capabilities while also contributing to stability and reducing memory management errors. It is like writing High-Level code like dynamic language code. However, it is essential to acknowledge one clear disadvantage of this approach: lower performance and increased memory usage compared to using specific C structures or arrays when implementing features.

To strike a balance, during language development, we identified performance bottlenecks—such as function call implementations—and replaced them with specific low-level implementations based on a pre-allocated array of structures. Such optimizations became necessary once we started supporting the Raspberry Pi Pico microcontroller.

While the Singleton Cache consumes less memory compared to the array of pointers, it introduces a challenge: every read operation from the list items could potentially trigger a write operation if the cache is updated. Such behavior is undesirable, especially when sharing lists among threads. In this scenario, we opt for the array of pointers to avoid relying on the singleton cache.

In Figure 4.9, we observe how the language grammar rules are implemented using PWCT. For each group of grammar rules, we define a specific function. We had a step that described each rule, allowing us to focus easily on specific rules using features like collapsing and expanding in the steps tree. Each group of steps associated with the same component can have an interaction page (a data entry form) that receives component parameters and controls the steps' generation and update processes. Notably, all components, including the "Call Function" component, are created within the PWCT environment itself. We have the flexibility to create new visual components or update existing ones. During the development of the Ring Compiler/VM, we exclusively used the standard components provided by PWCT. No new components were necessary because the available ones sufficed for implementing the required features. In the toolbar, there is a combobox for selecting the visual programming language. PWCT initially started with "HarbourPWCT" as the default visual language, but we specifically chose "CPWCT" to develop our project based on the C language.

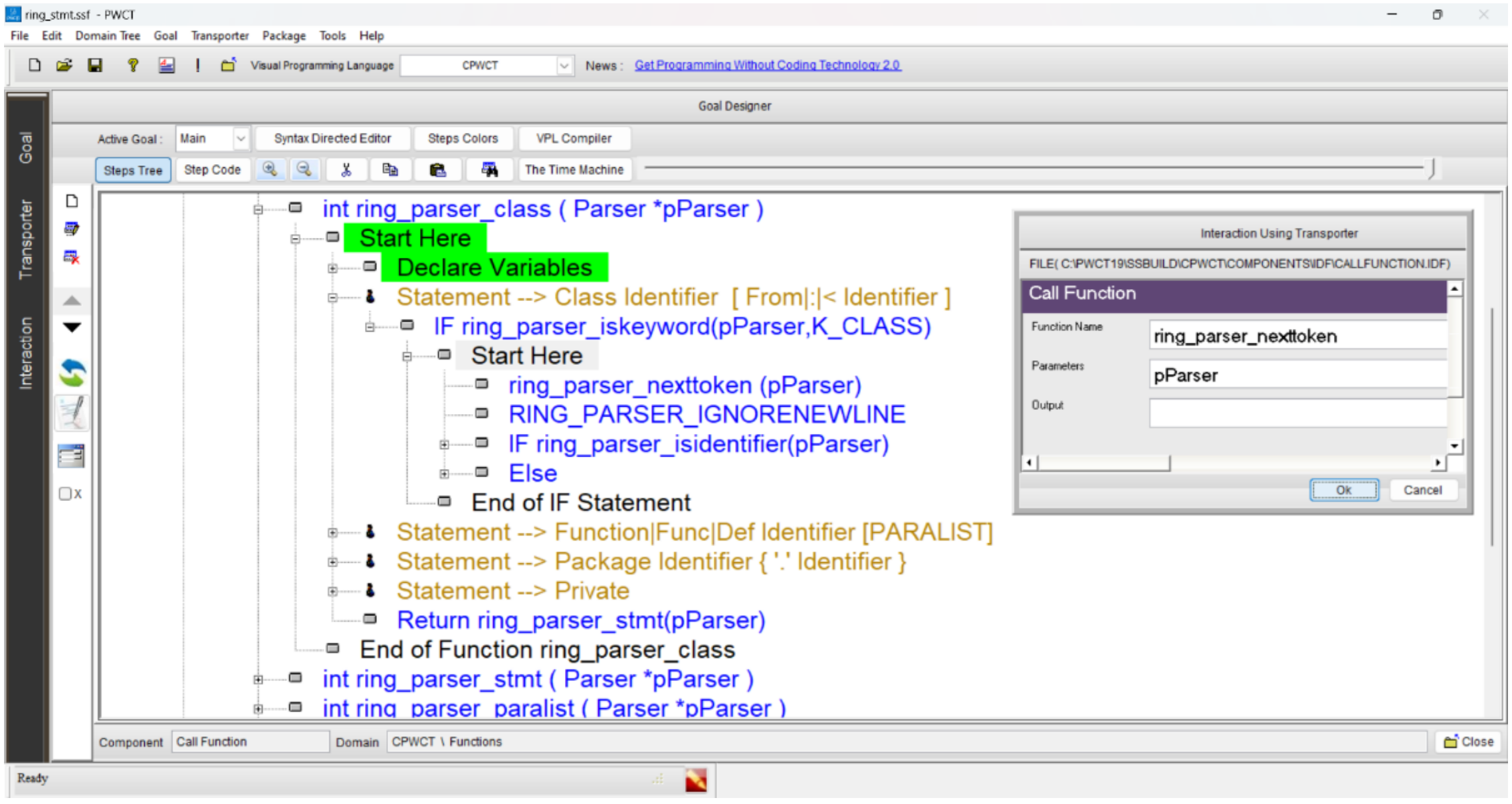

***Figure 4.9*** *Implementing the Ring language grammar using PWCT.*

Figure 4.9 shows a button labeled "The Time Machine." Clicking this button provides a menu of options that allow us to play the program like a movie, revealing the construction steps. Another crucial button is the "VPL Compiler." By using this button, we can examine the composition of different visual components, as demonstrated in Figure 4.10. An interesting feature in the results is the count of interactions (visual components) and the number of steps within the steps tree. These metrics provide insight into the abstraction level offered by the interaction pages of the visual components.

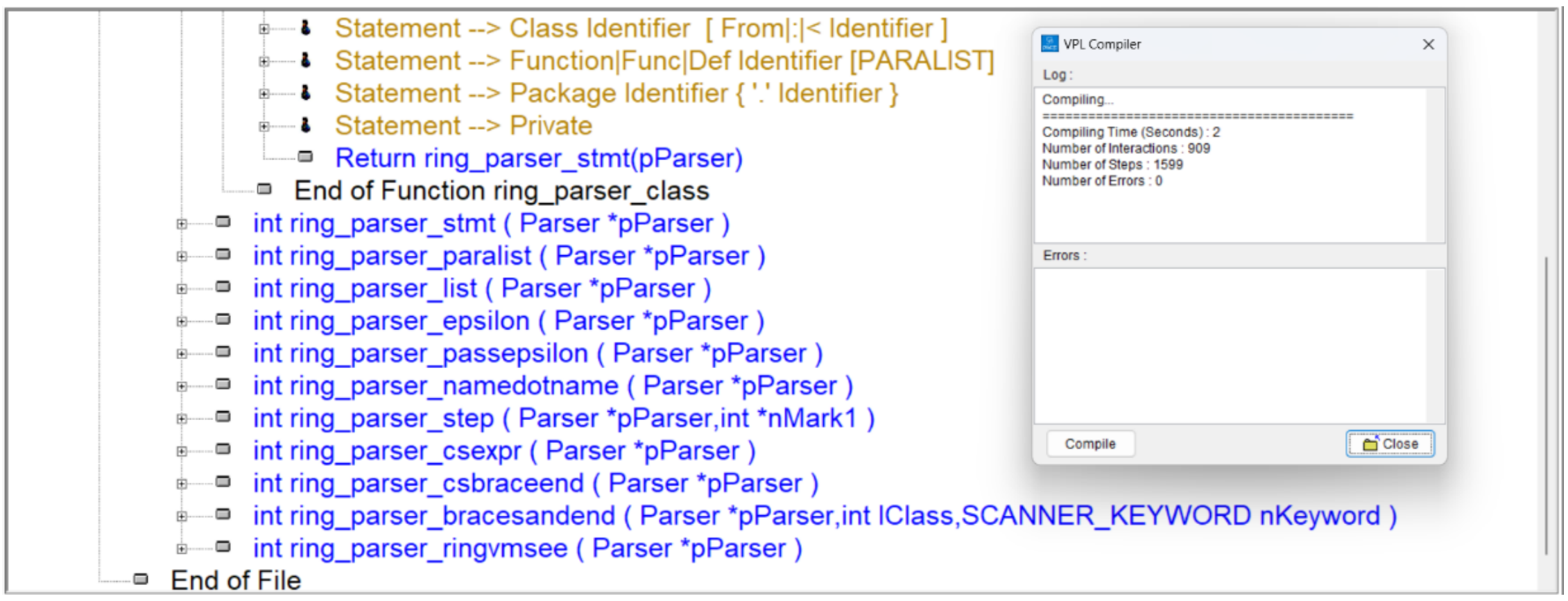

***Figure 4.10*** *Using the VPL Compiler to get statistics about the visual representation.*

An interesting question arises: why do we need a Visual Programming Language (VPL) Compiler if the visual language itself is designed to prevent errors? The answer lies in our ability to disable the Syntax Directed Editor, allowing us to manually arrange the generated steps for visual components to do quick organization and refactoring. However, this flexibility can sometimes lead to mistakes. Imagine a scenario where a component is inadvertently placed in the wrong location, and the programmer does not immediately notice the error. In such cases, the VPL compiler becomes invaluable—it can catch these composition errors and help ensure the program is correct.

The Ring compiler generates bytecode, where each instruction must contain an operation code and can include zero, one, or two arguments. This bytecode is stored by the Ring compiler as a Ring List, allowing the compiler to easily insert instructions during code generation. However, when this bytecode is passed to Ring VM, it undergoes a conversion process to a more suitable representation for execution.

This representation is stored as a single continuous block in memory (rather than multiple byte-code chunks). Notably, it includes extra space that can be utilized for new instructions produced at runtime by the Eval() function. By having this additional space, the need for frequent memory reallocation is reduced.

Within the Virtual Machine (VM), the bytecode representation employs fixed-size instructions. Specifically, the size of each instruction is 16 bytes for 32-bit microcontrollers (such as the Raspberry Pi Pico) and 24 bytes for 64-bit desktop environments. This bytecode is writable, allowing the VM to update instructions during runtime—caching certain values and even replacing instructions with faster alternatives that utilize pointers for variables, thus avoiding costly search processes.

In Figure 4.11, we observe the structure of the bytecode on the left side. On the right side, we find the VM instructions that have been added to enhance performance. Opting for a writable long-byte code format is somewhat unconventional; for instance, Python uses 2 bytes per instruction, while Lua uses 4 bytes [139,140]. However, our deliberate choice of a long-byte code format serves two key purposes:

- Simplicity of Implementation: Despite supporting a language with a substantial number of features (128 instructions), we aimed for a compact implementation. The writable long-byte code format allows us to achieve this without unnecessary complexity.
- Performance Optimization: By using a longer bytecode format, we gain flexibility. We can improve the performance of specific instructions without resorting to a just-in-time compilation of machine code or significantly increasing the overall implementation size.

When considering the disadvantages of using a writable long-byte code format, it is essential to address a few key points. First, this approach results in larger memory requirements, which increases the likelihood of cache misses—a factor that directly impacts performance. Additionally, storing the byte code in writable memory can be costlier, especially on microcontrollers like the Raspberry Pi Pico [141,142]. However, to mitigate the drawbacks associated with the larger bytecode size, the Virtual Machine (VM) incorporates a clever strategy: some instructions serve multiple purposes.

This technique is well-known and proves useful in common scenarios. By identifying instructions that are frequently used together, we can optimize their representation. In doing so, we avoid allocating unnecessary space for instructions that do not require the full extent of the provided memory [143].

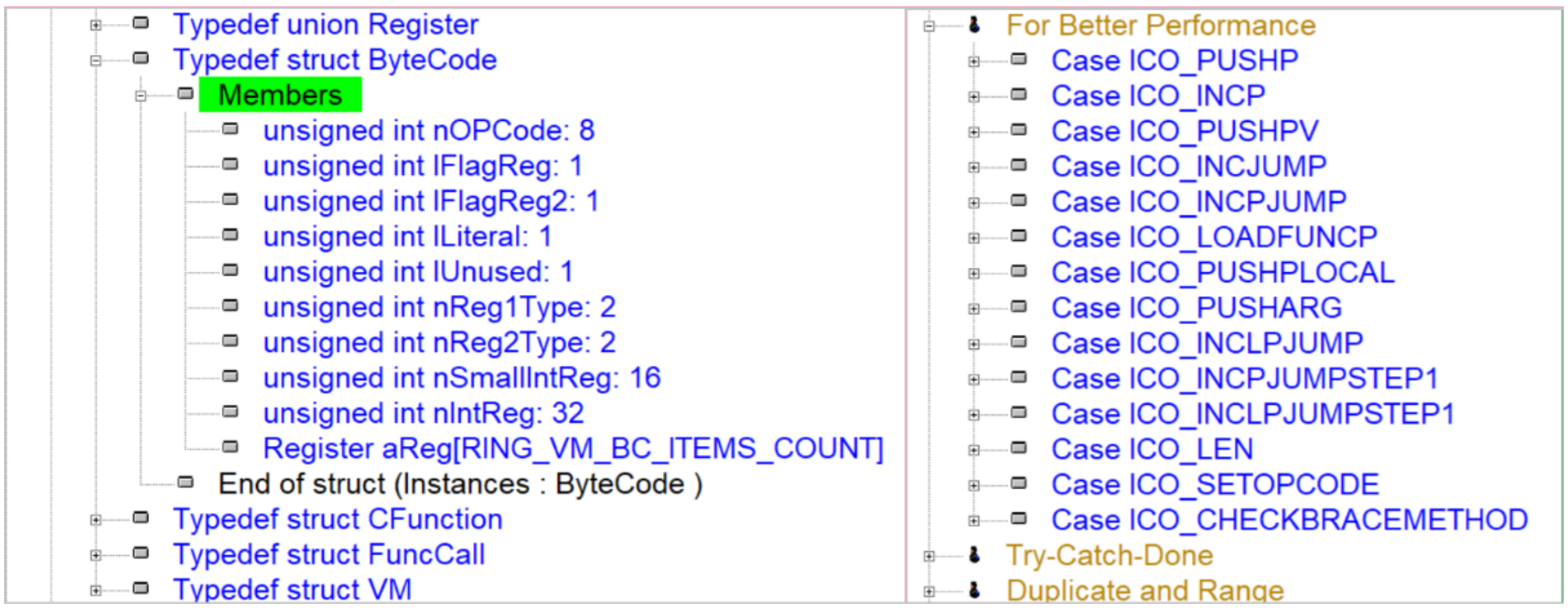

***Figure 4.11*** *Ring Virtual Machine implementation using PWCT.*

The Virtual Machine (VM) implements several useful features when embedded within projects. For instance, during program execution, we can suspend or resume the VM [144], allowing a running program to request VM suspension while preserving its state. Additionally, the VM supports having multiple language states—meaning that more than one instance of the Ring VM can coexist within the same application.

These features are of practical use in the “Try Ring Online” application. Within this application, when we write and run a program, it creates a new language state specific to our program. If the program requires input from the console, it halts the sub-virtual machine, signaling to the main VM that the console application is awaiting input. Users then type their input in a GUI provided by the main VM and click “Send”. These data are copied to a variable associated with Sub VM, and a resume operation follows. As a result, the console application in Sub VM can receive the input. This approach enables us to create a playground for the Ring language as a WebAssembly application without the need for threads.

In summary, all the modules related to the General Library, Loader, Compiler, and Virtual Machine are designed using visual programming through PWCT. We have 43 visual source files that generate 44 C source files and 28 C header files. Each visual source file could generate one or more textual files.

The Ring language keywords, standard functions, context-free grammar, compiler and runtime errors, and virtual machine instructions are published online on the Ring website (ring-lang.github.io/doc1.22/reference.html). To achieve a lightweight implementation while retaining a programming language with rich features and ensuring performance comparable to scripting languages, we made the following design decisions that address the third research question (RQ3) [139–144]:

1. Applying principles such as “Don’t Repeat Yourself” (DRY) and the “Keep it simple and stupid” (KISS) principle.
2. Using a single-pass compiler where the parsing and code generation are interleaved.
3. The ability to separate the compiler and the virtual machine and use any of them alone.
4. All the built-in functions are grouped in optional modules through preprocessor directives.
5. Ring Lists vs. C Structures: In most cases, we opted for Ring Lists over C structures.
6. Selective Use of C Structures: However, in specific features where performance impact matters significantly, we chose to use C structures. These targeted optimizations enhance critical parts of the language.
7. Flexible List Implementation: Our list implementation combines various data structures and optimization techniques, including Doubly Linked Lists, Deques (Double-Ended Queues), Singleton Caches, arrays of pointers, Hash Tables, and continuous memory blocks. This flexibility accommodates diverse use cases.
8. One block for Bytecode Storage: The bytecode resides in a single continuous memory block, avoiding fragmentation. Moreover, we intentionally allocate extra space within this block. This foresight reduces the need for frequent memory reallocation during runtime, especially when using the Eval() function.
9. Writable Long-Byte Code Format: The bytecode format uses a longer representation, which allows for performance improvements. During runtime, instructions can be dynamically replaced with faster alternatives, all without resorting to just-in-time compilation to machine code or bloating the implementation size.
10. The Virtual Machine does not use a global interpreter lock (GIL), which results in better performance when utilizing threads for CPU-bound tasks.

## 4.9 Chapter Summary

In this chapter, we presented the design and implementation of the Ring dynamic programming language. We explored its most important features and contributions, highlighting the advancements it brings to the field of dynamic programming languages. We presented Ring features related to syntax customization and extending object-oriented programming support with novel features that enable creating internal domain-specific languages that resembling external domain-specific languages like CSS and supernova. Also, we presented the visual implementation of the Ring Compiler/VM based on the PWCT visual programming language and listed the design decisions that are used to have a lightweight and multi-paradigm dynamic programming language.

In the next chapter, we will introduce the PWCT2 visual programming language. We will delve into its design, key characteristics, and main contributions.

# Chapter 5: The PWCT2 Visual Programming Language

## 5.1 Introduction

The primary aim of developing the PWCT2 visual programming language, as demonstrated in Figure 5.1, is to use the Ring programming language to develop a research prototype for the second generation of the PWCT visual programming language. PWCT2 offers several enhanced features, including a more flexible environment, time dimension and auto-run, rich colors and customization, rich comments using text, lines, images, and HTML, an enhanced form designer for GUI applications, support for importing Ring source code and interactive textual-to-visual code conversion, self-hosting of the PWCT2 environment, and cross-platform implementation that supports Windows, Linux, and macOS. These features aim to improve the overall functionality, user experience, and performance of the PWCT2 environment. This chapter addresses the fifth research question (RQ5).

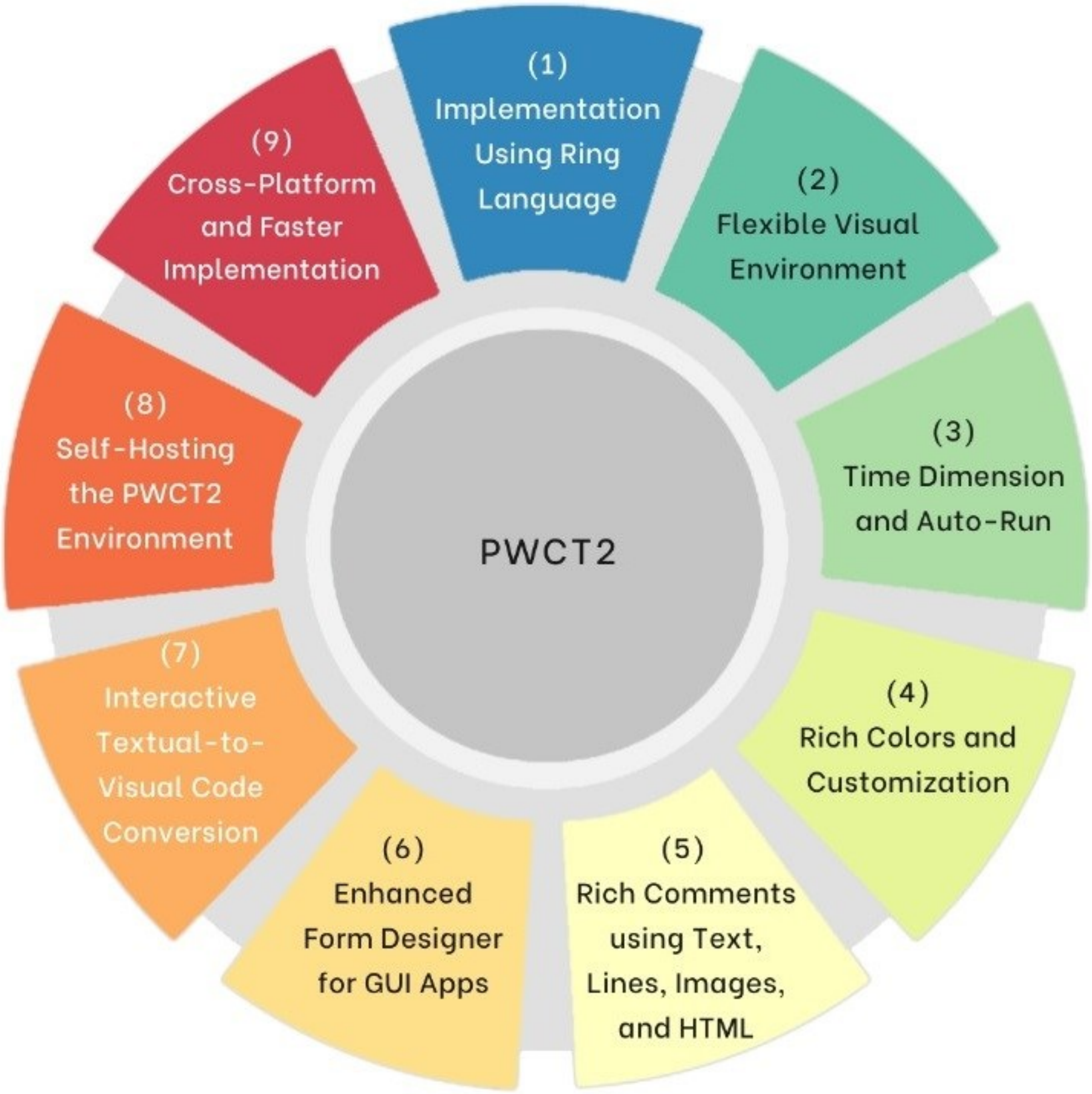

***Figure 5.1** The key features of the proposed visual programming language.*

In this chapter, we describe our system design and implementation. We highlight the important features of the proposed visual programming language, PWCT2, and present the system architecture, which is implemented using the Ring language.

## 5.2 Implementation using the Ring language

The architecture of the proposed visual programming language (shown in Figure 5.2) is divided into three main layers: Applications, PWCT Environment, and Ring Language. Each layer contains specific components that contribute to the overall functionality of the system. The bottommost layer, Ring Language, includes Development Tools, Libraries, Compiler, and Virtual Machine (VM). Development Tools provide various tools that assist in the development process, such as the package manager and Ring2EXE. Libraries consist of pre-written code libraries that developers can use to add functionality to their programs, saving time and effort by providing reusable code for common tasks. The compiler processes Ring textual source code and generates bytecode for the Ring Virtual Machine if the program is correct. If there are issues, it produces compile-time errors. The Virtual Machine (VM) provides a runtime environment for the programs written in Ring, executing the compiled code and managing the program's execution [91,145].

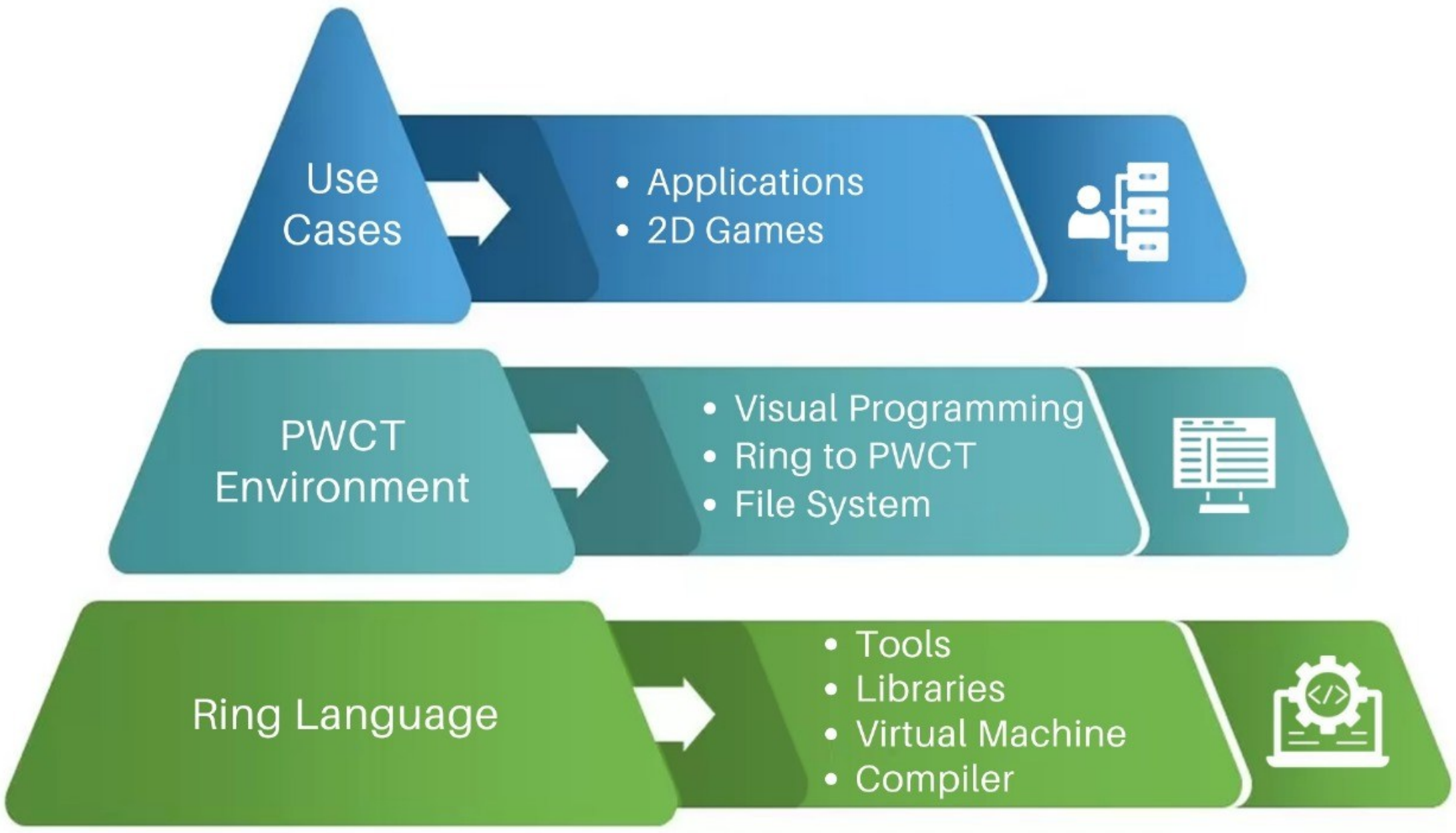

***Figure 5.2** The proposed self-hosting visual programming language architecture.*

The middle layer, PWCT Environment, consists of Visual Programming, the Ring to PWCT converter, and the File System (Visual Source). Visual Programming allows developers to create programs using visual elements rather than traditional text-based code, providing a more user-friendly development experience. This is achieved through tools such as the Steps Tree editor, time machine, components browser, interaction pages (data-entry forms), and form designer. The Ring to PWCT converter assists in importing the textual source code written in Ring and visualizing the structure and flow of the program by offering graphical representations of the program's logic. This enables us to continue program development using visual programming instead of relying on the textual source code. The File System (Visual Source) manages the visual source files within the PWCT environment, organizing and storing information about the visual elements used in the development process. PWCT can generate Ring source code from these visual source files, enabling the execution of these programs using the Ring compiler and virtual machine.

The topmost layer, Application Layer, has the various applications, including 2D games, that can be developed using PWCT based on the Ring programming language, ranging from simple utilities to complex software solutions. Overall, the architecture of the proposed system is designed to provide a comprehensive and user-friendly environment for visual programming based on the Ring programming language, making it accessible to both novice and experienced Ring developers.

## 5.3 Flexible Visual Environment

In this section, we introduce the design differences between PWCT (the first generation of Programming Without Coding Technology) and PWCT2 (the proposed new generation). In PWCT, visual programming is achieved through four sub-systems: Goal Designer, Components Browser, Interaction Pages, and Form Designer. The Goal Designer is used for designing modules and provides the Steps Tree Editor and the Time Machine. The Components Browser enables users to select specific components, each offering one or more interaction pages (data-entry forms). Entering data into these interaction pages generates or updates steps managed by the Goal Designer [34,66].

### 5.3.1 Single Main Window and Several Dockable Windows

In PWCT [66], the visual programming sub-systems are not designed to be used simultaneously on one screen; each sub-system uses a separate window. For instance, the Goal Designer and Components Browser, or the Components Browser and Interaction Pages, cannot be viewed at the same time. In the proposed visual programming language PWCT2, this design has been revised to use a single main window for all sub-systems, as shown in Figure 5. Each sub-system now occupies a separate dock-able window [146,147], allowing users to view all sub-systems simultaneously, which avoids the complete redrawing of the screen when switching from one to another. We also added two dock-able windows: one for Project Files, where users can quickly navigate to specific folders and open visual source files, and another for the Output window, where users can see the program output, send input to the running program, or terminate it at any time.

The previous generation of PWCT did not provide these features (project files/output window) and relied on operating system dialog boxes to open files and the command prompt window to display program outputs [12]. Another feature added by PWCT2, which does not exist in PWCT, is the ability to open multiple interaction pages simultaneously. This simplifies the process of reading and updating programs. The Steps Tree Editor provides a feature to open all the interaction pages at once through a keyboard shortcut.

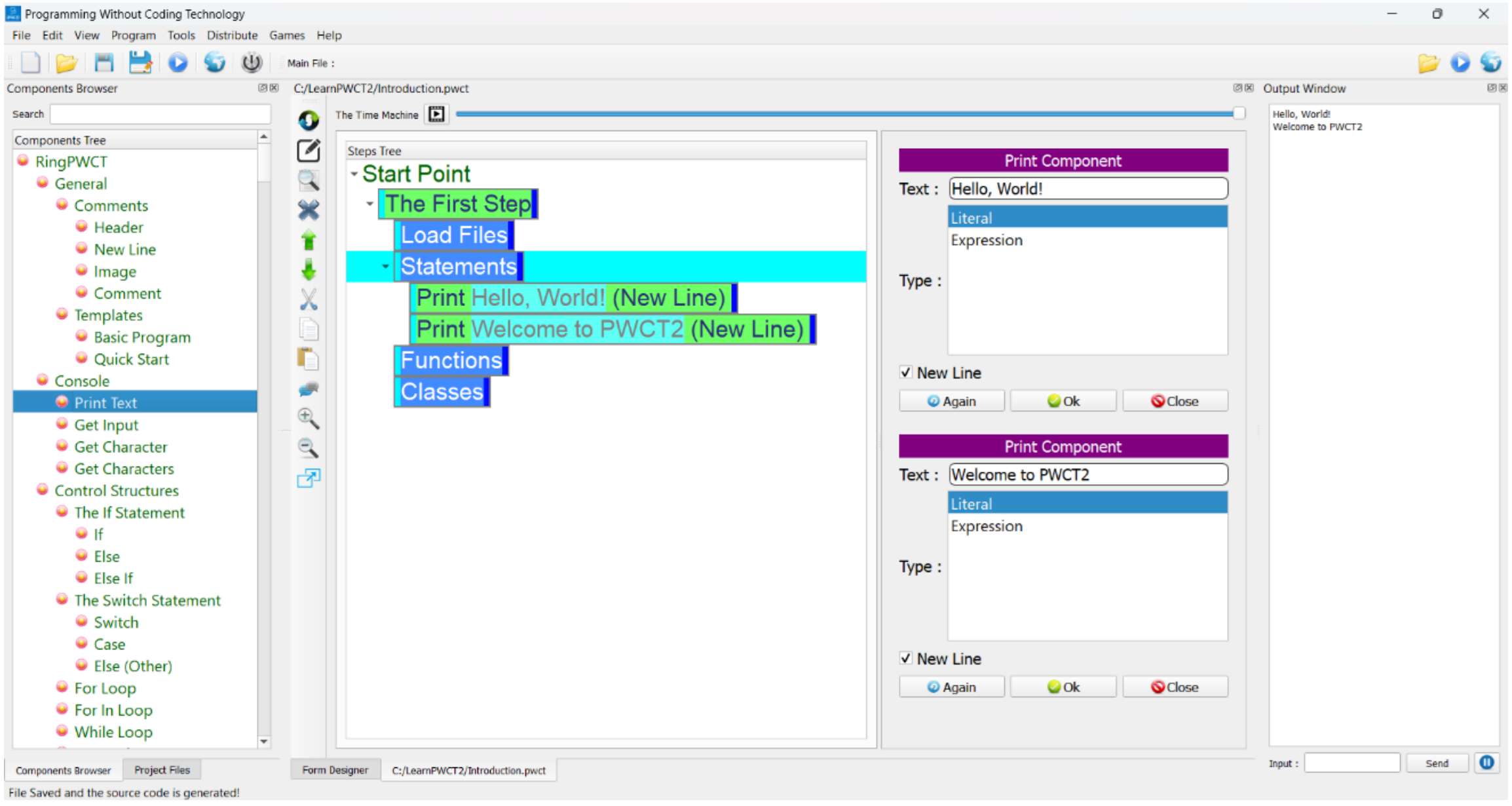

*Figure 5.3* *PWCT2 uses a main window and dock-able windows.*

The program illustrated in Figure 5.3 is a simple example that prints two lines of text: "Hello, World!" and "Welcome to PWCT2.". The proposed visual programming language makes it straightforward to create and modify the program visually. Users can add, remove, or change steps by interacting with the Steps Tree and the Print Component.

For instance, to change the text being printed, users can simply edit the text in the Print Component interaction page and click "Ok." Additionally, new steps can be added by selecting the desired components from the Components Browser.

The Components Browser (left panel) lists various visual components that can be used in the program. These components belong to the RingPWCT visual programming language. Categories include Comments, Templates, Console, Control Structures, and more. Each category contains specific components, such as "Print Text", "Get Input", "The If Statement", and "For Loop". We can find and select a component using the mouse or through a search process by the component name or part of the name.

The Steps Tree (middle panel) displays the structure of the program in a hierarchical tree format. The program starts with a "Start Point" and includes steps generated by the Basic Program component, such as "The First Step", "Load Files", "Statements", "Functions", and "Classes". Using the Basic Program component is optional, and these generated steps are just comments for organization. Additionally, there are steps generated by the Print Text component, like "Print Hello, World! (New Line)" and "Print Welcome to PWCT2 (New Line)". In general, each step represents a specific action or command in the program, or it could be just a comment to provide a better understanding of what the program does or its structure.

### 5.3.2 Flexible Steps Tree Editor

The Steps Tree editor in PWCT2 is designed for flexibility and provides many features that do not exist in PWCT, including drag-and-drop functionality to move steps from one location to another instead of using cut and paste. This enhanced flexibility significantly improves the user experience by simplifying the process of organizing and modifying steps. In PWCT, when we add a step as a child to another step, the parent step must be of a type that allows children, and the added step will be at the end of the children. This is a limitation if we want to insert a new step between two other children, requiring the steps to be added first and then moved to the desired location.

In PWCT2, the insertion process is supported, where selecting a step that does not support children and then adding a new step inserts it after the selected step. This change enhances the capability to modify the program's structure on the fly, making development more intuitive.

The Steps Tree editor provides common features expected from editors, such as Find and Replace, Go-To, and Print to PDF, ensuring that users have access to essential editing tools. In Figure 5.4, we demonstrate how to insert steps between other steps by inserting the “Print TWO (New Line)” step between the steps that print the “ONE” and “THREE” messages on the screen. In the second section of the figure, we show how to use the Find and Replace window to find steps that contain specific text. These improvements collectively contribute to a more user-friendly development environment.

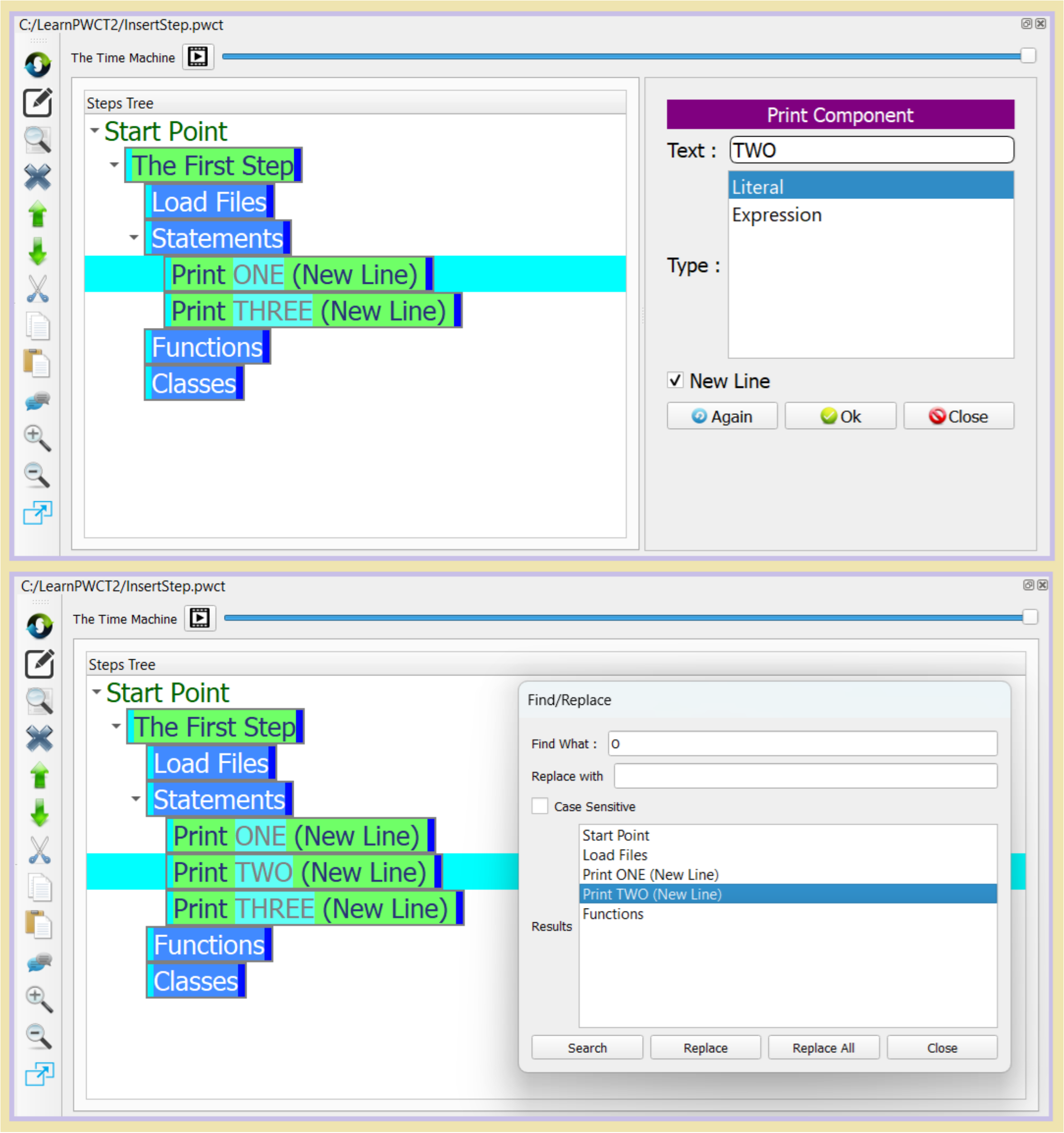

***Figure 5.4*** *Inserting steps in the Steps Tree and using the Find and Replace window.*

### 5.3.3 RingPWCT Components in the Components Browser

In PWCT [66], the environment includes multiple visual programming languages, such as HarbourPWCT, CPWCT, PythonPWCT, and SupernovaPWCT. Each visual language comes with a group of components classified into different domains, generating textual source code in specific textual programming languages like Harbour, C, Python, and Supernova.

In our research prototype of PWCT2, we focus on supporting the Ring programming language through the RingPWCT visual programming language, which provides visual components that generate textual source code in the Ring programming language.

Table 5.1 provides an overview of the 394 visual components available in the RingPWCT visual programming language within PWCT2.

The table categorizes the components into different domains, specifying the number of components in each domain and providing an example for each domain. For instance, the General domain includes six components, with “Quick Start” as an example, while the Console domain comprises four components, with “Print Text” being one of them. Other notable domains include Control Structures, with 13 components like “For-In Loop”, and GUI, which has the highest count of 88 components, exemplified by the “Window Class”.

The table also highlights various other domains such as Functions, Program Structure, Lists, Strings, Date and Time, and Math, among others. Each domain contains a specific number of components tailored to different programming tasks. For example, the Date and Time domain includes seven components like “Add Days”, while the Database domain, one of the most extensive, contains 34 components such as “ODBC Connect”. This comprehensive categorization helps users navigate and utilize the diverse set of tools available in RingPWCT to enhance their programming experience within PWCT2.

*__Table 5.1__ The RingPWCT visual programming language components.*

| ID | Domain | Components Count | Example |
|---|---|---|---|
| 1 | General | 6 | Quick Start |
| 2 | Console | 4 | Print Text |
| 3 | Control Structures | 13 | For-In Loop |
| 4 | Variables and Operators | 17 | Assignment |
| 5 | Functions | 3 | Define Function |
| 6 | Program Structure | 2 | Load Source File |
| 7 | Lists | 15 | New Empty List |
| 8 | Strings | 16 | Get String Length |
| 9 | Date and Time | 7 | Add Days |
| 10 | Check Data Type | 3 | Check Character |
| 11 | Math | 1 | Math Functions |
| 12 | Files | 29 | Read File to String |
| 13 | System | 12 | Get System Variable |
| 14 | Dynamic Code | 3 | Eval |
| 15 | Database | 34 | ODBC Connect |
| 16 | Security and Internet Functions | 11 | Download |
| 17 | Object-Oriented Programming | 10 | Define Class |
| 18 | Functional Programming | 3 | Anonymous Function |
| 19 | Reflection | 29 | Globals Info |
| 20 | Standard Library | 71 | Stack Class |
| 21 | Web Library | 12 | WL WebPage Class |
| 22 | LibCurl Library | 5 | LibCurl Easy Init |
| 23 | GUI | 88 | Window Class |

### 5.3.4 Advanced Visual Components and Templates

In PWCT [12], visual components are based on a specific scripting language designed for the PWCT environment. This scripting language is intended to be easy to use and increase productivity by providing specific commands that guide the steps generation process based on data entered in the interaction pages. However, this scripting language is very limited and provides simple concepts related to variables, if-statements, steps/code generation, and rules for relationships among components. It does not have loops, functions, or the ability to be extended without modifying its interpreter [148]. These limitations prevent the development of rich and powerful components that could perform advanced tasks during steps generation. Components are designed under the assumption that the component generating the steps will also be responsible for updating these steps. In other words, a component cannot generate steps that belong to other components. For example, to create a template of steps that belong to different components, a specific visual source file with these generated steps must be created. The user can then start a new visual source file from these templates. This means that these templates must be used at the start of creating new visual source files, and multiple templates cannot be used in the same visual source file without creating a new template that integrates them.

PWCT uses two different programming languages: Visual FoxPro for developing the PWCT environment and a scripting language called RPWI designed for developing visual components. In the proposed visual programming language (PWCT2), the Ring textual programming language is used for developing both the PWCT2 environment and the visual components. Using Ring for developing the environment components enables us to create advanced components and avoids the known limitations of RPWI. For example, we have the Quick Start component, which can be used to generate steps that belong to multiple components, as shown in Figure 5.5.

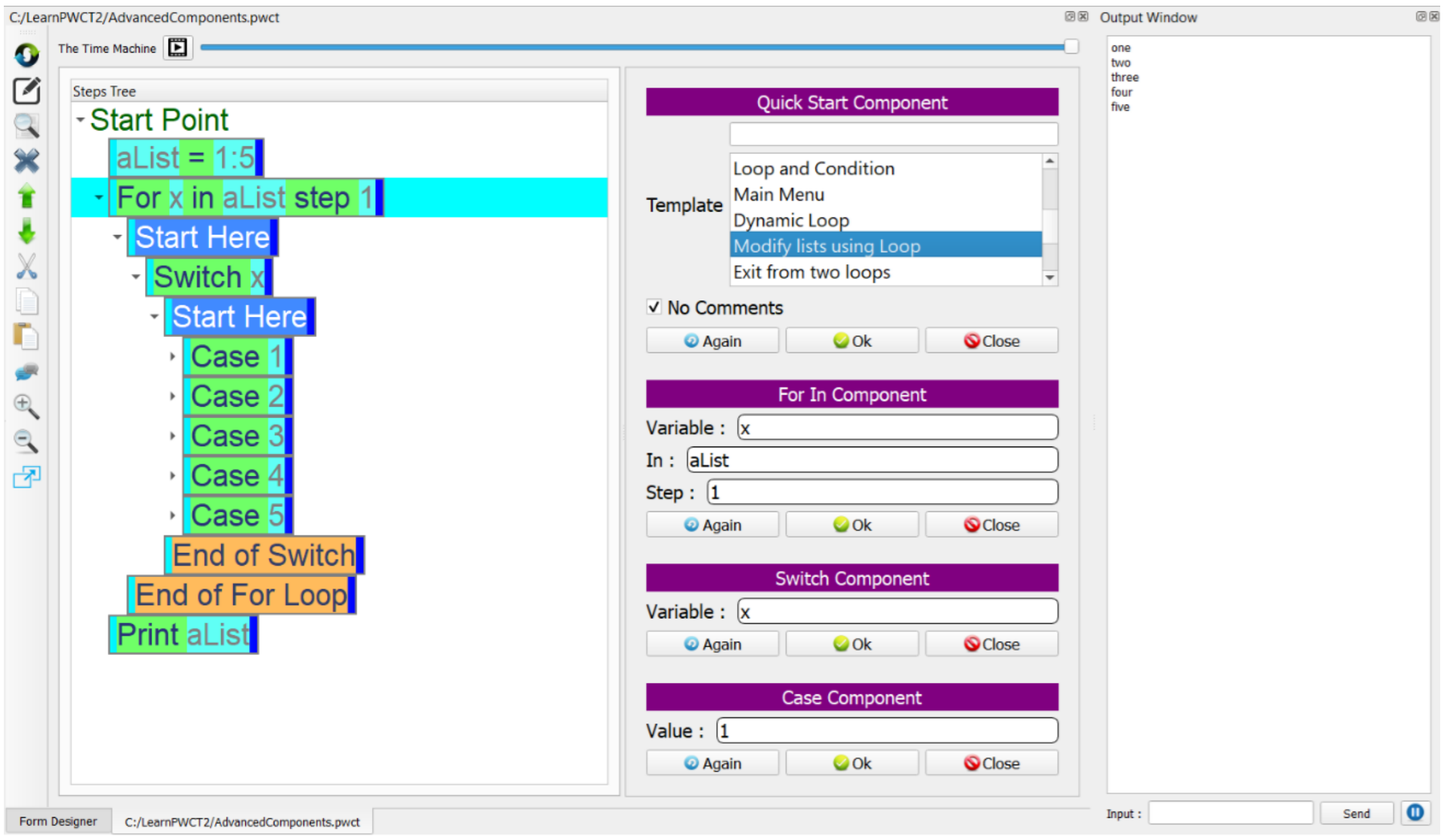

***Figure 5.5*** *A component that generates steps that belong to other components.*

The Quick Start component contains many templates that we can use. One of these templates is the Modify Lists Using Loop template. Selecting this template generates multiple steps that belong to different visual components, such as the For In component, Switch component, Case component, etc. The Quick Start component can also generate comments and provides a No Comments checkbox to avoid generating comments if desired.

## 5.4 Time Dimension and Auto-Run

In PWCT, users can use the Time Machine to change the time position and go backward in time to see the program at a specific point in the past and check the order of steps added to the program. We also have the option to play the program as a movie and see how the visual components are used step by step to create the program. Additionally, we can run the program at a specific point in the past and see the program output at that point. In PWCT2, we support the Auto-Run feature as shown in Figure 5.6, where changing the time position or adding new steps to the program will directly execute the program and display the output in the output window.

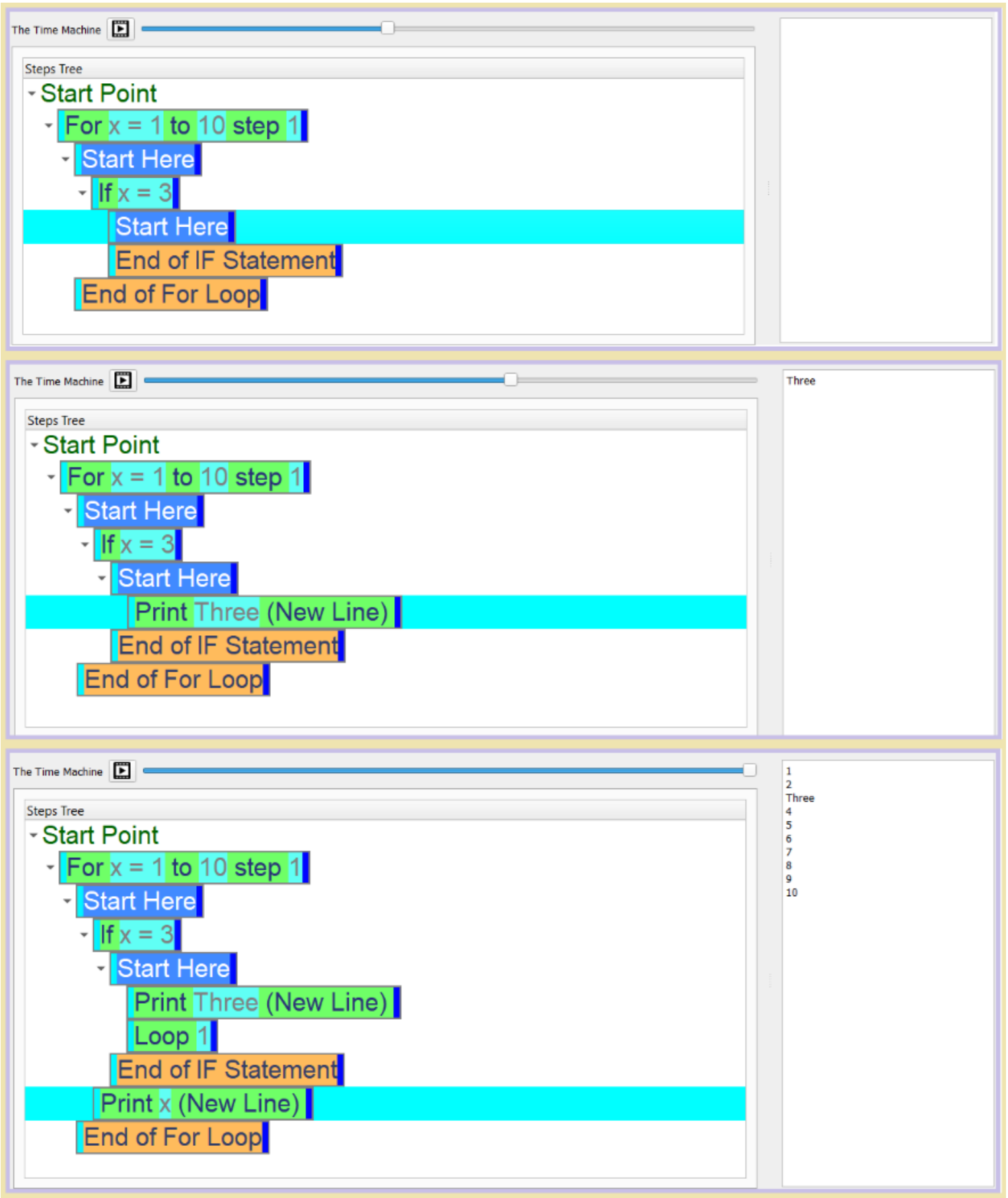

***Figure 5.6*** *Using the Time Machine and the Auto-Run feature.*

The program contains a for-loop and a conditional statement. The first position in the figure shows the initial state with a For loop running from x = 1 to 10 with a step of 1 and an If statement checking if x equals 3. The second position captures the execution of the If statement when x equals 3, printing “Three” on a new line. The third position demonstrates the continuation of the loop, printing the value of x on a new line for each iteration. The output on the right side of the third screenshot lists the numbers 1 to 10, with “Three” replacing the number 3, highlighting the effect of the conditional statement.

## 5.5 Rich Colors and Customization

Unlike PWCT, which provides one style for the environment (White Style), PWCT2 enables us to select the PWCT environment style (White, Dark, Blue, etc.). Like in PWCT, we have a Customization window to determine the colors used in the Steps Tree editor based on the step type (see Table 5.2). Each visual component can generate one or more steps in the Steps Tree after entering the required data on the interaction page. These generated steps might belong to the same component or be related to other components if the original component is a template. We can use the same component again by clicking the ‘Again’ button on the interaction page.

***Table 5.2*** *Different types of Steps inside the Steps Tree.*

| ID | Step Type | Description |
|---|---|---|
| 1 | Start Point | The program root (one for each visual source file) |
| 2 | Comment | Just a comment and does nothing during runtime |
| 3 | First | The first step generated by the component |
| 4 | Allows Interaction | The step could include sub steps |
| 5 | Leaf | The step cannot include sub steps |

PWCT2 can use multiple colors in each step to highlight the data entered by the user from the text generated by the visual component, whereas PWCT uses one text color and one background color per step. This improvement is inspired by Scratch [149]. Additionally, we can show or hide the dock-able windows as needed. As shown in Figure 5.7, we focus on the Steps Tree editor and interaction pages after setting the style to Dark and customizing the Steps Tree colors using the Customization window.

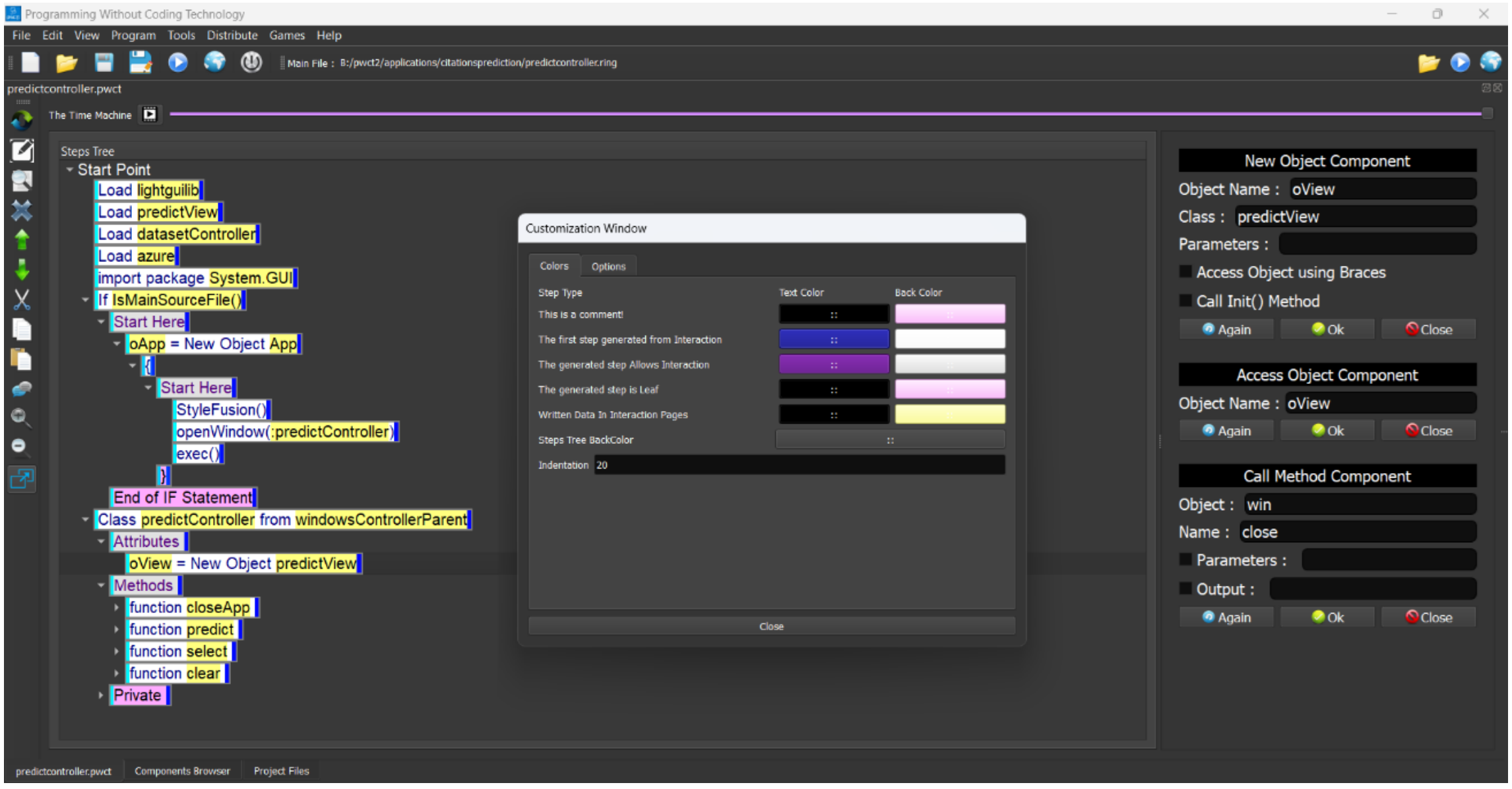

***Figure 5.7*** *Using the Customization window to select the Steps Tree colors.*

Additionally, the Customization window enables us to set the Steps Tree background color and adjust the indentation level in the Steps Tree. In the Options tab, we also have advanced options that can enable the Auto-Run feature and control the environment's behavior in different situations, such as opening the interaction pages in the Steps Tree dock-able window or in separate windows.

In PWCT [12], users can open only one visual source file at a time for each instance of PWCT. Opening multiple visual source files simultaneously requires running multiple instances of PWCT. In PWCT2, we have the Project Files dock-able window, as shown in Figure 5.8, where we can easily select and open any visual source file or form file. Additionally, using the options in the Customization window, we can choose to open files in new tabs, allowing us to work with and view multiple files simultaneously. This improvement is inspired by popular editors and IDEs like Visual Studio and NetBeans [150,151]. The Customization window is open in the center, displaying two tabs: "Colors" and "Options". The "Options" tab is selected, showing a list of customization options with checkboxes. The options listed in the Customization Window include Auto Run, Open files in new tabs, Show the Time Machine options, Steps Tree—Hide Step Code Tab, Show Steps Tree Lines, Light Tree Lines, Steps Tree—Show Nodes Icon, Open interaction pages in new windows, Allow Syntax Errors in Interaction Pages, Avoid Components Browser, Avoid Components Browser Auto-Complete, Components Browser—Always Show Search Window, Reflect changes in font size to other windows, and Borders around steps (in supported styles).

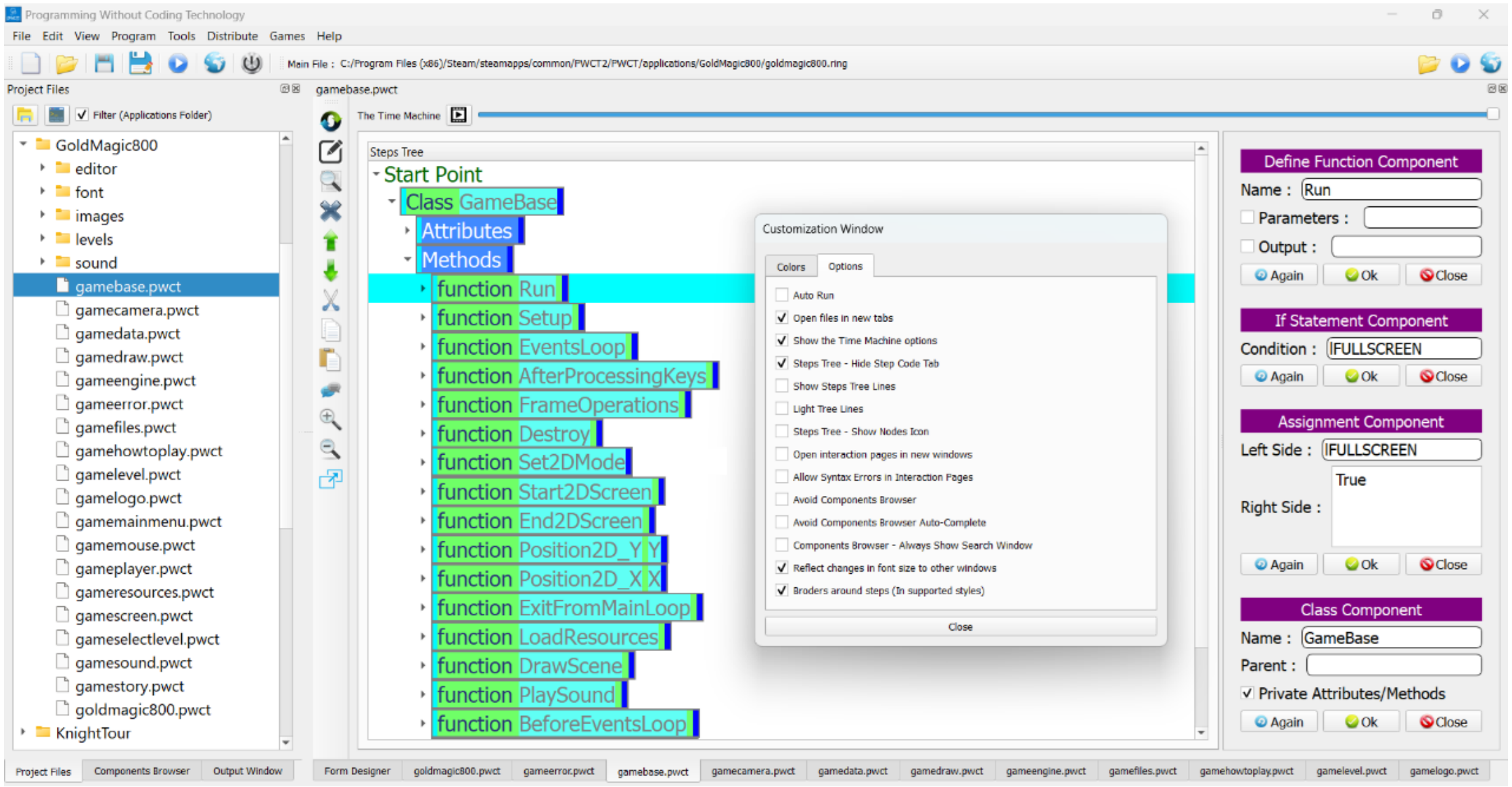

***Figure 5.8** Opening multiple visual source files.*

For projects with many files, we have the option of a Main File (available in the toolbar), which we can run at any time to start the project, even if we are focusing on another sub-file within the project. Switching between files is easy using tabs that include the file name and are located at the bottom of the window. We can also view two files side by side in the PWCT2 environment by moving the dock-able window of the file. On the left side, there is a “Project Files” dock-able window displaying a list of files and folders related to a project named “GoldMagic800”, which is a puzzle game that contains 44 levels [152].

The highlighted file is “gamebase.pwct”. In the center, there is a “Steps Tree” dock-able window showing a hierarchical list of methods under the “Class GameBase”. On the right side, there are interaction pages for defining function components, if-statement components, assignment components, and class components. The user can determine which interaction pages to open simultaneously and use them to generate new steps and create programs without having to switch back to the components browser window.

## 5.6 Rich Comments using Text, Lines, Images and HTML

PWCT [12] enables adding comments to our Steps Tree, referred to as “user steps”, which use text to describe something or add information about the program. In PWCT2, we use different types of comments, not just text. We can add lines, images, and headers with specific fonts and colors. In Figure 5.9, we have a program that uses a Raspberry Pi Pico to control an LED, making it blink, with the ability to interact using a switch [153].

Using rich comments, we added a header that represents the sample title, followed by a horizontal line, and then an image of the circuit. Comments could include HTML, as demonstrated in Figure 5.10. This feature could be used for adding tables from both local and online web pages to our program. This improvement is inspired by the Envision visual programming system [154].

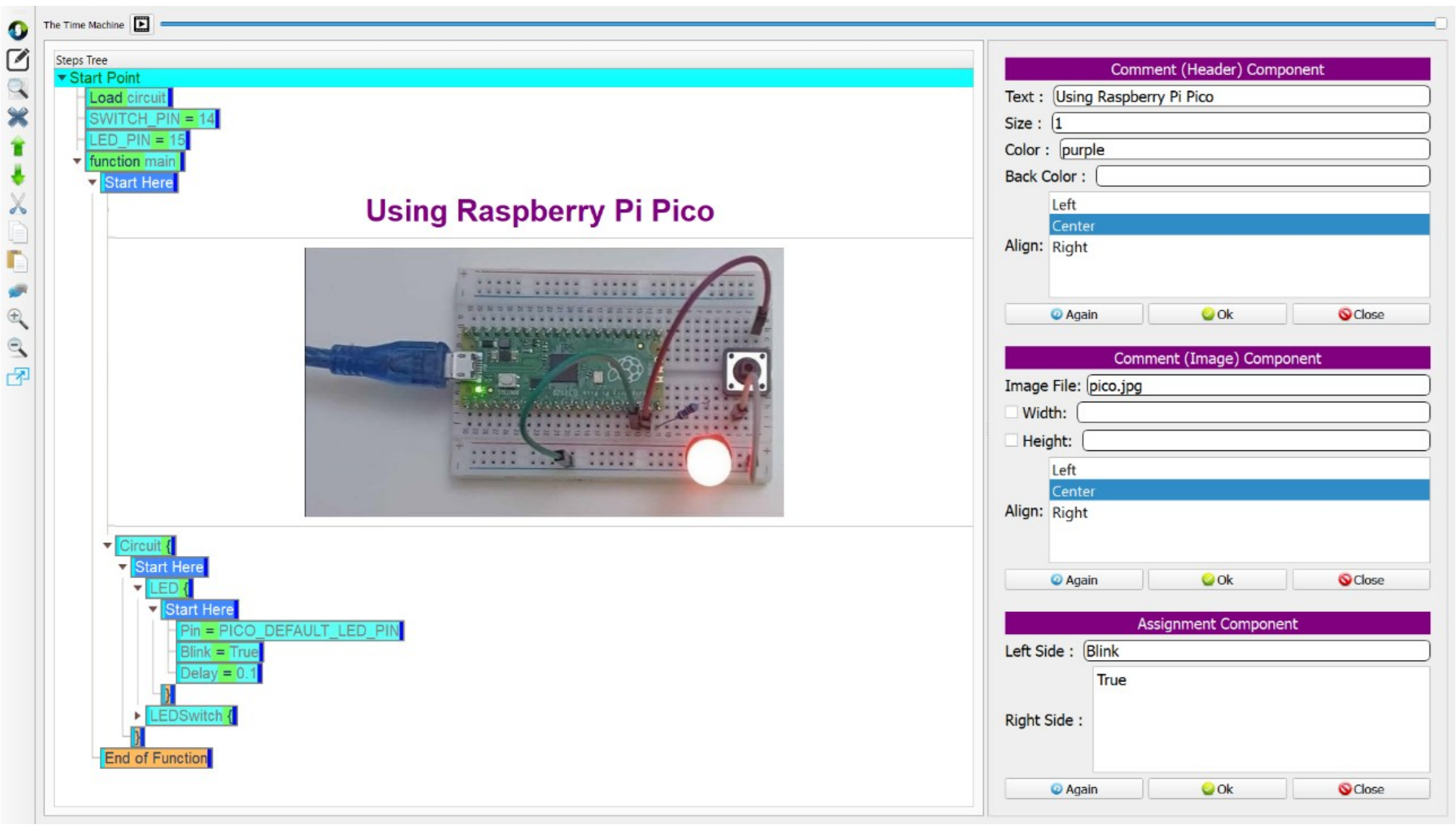

***Figure 5.9*** *Using rich comments (Lines, Images, and Headers).*

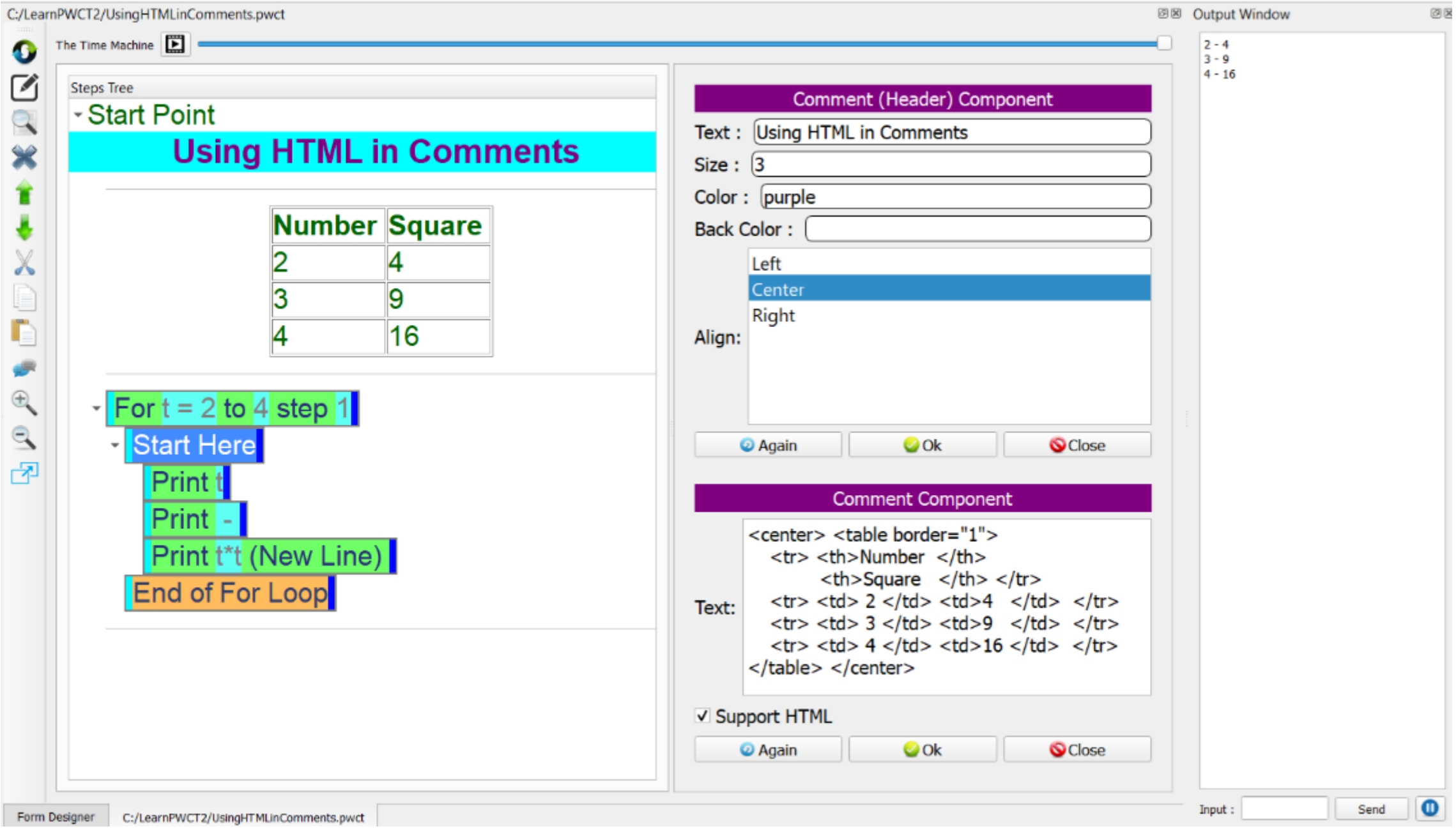

***Figure 5.10*** *Using HTML in comments.*

## 5.7 Enhanced Form Designer for GUI Apps

The Form Designer is a tool that simplifies the process of creating user interfaces by allowing users to easily add UI controls to the form, adjust properties, and define events such as click actions for buttons. In PWCT, the Form Designer, used for GUI applications, differs from popular form designers like those in Visual Studio. Instead of using a Toolbox and Properties window, the PWCT Form Designer utilizes the Components Browser and Interaction Pages, making it tightly integrated with the Goal Designer and not suitable for use as an external tool [155]. In PWCT2, we revised this design and provided a reusable Form Designer, which is successfully used in both PWCT2 and Ring Notepad (the IDE provided by the Ring programming language). In PWCT2, the Form Designer uses the Model-View-Controller (MVC) design pattern [132,156,157]. The Form Designer, as shown in Figure 5.11, allows users to visually design the user interface of their application. It includes the Toolbox, Form Layout, and Properties Panel. The Toolbox contains various UI elements, like labels, buttons, text edits, checkboxes, etc., that can be added to the form. The Form Layout is the main area where UI elements are placed. For example, in the figure, we notice it includes UI controls such as Title, Author, Abstract, and Output. This layout is related to an application that predicts the citation count of research papers in the field of Otology. The Properties Panel shows the properties of the selected object (e.g., a button named btnSelect). Properties include Name, Position (X and Y), Size (Width and Height), Text Color, Background Color, Font, Text, Image, and Click Event.

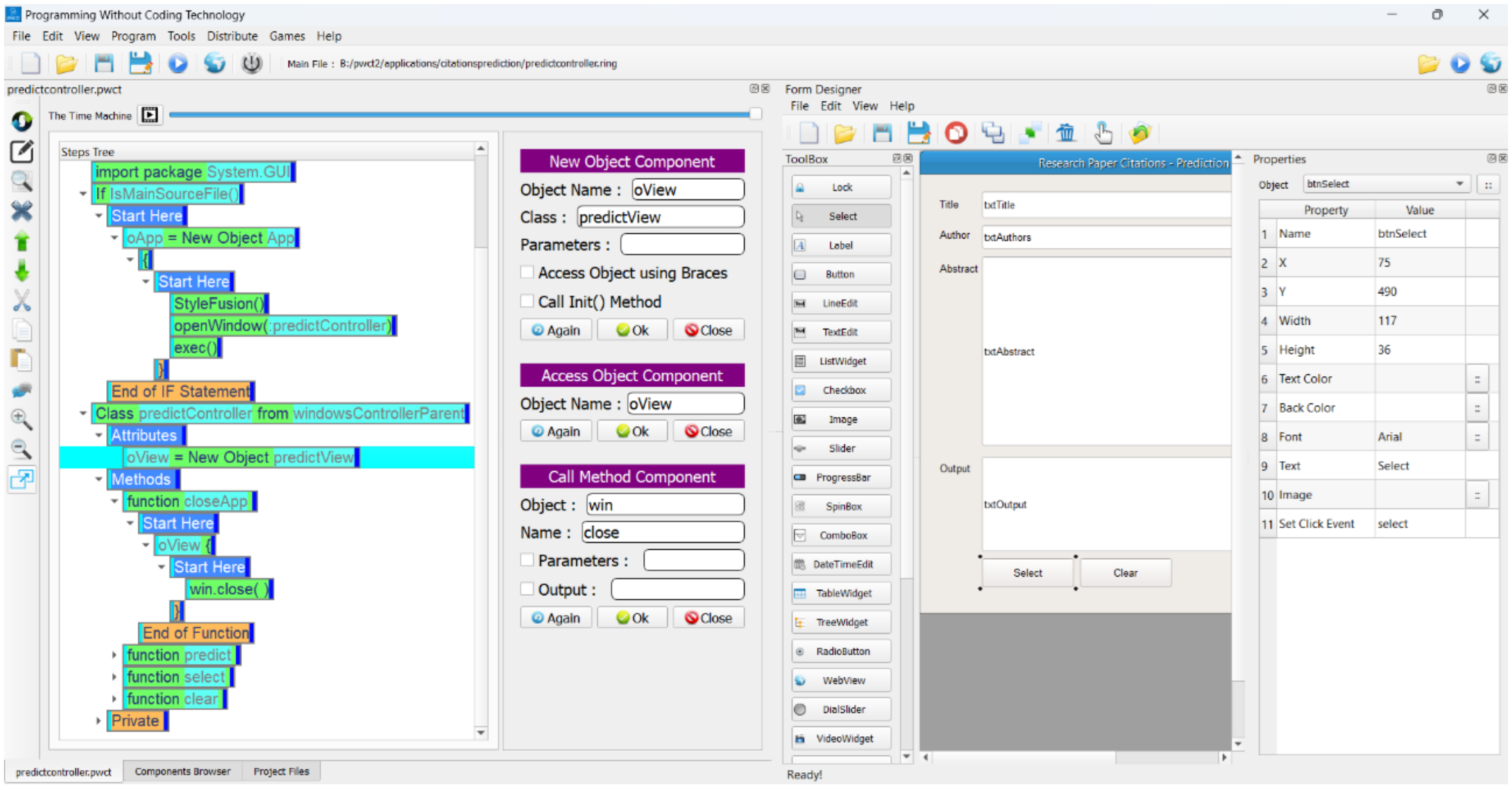

***Figure 5.11*** *Using the PWCT2 Form Designer.*

The Steps Tree (left panel) displays a hierarchical view of the steps involved in the Controller visual source file, including importing the System.GUI package, checking if the file is the main source file, creating and accessing objects, defining the class, and more. The New Object Component (top middle panel) allows users to create a new object from a specific class. The Access Object Component (middle panel) is used to access an existing object within the program and allows us to use the object's attributes and methods. The Call Method Component (bottom middle panel) is used to call a specific method supported by a specific object.

## 5.8 Interactive Textual-to-Visual Code Conversion

PWCT2 comes with a feature called interactive Textual-to-Visual code conversion, as demonstrated in Figure 5.12. The first section of the image illustrates the user writing Ring code, showcasing the textual input of programming constructs. In this section, the code creates an object from the point class, sets the object's attributes, and defines the point class using the Ring programming language.

In the second section, this Ring code is automatically converted into a visual representation using Ring2PWCT. The visual interface displays the code elements as graphical components, arranged in a logical flow that mirrors the structure of the original Ring code. This conversion allows users to interact with and manipulate the program visually, making it easier to understand, modify, and debug without delving into textual syntax. The transition from text to visual representation bridges the gap between textual programming and visual programming, enhancing the user's experience and productivity.

To start using this feature, it is sufficient for the keyboard Focus to be active at the Steps Tree at any step. Once the user starts typing, the components browser will be activated, where PWCT2 expects that the user is searching for a visual component to use. If the visual component does not exist, PWCT2 expects that the user will type Ring source code and highlight this code. If the user presses ENTER, PWCT2 will then use Ring2PWCT to generate the visual representation. The generated steps will be inserted into the selected location in the Steps Tree, enabling us to use this feature at any location in our program.

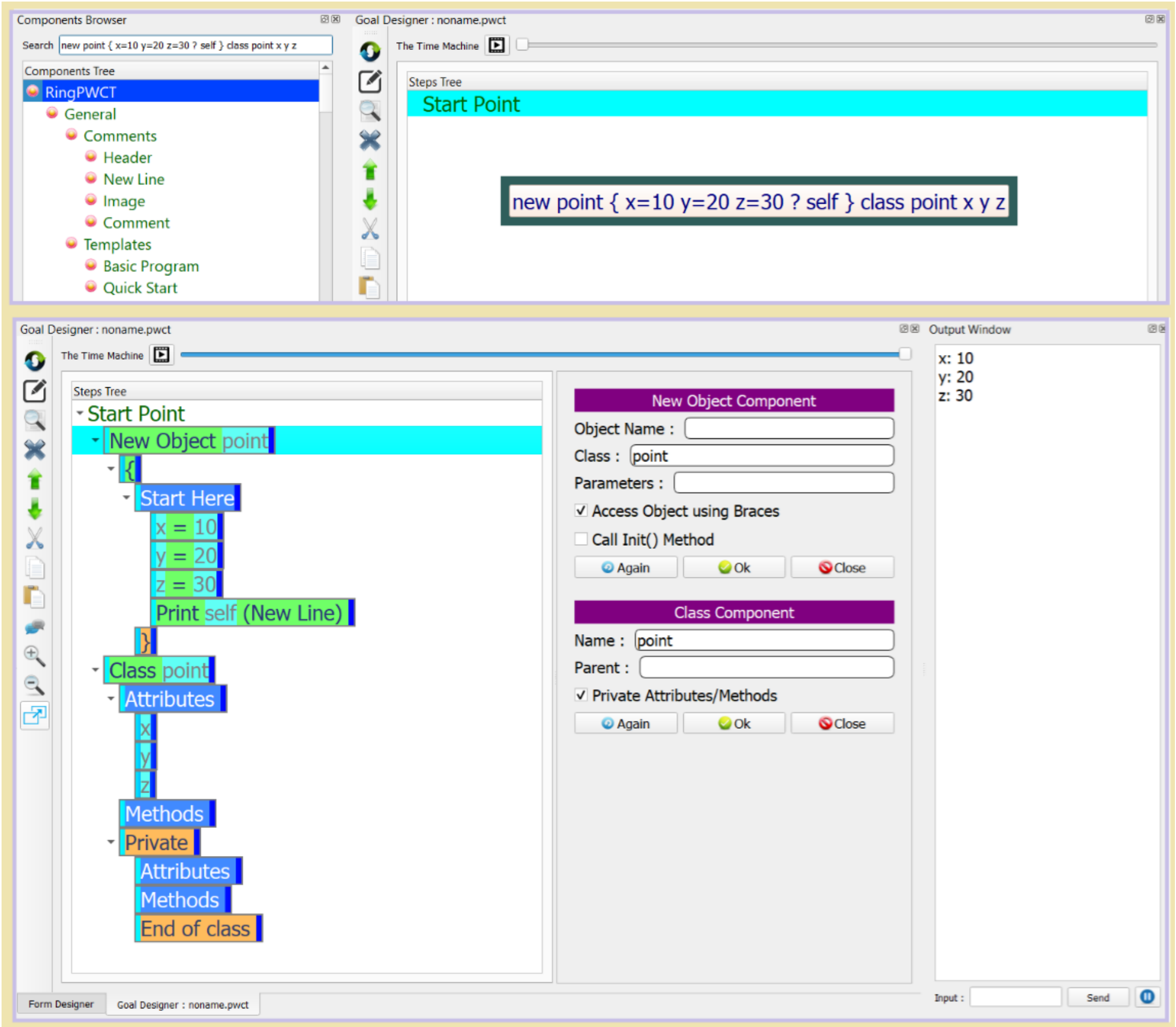

***Figure 5.12*** *Interactive Textual-to-Visual code conversion (Ring2PWCT).*

This feature can be used to mix textual programming (for writability) and visual programming (for readability) in the same project [158,159]. Additionally, the code could be generated using large language models (LLMs) such as Copilot [160] as demonstrated in Table 5.3. Furthermore, this feature enables us to import current projects written in Ring so that we can continue developing them using PWCT2.

The implementation of Ring2PWCT involves some of the same phases as compiling, such as scanning and parsing [24,47]. Instead of low-level code generation, we generate a visual source file. Alternatively, it can update the current file if Ring2PWCT is used to insert steps into the existing visual source file instead of creating a new one. Ring2PWCT is generally considered a form of translation, specifically source-to-source translation, because it converts source code from one high-level language to another (in this case, from Ring code to PWCT visual code) [161].

*Table 5.3* *Using Copilot and PWCT2.*

<table>
<tr><th>Step</th><th>Description</th><th>Image</th></tr>
<tr><td>1</td><td>Use Copilot AI to generate textual source code in the Ring programming language. The task is to write a method in Ring code to be added to a class called Point, which has three attributes: x, y, and z. The method should ask the user for the values of these attributes using the getNumber() function and print a message asking the user to enter the value before using getNumber().</td><td>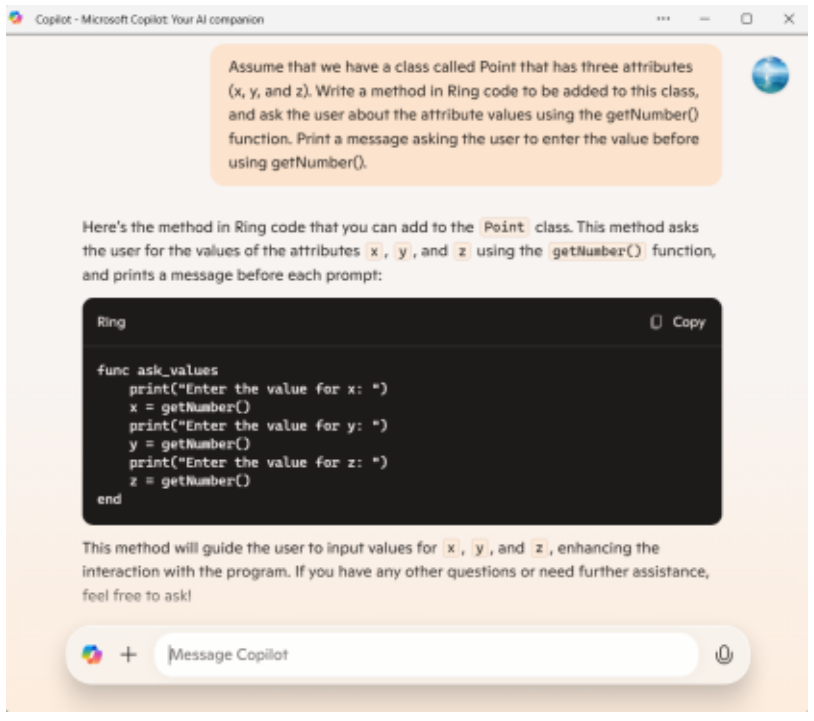
</td></tr>
<tr><td>2</td><td>In this step, we switch to PWCT2, select the parent step in the Steps Tree Editor, and finally paste the generated source code. The parent step is called “Methods”, which exists in the Point class. Once we paste the textual source code, PWCT will display the pasted code in a popup rectangle to indicate that this is a textual-to-visual code conversion operation (not a search operation by the visual component name). Pressing ENTER will start the conversion process.</td><td>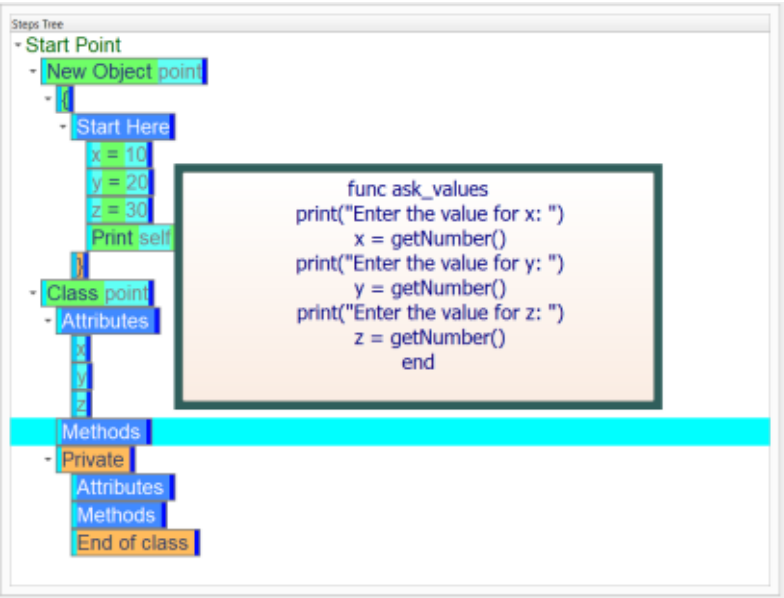
</td></tr>
<tr><td>3</td><td>After the textual-to-visual code conversion process using Ring2PWCT, we will see the generated steps in the Steps Tree. We can continue development using the visual programming features provided by PWCT2, such as the Steps Tree Editor and Time Machine.</td><td>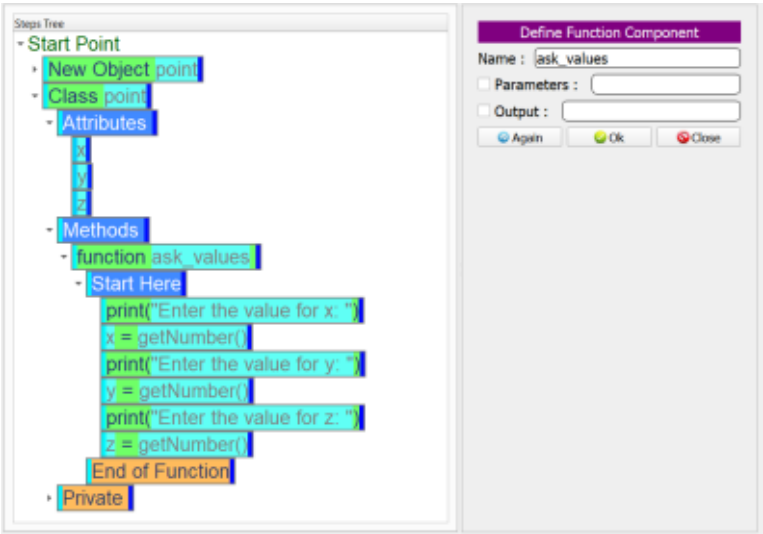
</td></tr>
</table>

Ring2PWCT is designed to pass errors when possible. For example, using the Ring code (for x = 1 to 10 if x = 3 ? “Three”) as input can be converted to visual code (similar to the first section in Figure 8), even if keywords such as EndFor and EndIf are missing.

## 5.9 Self-hosting the PWCT2 environment

The PWCT visual programming language is developed using Visual FoxPro, where enhancing the development environment requires using VFP textual code. Although the environment includes a domain-specific language (RPWI) for visual component development, it remains textual code, and using visual programming to improve PWCT itself is not supported. Therefore, self-hosting PWCT2 could be an attractive feature because it allows the development and customization of the PWCT2 environment using its own visual programming tools.

This approach enables us to improve and refine PWCT2 through visual programming instead of traditional textual coding, making the process more user-friendly and accessible, as shown in Table 5.4, which demonstrates converting a class from textual code to visual code.

While PWCT2 is designed to support interactive textual-to-visual code conversion, this feature is intended to be used within the PWCT environment inside the Steps Tree Editor. Based on this feature, we developed a command-line tool that can take a Ring source file as input and produce a visual source file as output.

Since PWCT2 and its visual components are written in Ring, we used this tool (Ring2PWCT) to convert the PWCT2 source code files from Ring to visual source files, allowing us to use PWCT2 to continue developing itself, addressing the sixth research question (RQ6). In Table 5.4, we present an example of using Ring2PWCT to convert one of the source code files related to the PWCT2 implementation using the Ring language.

This file contains the View class for the (Print Text) visual component. This class inherits from the ComponentViewParent class, which contains common attributes and methods useful for developing new visual components inside PWCT2.

*__Table 5.4__ Using Ring2PWCT to convert the PrintComponentView class.*

<table>
<tr><th>Attribute</th><th>Value</th></tr>
<tr><td>Class Name</td><td>PrintComponentView</td></tr>
<tr><td>Parent Class</td><td>ComponentViewParent</td></tr>
<tr><td>Textual Source<br>(Ring)</td><td>class PrintComponentView from ComponentViewParent<br><br># "Print Component"<br>Title(T_CT_PRINT_IP_TITLE)<br># "Text : "<br>myText = TextBox(T_CT_PRINT_IP_TEXT , :text)<br># "Type :"   ["Literal","Expression"]<br>DisableListBoxSort()<br>myList = ListBox(T_CT_PRINT_IP_TYPE, :type , T_CT_PRINT_IP_TYPELIST)  {<br>setCurrentItemChangedEvent( Method(:TypeChangeAction) )<br>}<br># New Line<br>CheckBox(T_CT_PRINT_IP_NEWLINE,:newline) {<br>setcheckstate(2)<br>}<br>PageButtons()</td></tr>
<tr><td>Visual Source<br>(PWCT2)</td><td>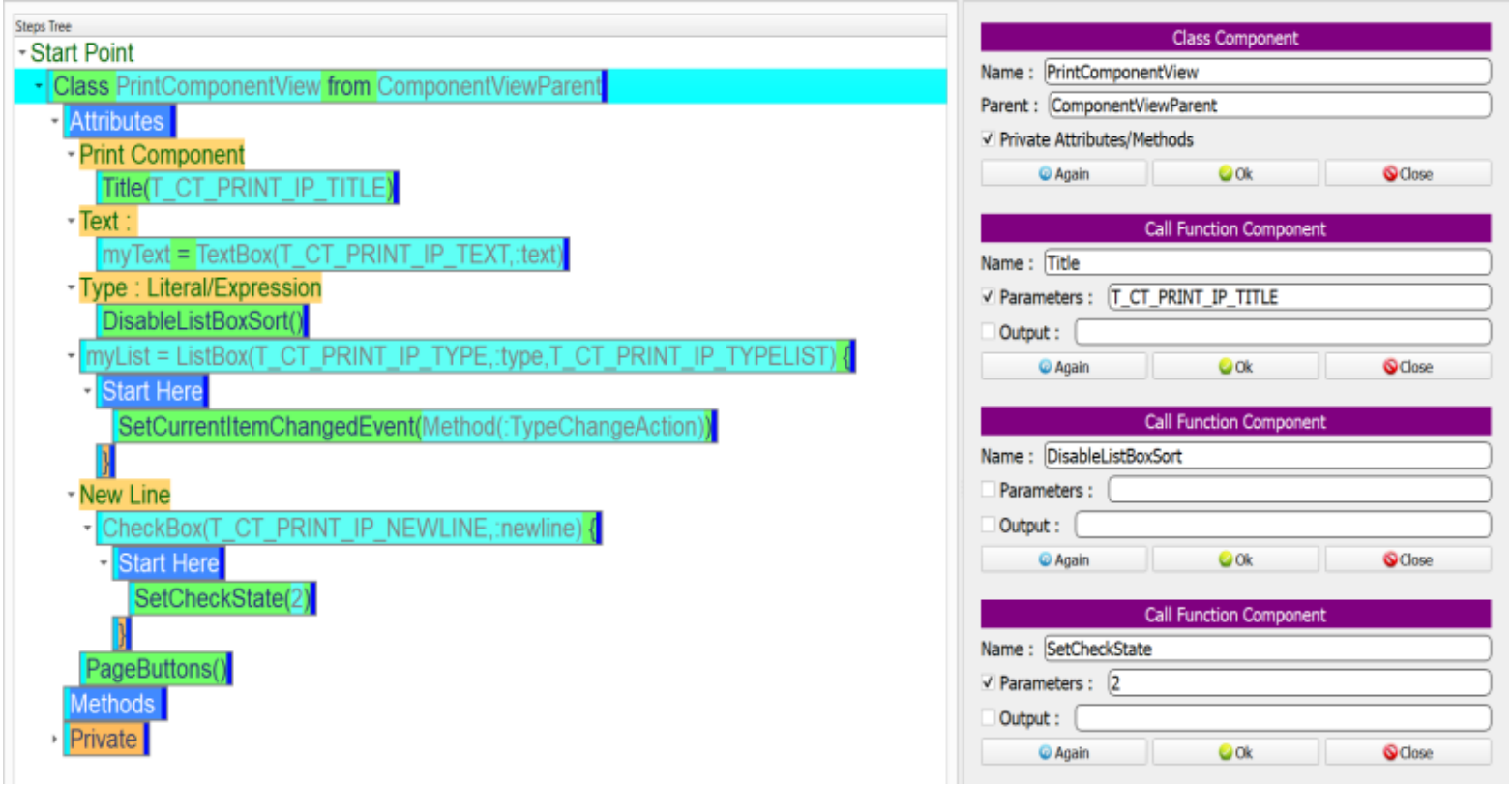
</td></tr>
</table>

## 5.10 Cross-platform and Faster Implementation

PWCT was developed as a Microsoft Windows product, with no native support for other operating systems like Linux and macOS. Additionally, PWCT is 32-bit software. Since PWCT2 is developed using the Ring programming language, which is a lightweight, versatile, and cross-platform language, it was possible to create a cross-platform and 64-bit version of PWCT2 based on Ring and the Qt framework, supported by the RingQt extension. In Table 5.5, we present the modules, along with the count of files and lines of code. Figure 5.13 demonstrates using PWCT2 on macOS to develop the Tetris game using the Ring game engine for 2D games, which is based on the Allegro game programming library [162,163]. The game engine provides classes that can be used to quickly prototype simple 2D games. These classes include Game, Text, Sprite, Animate, Map, etc.

*Table 5.5* PWCT2 modules.

| ID | Module | Files | LOC | Comment |
|---|---|---|---|---|
| 1 | Environment | 5 | 2649 | 300 |
| 2 | General Functions | 9 | 524 | 122 |
| 3 | Translation | 3 | 584 | 20 |
| 4 | Goal Designer | 27 | 4908 | 1473 |
| 5 | Components Browser | 5 | 8876 | 70 |
| 6 | VPL Components | 1185 | 57,612 | 7167 |
| 7 | Component Parent Classes | 3 | 739 | 283 |
| 8 | Form Designer | 52 | 9487 | 312 |
| 9 | File System (Visual Source Files) | 6 | 368 | 415 |
| 10 | Tools | 59 | 6484 | 546 |

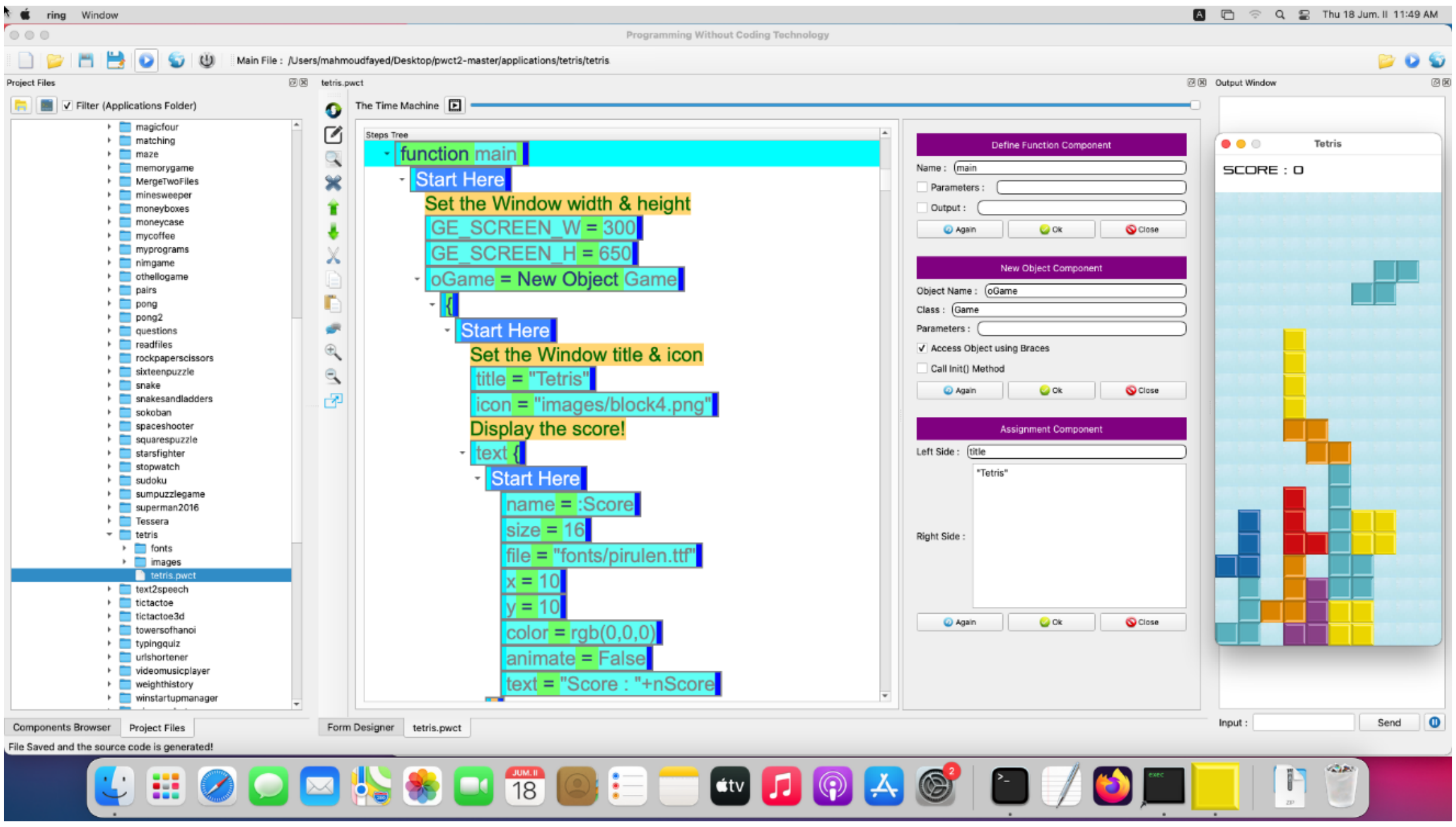

***Figure 5.13*** *PWCT2 for macOS.*

The PWCT2 project is one of the early advanced projects developed using Ring (see Figure 5.14). The Environment module provides the main window and creates the different dock-able widgets. The General Functions module encompasses general functions required by the software. The Translation module handles various translation functions. The Goal Designer modules contain the Steps Tree Editor and the Time Machine. The Components Browser allows users to navigate through available components and select a component to use.

The VPL Components module, being the largest, includes all the RingPWCT visual programming language components necessary for creating and managing programs. The Component module provides the foundational elements required for the different visual components, including interaction pages and generating visual steps. The Form Designer module provides functionality for designing forms within the environment. The File System module manages the visual source files. Finally, the Tools module contains various utilities, including the textual-to-visual code converter (Ring2PWCT).

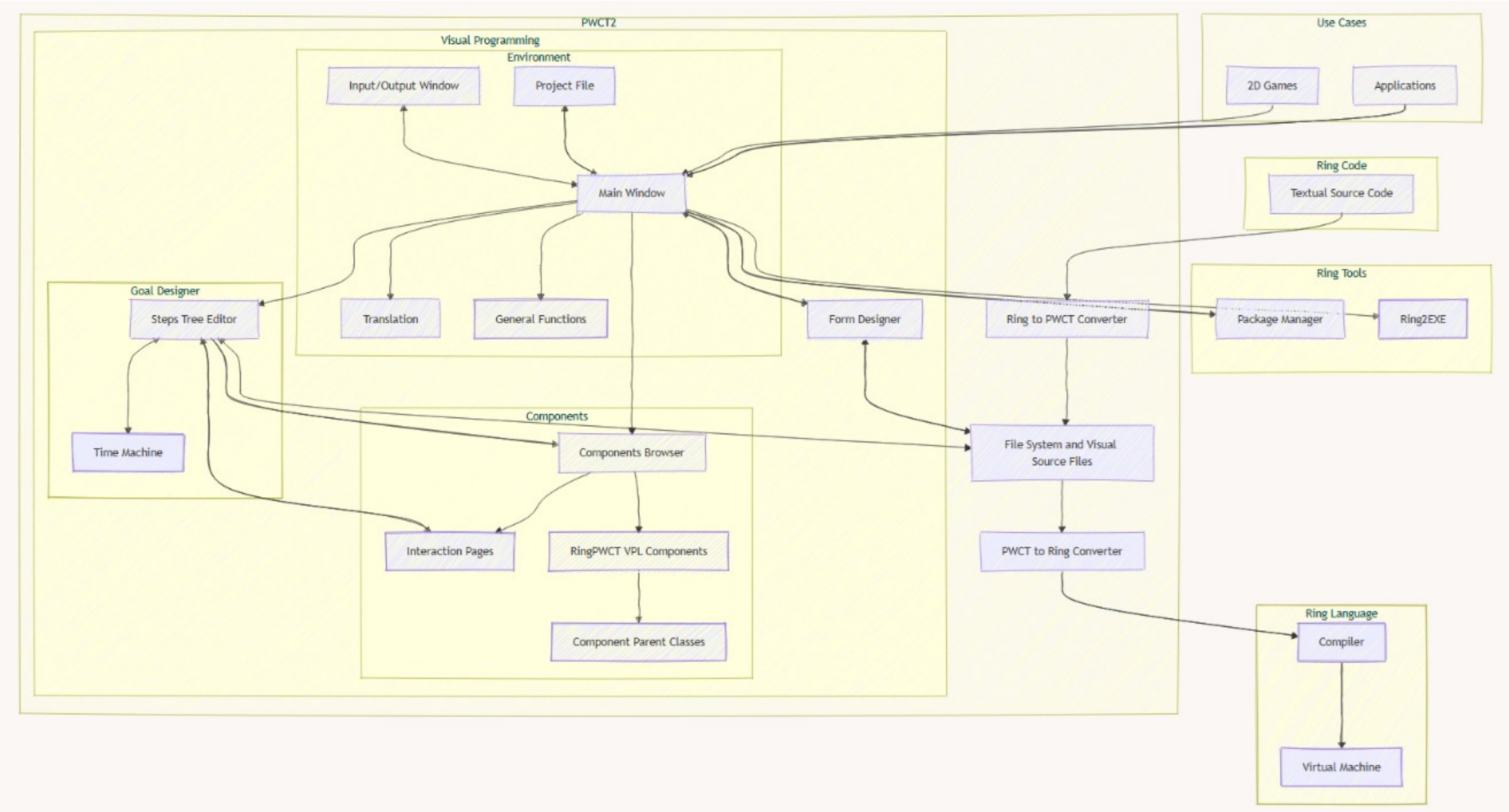

***Figure 5.14*** *The PWCT2 system developed using the Ring programming language.*

During the development of PWCT2, we made specific design decisions that resulted in improved performance compared to PWCT. These design decisions helped enhance various aspects of PWCT2, with notable improvements in code generation time and storage requirements for the visual source files.

These design decisions are as follows:

- The Steps Tree is stored in the visual source files in the correct order of control flow, with PWCT2 adding steps at the end of the file or inserting them based on their actual position. In contrast, PWCT always adds new steps at the end of the file, requiring the Steps Tree to be ordered during the code generation process.
- Storing the visual source in memory through Ring Lists during development and saving to storage only when needed, instead of using database files and storage on the hard disk during development as in PWCT. It is known that accessing computer memory is faster than accessing the hard disk drive.

- Using the Ring programming language instead of Visual FoxPro. Since Ring development is active, the latest versions of the Ring Compiler/VM are produced using the latest versions of C/C++ compilers, which also benefit from the performance improvements in these tools.
- Using the Qt framework for the GUI environment through RingQt. The framework is written in C++ and provides better performance with each update and direct support for the TreeView control instead of using ActiveX control as in PWCT.

## 5.11 Arabic PWCT2

Since the PWCT2 visual programming language is designed to support translation, besides the default English version, we provide a complete Arabic version, as demonstrated in Figure 5.15. In this figure, we see a program that prints numbers from one to ten. We print a text message before printing the numbers and another one after printing half of the numbers.  The translation covers various components in the system, such as the main window and sub-windows. Additionally, each of the 394 components in the RingPWCT visual programming language is translated, including the interaction pages' user interface and the generated steps inside the steps tree. Furthermore, the form designer is translated to use Arabic names for all user interface elements, including the toolbox and properties window.

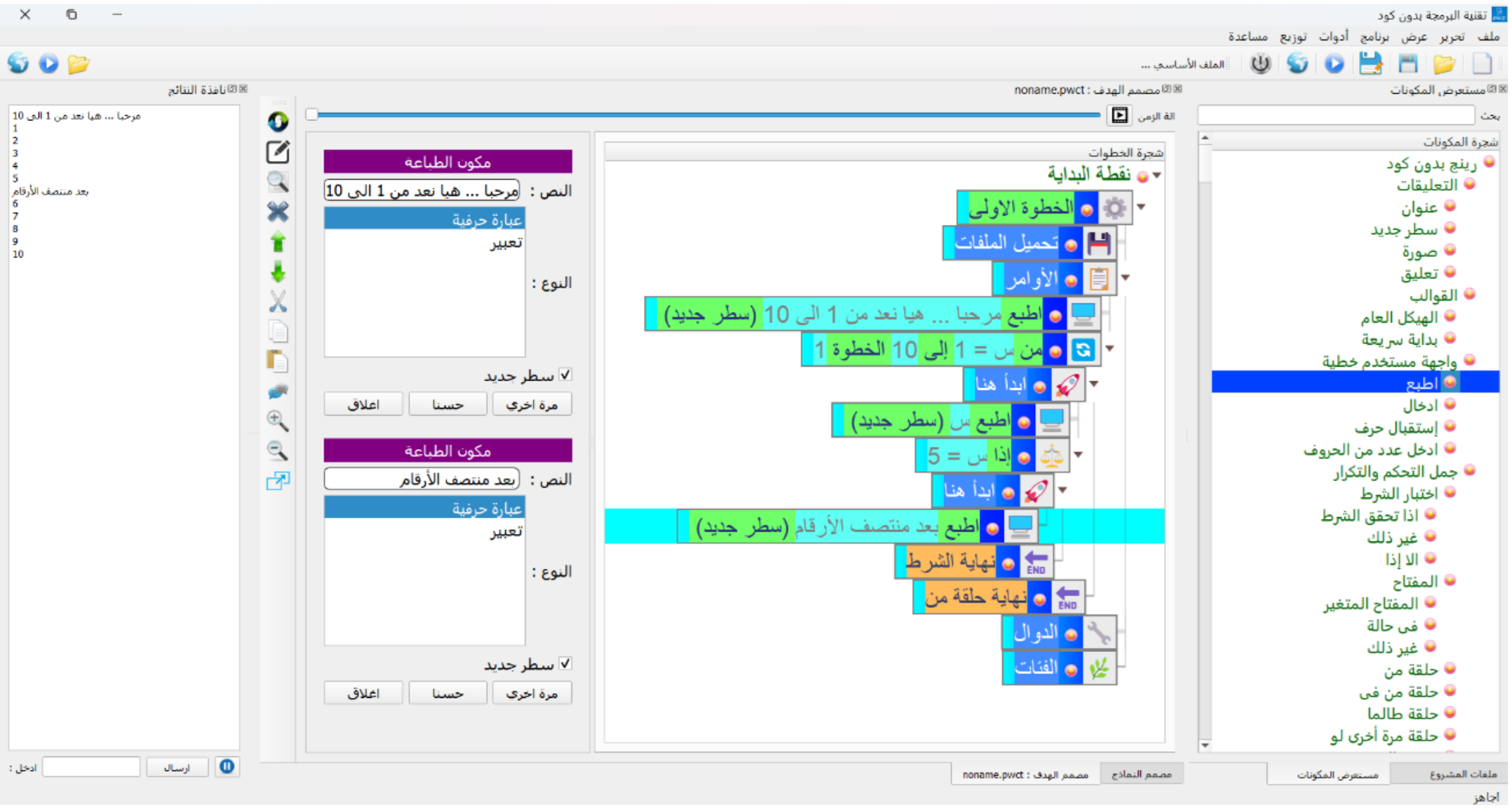

***Figure 5.15*** *Arabic translation for the PWCT2 visual programming language.*

## 5.12 Chapter Summary

In this chapter, we introduced the PWCT2 visual programming language, outlining its design and key features. We examined how PWCT2 builds upon existing visual programming languages like PWCT and addressed the unique contributions it brings to the field. Through this exploration, we have highlighted the strengths of PWCT2, demonstrating its potential to enhance visual programming and software development based on the Ring programming language.

In the next chapter, we will present the experiments and results that evaluate the effectiveness and performance of both the Ring dynamic programming language and PWCT2 visual programming language. This analysis will provide empirical evidence to support the claims and contributions discussed in the preceding chapters.

# Chapter 6: Experiments and Results

## 6.1 Introduction

In this chapter, we present the various results related to our study. First, we introduce results related to the Ring programming language then we introduce the results related to the PWCT2 visual programming language. The results include different aspects like use-cases, users feedback, performance evaluation, etc.

## 6.2 Results related to Ring

In this section we introduce results related to the Ring programming language. At first, we provide information about the early users, followed by download statistics. Next, we discuss multiple use cases. Additionally, we delve into our findings concerning Ring's visual implementation using the PWCT visual programming language. Then, we present the results related to Ring's lightweight implementation, followed by the performance benchmarks.

### 6.2.1 Early Users and the Programming Language Used Prior to Ring

Once we launched the Ring website in 2016, we posted a message in the Ring Group seeking users interested in trying or testing the language and contributing by reporting bugs. In the public group, interested users shared their age, gender, country (location), and the programming languages they used prior to Ring. We noticed 43 messages, with 42 males and 1 female. Most of the users are between the ages of 20 and 35, and 81% reported that they were using C++, PHP, C#, Java, or Python, as demonstrated in Figure 6.1. We noticed that 28 users (65%) were using statically typed languages, while 15 users (35%) were using dynamically typed languages. This diverse usage background reflects the rich experience of our users with different programming languages, leading to various feature requests in different directions. Developers who used C# requested the development and addition of the Form Designer to our code editor (Ring Notepad), which was developed and added to Ring in version 1.3. Developers with a C/C++ background asked for features related to C/C++ extensions, leading to the revision and improvement of the Ring API and the addition of tutorials on using it. Additionally, developers who used PHP for web development requested better support for web development, which led to the addition of the Apache web server to Ring Notepad in Ring 1.6. Those users helped us discover and fix many issues. They also improved the Ring documentation by adding the Frequently Asked Questions (FAQ) chapter. Over time, they contributed over 800 samples of the Ring language to the RosettaCode website.

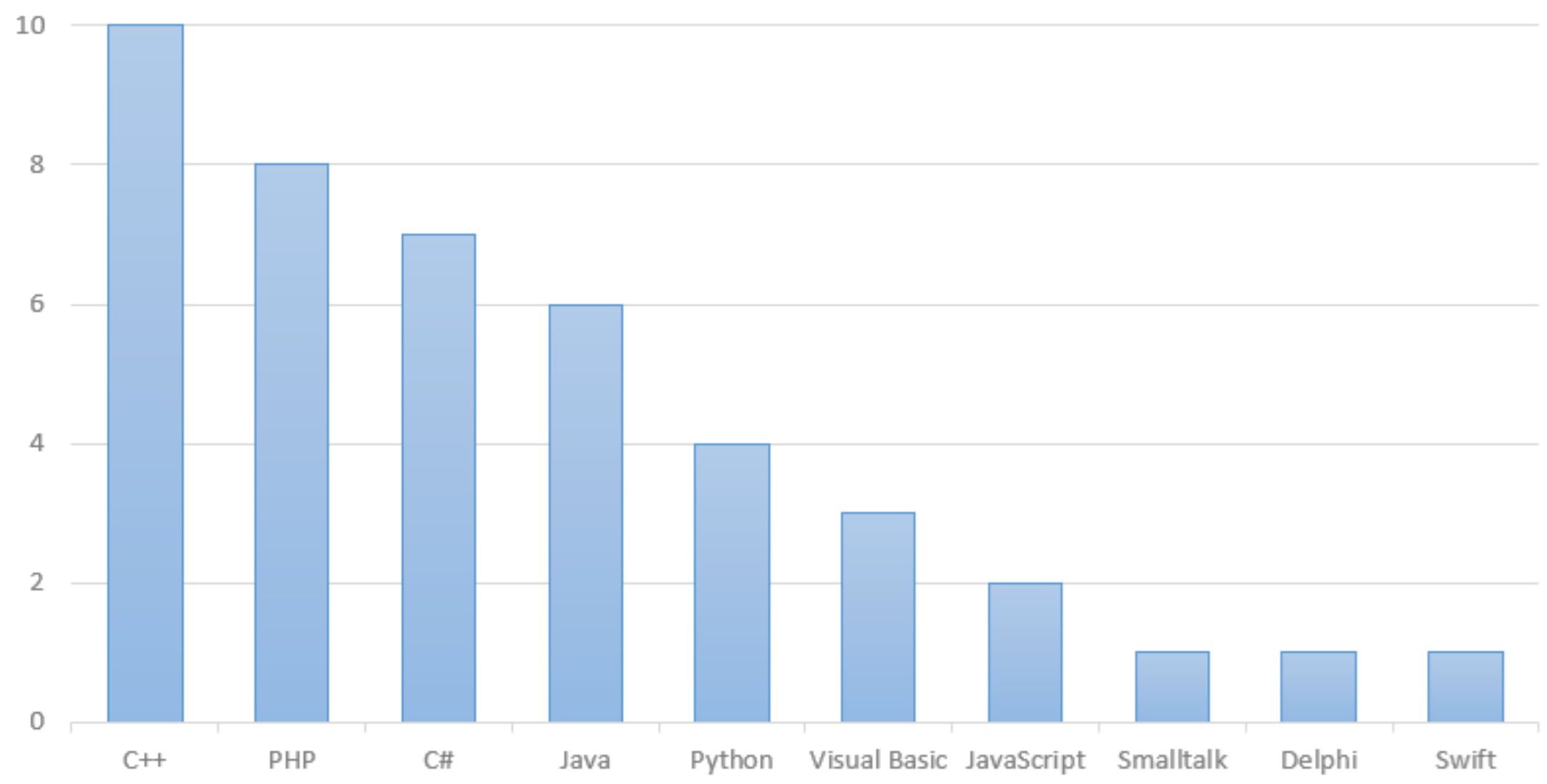

***Figure 6.1*** *Early users and the language used prior to Ring.*

### 6.2.2 Feedback from Online Course

We presented a free online course consisting of 18 videos in Arabic that introduced the Ring programming language (covering input/output, control structures, procedural programming, and object-oriented programming). The course is available on YouTube (youtube.com/playlist?list=PLpQiqjcu7CuFc027iGHaBLPCZHuzCHkBC). We then invited interested learners to watch the course and submit the samples they wrote during their learning through GitHub so we could track their progress. We received samples from 76 participants.

In Table 6.1, we present the course content, while in Table 6.2, we introduce the statistics about the course. Twenty participants (26.3%) were not interested and finished fewer than two lessons, while 56 participants (73.7%) were interested and finished two or more lessons. Of those 56 participants, 23 (30% of the total) finished the course. We noticed that two participants became active contributors to Ring language samples and applications. The contributors help us test, report bugs, and add samples, applications, and tutorials. As of 2024, Ring is distributed with hundreds of samples and over 70 applications/games, each ranging from a few hundred to a few thousand lines of Ring code. With respect to the female participants, four of them completed the course, one completed just one lesson, and the last one completed three lessons.

*Table 6.1* Course Content.

| Lesson | Description | Duration (H:M:S) |
|---|---|---|
| 1 | Installing Ring and writing the Hello World program | 0:30:43 |
| 2 | Input/output, data types, strings, and numbers | 0:36:33 |
| 3 | Arithmetic/Logical operators and the if Statement | 0:58:20 |
| 4 | Lists, nested lists, and loops | 1:01:71 |
| 5 | While Loop | 0:48:34 |
| 6 | Defining and using Functions | 0:45:30 |
| 7 | Standard functions | 0:28:39 |
| 8 | Using Eval() | 0:23:01 |
| 9 | Internet Library | 0:38:47 |
| 10 | Database and SQL | 0:33:13 |
| 11 | Classes and Objects | 0:47:07 |
| 12 | Declarative Programming | 0:52:29 |
| 13 | Domain-Specific Languages | 0:19:46 |
| 14 | Domain-Specific Languages (Part 2) | 0:46:33 |
| 15 | Functional Programming | 0:42:37 |
| 16 | Reflection and Meta-Programming | 0:27:57 |
| 17 | Memory Management and variables scope | 0:58:28 |
| 18 | Interactive Debugger | 0:22:08 |

*Table 6.2* Statistics from Online Course.

| Variable | Value |
|---|---|
| Male | 70 |
| Female | 6 |
| Completed less than two lessons | 20 |
| Completed more than one lesson | 56 |
| Completed the course | 23 |
| Contributors | 2 |

### 6.2.3 Feedback After a One-Hour Lecture

We presented a one-hour lecture about the Ring language to third-year students at the College of Computer and Information Sciences at King Saud University in Saudi Arabia. The lecture was presented twice: the first time to 35 students and the second time to 25 students. All 60 students were male. They had studied multiple courses related to programming, including Introduction to Programming and Object-Oriented Programming. They used Java during these programming courses. After the one-hour lecture, we told them, “If you are interested in the Ring language, try to download and install it, write some simple programs, and see if you become more interested in learning about the language”. As shown in Figure 6.2, out of the 60 students, 44 were interested, and all of them successfully installed the language and tried writing some programs using it. One of the students said, “Why don’t we learn Ring instead of Java? It seems easier”. Another student said, “This language looks like Python”.

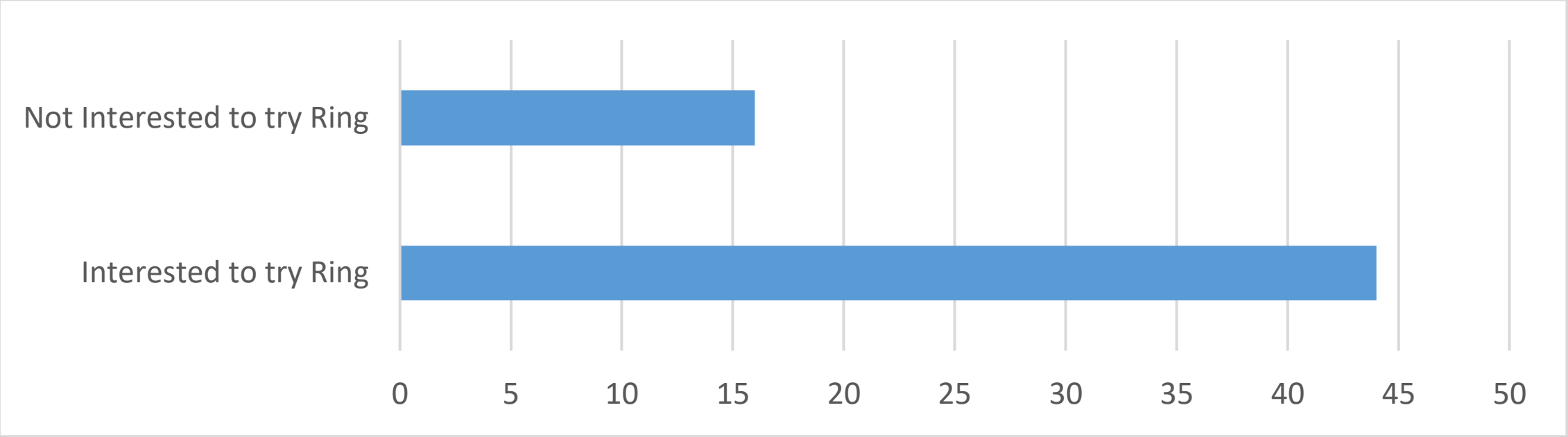

***Figure 6.2*** *Feedback from students about Ring language after a one-hour lecture.*

### 6.2.4 Downloads Statistics and Users Group

Ring, as an open-source programming language, is hosted on GitHub. Users have two options to get the language: they can clone the source code or download a precompiled binary release for Windows, Linux, or macOS.

The project has garnered more than 1200 stars from developers worldwide. To foster discussions about the language, Ring maintains an official Google Group (over 450 members). The group contains conversations covering various aspects of the language across more than 2800 topics [164].

External services tracking GitHub downloads indicate that the project has been downloaded over 18,000 times. Furthermore, a mirror exists for the project files hosted on Sourceforge. This mirror tracks download counts and their associated countries. Impressively, the downloads from this mirror have surpassed 62,500.

In Figure 6.3, we present the operating systems used during downloads, while in Figure 6.4, we present the countries that have the most downloads [165]. We expect that each programming language could be more popular in specific countries due to marketing reasons and the availability of educational resources.

Many YouTube videos about the Ring language are presented in Arabic by Egyptian developers and YouTubers. This could also be one of the reasons Egypt has more users than other countries.

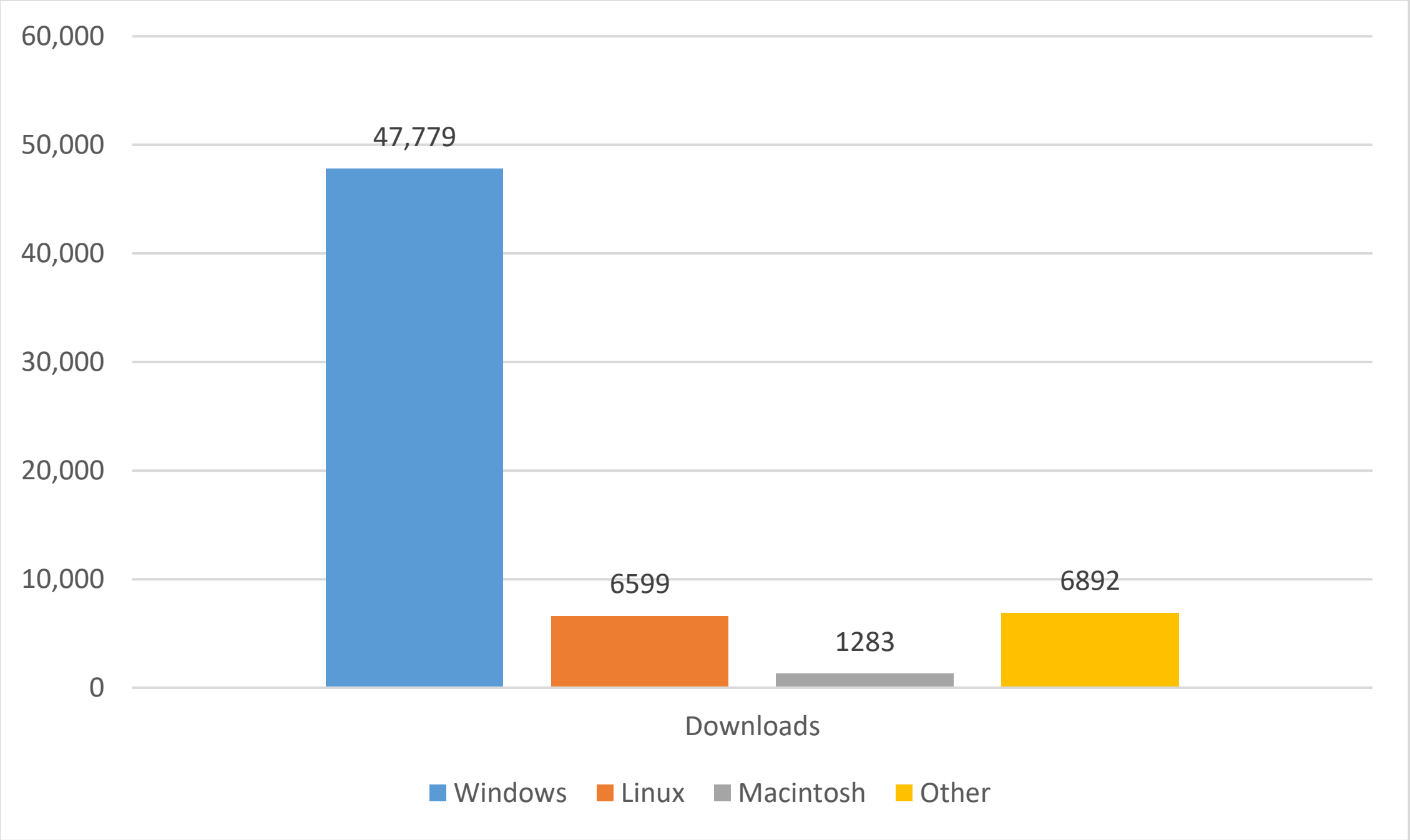

***Figure 6.3*** *Ring downloads statistics grouped by the Operating System.*

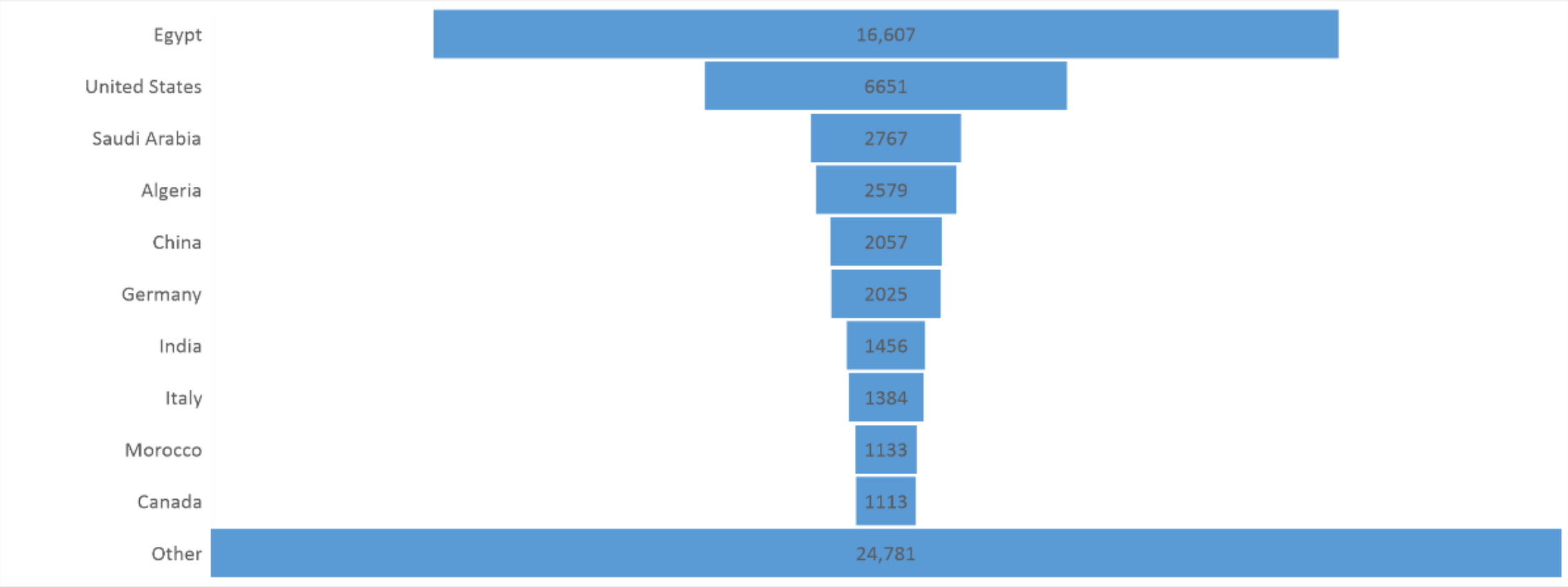

***Figure 6.4*** *Ring downloads statistics grouped by the Country.*

### 6.2.5 Use Cases and Printed Books

In Table 6.3, we present some of the use-cases of the proposed programming language and environment. These uses-cases are related to different domains like Front-end applications for Machine Learning models, Games development, Text/Data processing, and Web development. We selected just one or two use-cases for each domain to avoid unnecessary duplication. For more applications, the Ring language is distributed with over 80 applications/games/tools.

***Table 6.3*** *Some use cases for the Ring programming language and environment.*

| Ref. | Type | Domain | Description |
|---|---|---|---|
| [166] | Research Paper | Front-end apps for ML Models | Predicting citations count |
| [167] | Research Paper | Front-end apps for ML Models | Predicting game result |
| [91] | Printed Book (USA) | Games Development | Shooter Game |
| [152] | Steam Game | Games Development | Puzzle Game |
| [168] | Research Paper | Text/Data Processing apps | Predicting impedance |
| [169] | Printed Book (Egypt) | Text/Data Processing apps | Arabic Poetry Analysis |
| [170] | YouTube Videos | Desktop/Web development | Free course |
| [171] | Research Paper | LLMs Training | Dataset preparation |

The first two use-cases involve utilizing the form designer and the standard libraries such as GUILib, InternetLib, and JSONLib, to develop front-end applications for machine learning models [166,167]. These applications could offer a user-friendly interface that receives input from users. The input is then transmitted to the machine learning model over the internet, and the resulting prediction is returned in JSON format. Afterward, the application processes this data and displays the outcome. Additionally, the GUI (Graphical User Interface) could include features such as data visualization, statistics, or a display for the dataset using the grid control, as demonstrated in Figure 6.5.

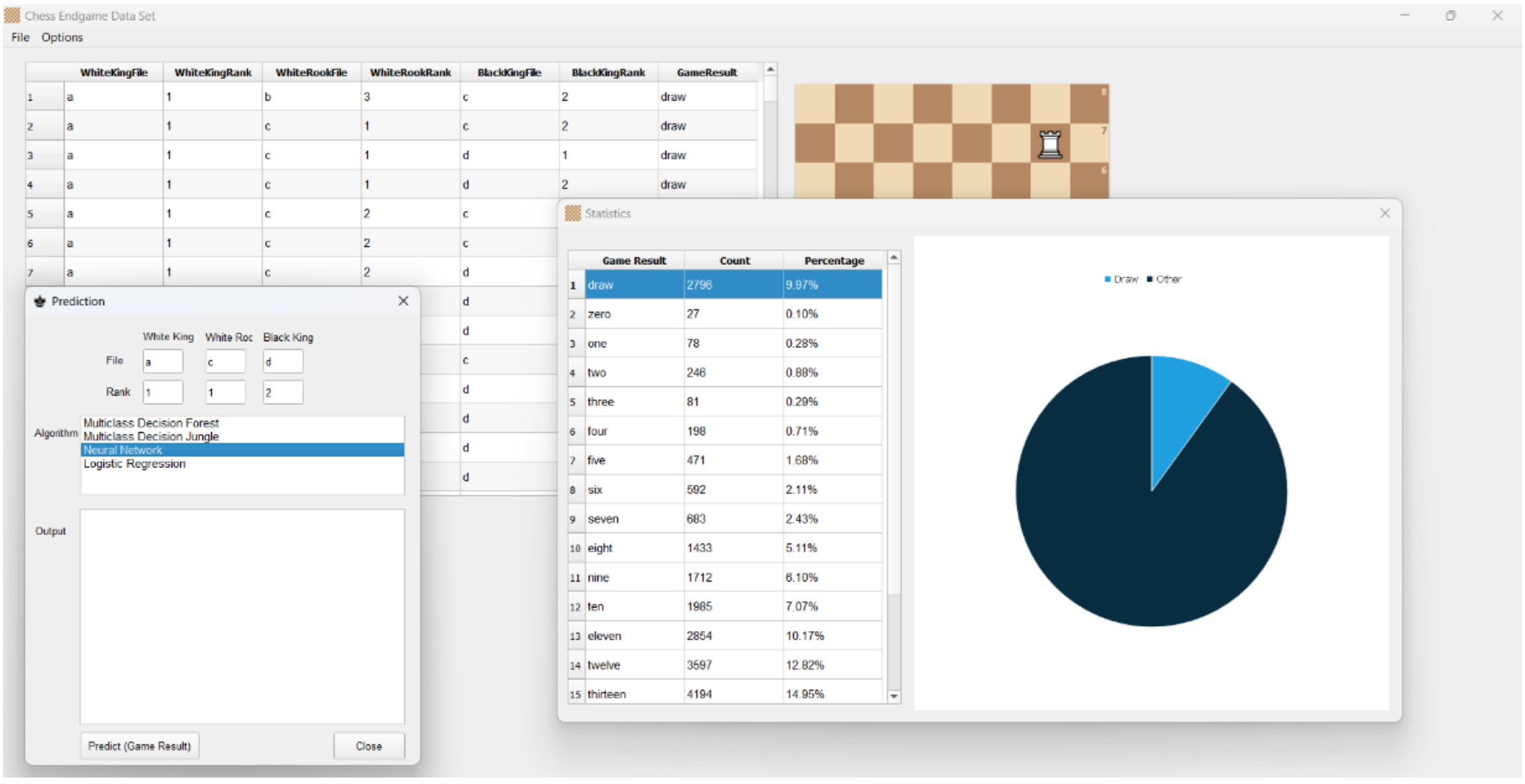

***Figure 6.5*** *A GUI application developed using the Ring language.*

The third use-case is about using the Ring programming language for 2D game development. This is explained through a printed English book (In the USA). The book contains nine chapters and is over 600 pages. The source code is available online through a GitHub project [91].

Figure 6.6 shows a puzzle game available on the Steam platform, written in Ring code and utilizing the Allegro and OpenGL libraries. The game, titled “Gold Magic 800”, comprises 44 levels [152]. These levels have been meticulously designed by a specific level designer, which was also developed using Ring. Notably, the Level editor employs the Qt library. This game serves as an excellent example of how different libraries provided by the Ring language can be seamlessly mixed within the same project.

In [168], the Ring language is used to prepare a dataset before using it to train a machine learning model. Another use-case is developing a Ring program that analyzes Arabic poetry. The application contains over 3000 lines of Ring source code and is explained in detail in a printed Arabic book (In Egypt) [169].

In [170], A YouTube channel with over 350 K subscribers provided over 500 videos about the Ring programming language. These videos start by explaining the language fundamentals and how to apply the different programming paradigms using it. The videos cover desktop and web development, too. In [171], the authors used Ring language samples and documentation to train LLMs how to write Ring programs.

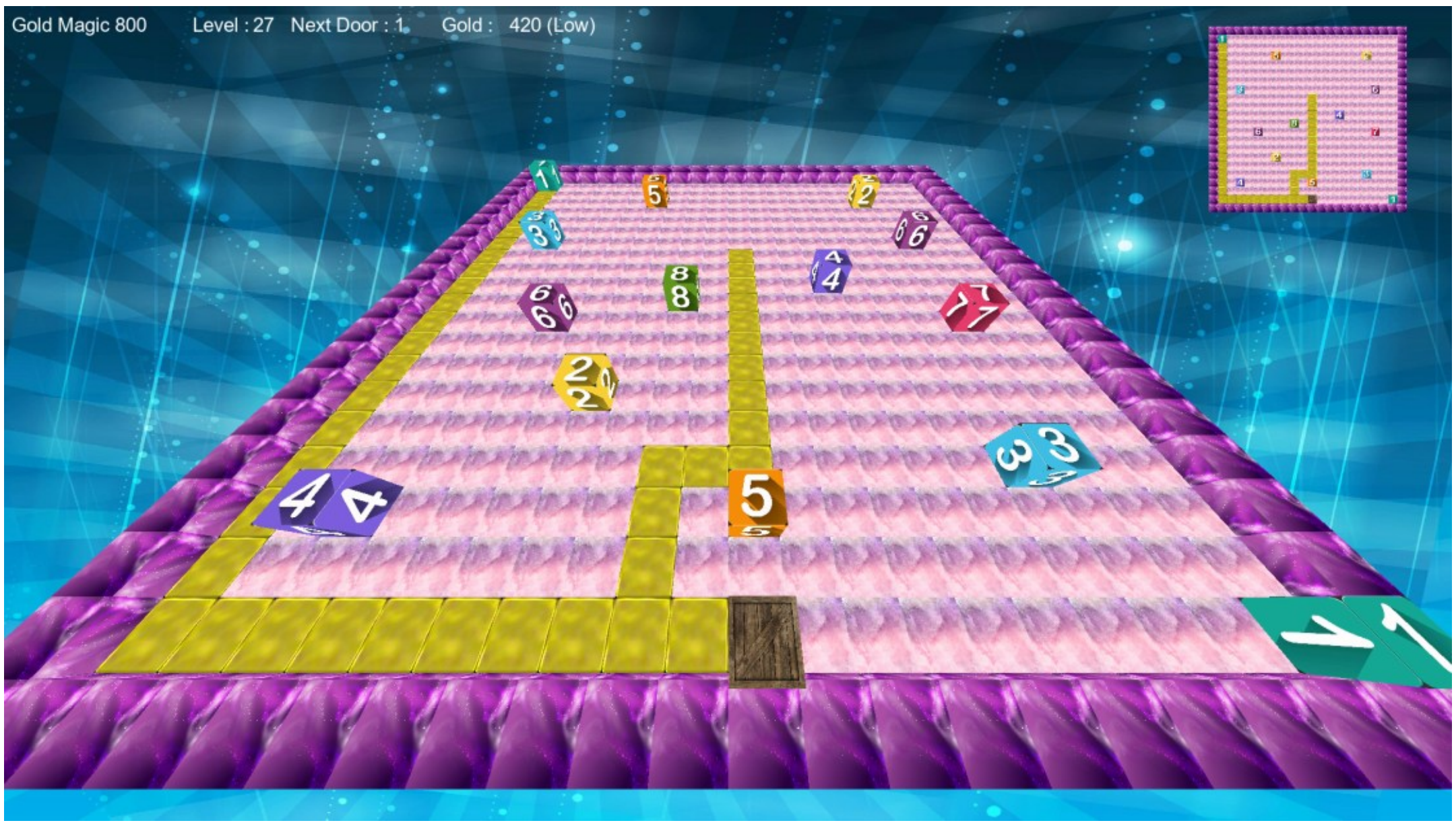

***Figure 6.6*** *The GoldMagic800 game—A puzzle game developed using RingAllegro.*

### 6.2.6 Visual Implementation

In Table 6.4, we present the results of our visual implementation. The Table includes details for each visual source file: the amount of storage used on the hard disk, the memory used by PWCT after loading the file, the number of visual components, the count of steps within the steps tree, the lines of code in the generated source files (.c), and the lines of code in the generated header files (.h) if the visual source also generates such files. Finally, we provide the total lines of source code (without comments/blank lines) generated by the visual source file.

***Table 6.4*** *Results of using PWCT to implement Ring compiler and virtual machine.*

| Modules | File Name | Storage (MB) | Mem. (MB) | Components | Steps | Visible Steps | Comment | LOC |
|---|---|---|---|---|---|---|---|---|
| Loader | ring | 2.07 | 21.5 | 211 | 320 | 287 | 35 | 236 |
| | state | 7.7 | 32 | 513 | 841 | 720 | 77 | 640 |
| General Library | general | 4.32 | 24.6 | 211 | 388 | 321 | 18 | 287 |
| | hashtable | 2.27 | 35 | 189 | 322 | 268 | 16 | 251 |
| | item | 4.04 | 24.3 | 231 | 440 | 356 | 54 | 301 |
| | items | 0.54 | 19.3 | 52 | 87 | 74 | 3 | 63 |
| | list | 10.85 | 40.1 | 969 | 1798 | 1432 | 118 | 1378 |
| | string | 3.94 | 26.6 | 271 | 497 | 399 | 18 | 383 |
| | hashlib | 1.72 | 25.5 | 54 | 81 | 70 | 3 | 59 |
| Compiler | codegen | 5.74 | 28.4 | 425 | 700 | 588 | 68 | 543 |
| | expr | 14.52 | 46.8 | 705 | 1263 | 1059 | 155 | 918 |
| | objfile | 5.98 | 29.1 | 524 | 934 | 757 | 85 | 606 |
| | parser | 3.61 | 23.9 | 327 | 460 | 415 | 49 | 372 |
| | scanner | 10.18 | 37.9 | 790 | 1318 | 1097 | 75 | 1006 |
| | stmt | 11.59 | 41.2 | 913 | 1603 | 1376 | 278 | 1132 |
| Virtual Machine | vm | 14.18 | 48.1 | 1498 | 2285 | 1992 | 278 | 1655 |
| | vmapi | 7.23 | 31.6 | 522 | 874 | 744 | 103 | 673 |
| | vmduprange | 0.8 | 21.2 | 70 | 127 | 104 | 6 | 94 |
| | vmerror | 1.5 | 20.7 | 139 | 265 | 220 | 37 | 186 |
| | vmeval | 2.48 | 21.8 | 233 | 428 | 371 | 81 | 295 |
| | vmexit | 1.82 | 21.1 | 83 | 167 | 132 | 15 | 119 |
| | vmexpr | 12.28 | 43.8 | 986 | 1817 | 1445 | 57 | 1383 |
| | vmext | 10.5 | 36.2 | 34 | 72 | 59 | 9 | 43 |
| | vmfuncs | 7.78 | 32.7 | 498 | 1000 | 838 | 206 | 675 |
| | vmgc | 11.64 | 41.9 | 922 | 1746 | 1422 | 240 | 1264 |
| | vmjump | 1.97 | 21.3 | 119 | 231 | 183 | 11 | 170 |
| | vmlists | 7.21 | 30.7 | 379 | 674 | 549 | 36 | 513 |
| | vmoop | 11.13 | 39.9 | 820 | 1497 | 1268 | 215 | 1081 |
| | vmperformance | 3.45 | 25.6 | 192 | 332 | 273 | 24 | 255 |
| | vmstackvars | 6.85 | 30.4 | 487 | 997 | 770 | 81 | 697 |
| | vmstate | 5.5 | 27.2 | 385 | 619 | 578 | 146 | 435 |
| | vmstrindex | 0.69 | 19.6 | 49 | 78 | 67 | 2 | 61 |
| | vmthread | 1.59 | 44.4 | 148 | 262 | 219 | 32 | 191 |
| | vmtrycatch | 0.76 | 19.7 | 20 | 38 | 33 | 5 | 25 |
| | vmvars | 8.51 | 33.8 | 362 | 685 | 569 | 91 | 497 |
| | vminfo_ext | 6.06 | 28.1 | 289 | 443 | 389 | 24 | 360 |
| | dll_ext | 10.74 | 37.1 | 89 | 147 | 123 | 4 | 112 |
| | file_ext | 9.57 | 36.9 | 688 | 1235 | 991 | 26 | 961 |
| Built-in Functions | genlib_ext | 22.75 | 66.6 | 1732 | 2965 | 2438 | 169 | 2308 |
| | list_ext | 5.65 | 28.5 | 531 | 982 | 782 | 35 | 740 |
| | math_ext | 3.93 | 25.2 | 336 | 649 | 489 | 3 | 497 |
| | os_ext | 5.25 | 26.8 | 313 | 572 | 464 | 23 | 427 |
| | refmeta_ext | 8.05 | 33.5 | 636 | 1075 | 886 | 26 | 851 |

Each visual source file belongs to one of the modules, such as Loader, General Library, Compiler, Virtual Machine, or the built-in functions. PWCT stores each visual source file in two files: *.SSF and *.FPT. The storage size listed in the table represents the summation of the file sizes of both files. The “components” column includes the total number of components used within the visual source file, even accounting for repeated usage of the same components. Each component corresponds to an interaction page (data-entry form) and may generate one or more steps.

We present a summary of the results in Table 6.5. Also, we highlight the results for each module in Figure 6.7. In this Figure, we notice that the Virtual Machine is the largest module while the optional "built-in functions" is the second largest module.

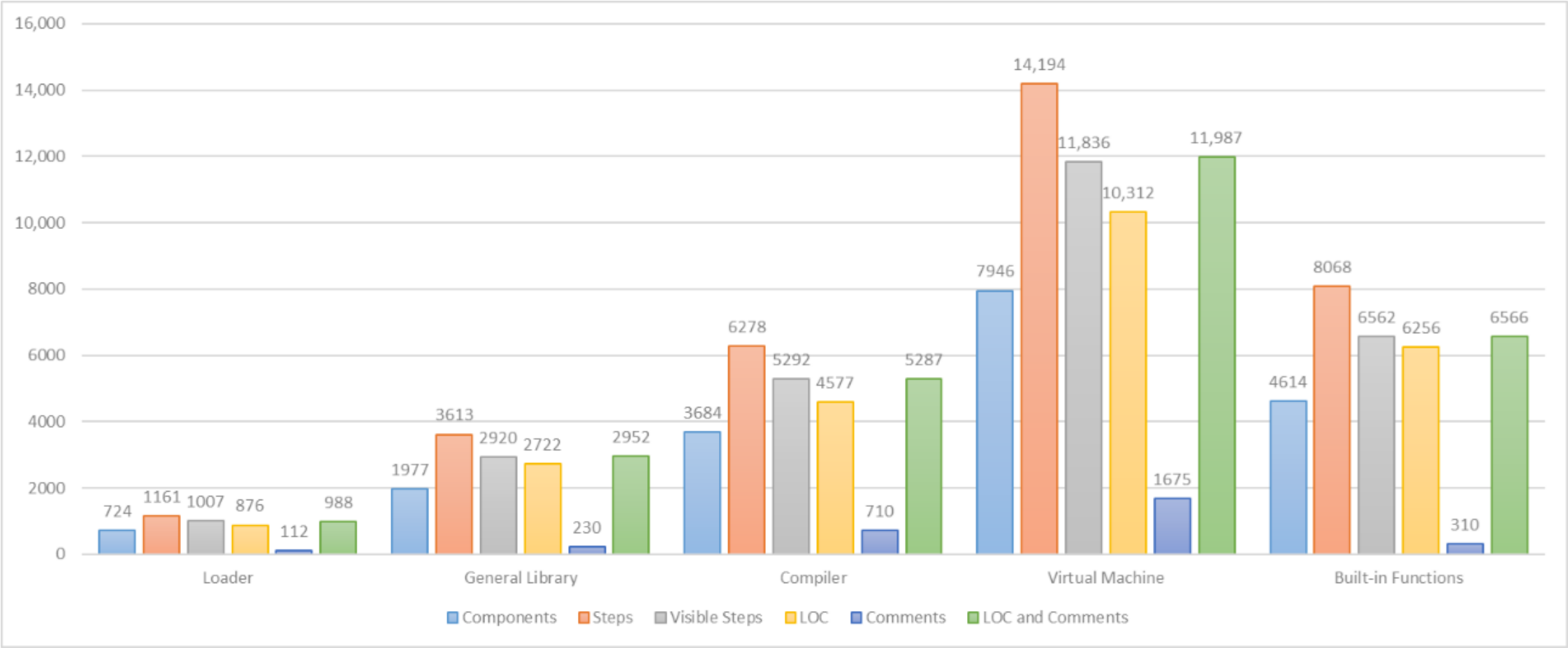

***Figure 6.7*** *Visual implementation size for each module.*

***Table 6.5*** *Summary of visual implementation size.*

| Criteria | Total |
|---|---|
| Modules | 5 |
| Visual Source Files | 43 |
| Storage Size (MB) | 278.95 |
| Memory (MB) | 1350.6 |
| Visual Components | 18,945 |
| Steps | 33,314 |
| Steps (Visible) | 27,617 |
| Lines of Code (LOC) | 24,743 |
| Comments | 3037 |
| LOC including comments | 27,780 |

In Figure 6.8, we present the loading time required to display the visual representation and the code generation time for each visual source file. These values were measured 10 times for each file. The cell colors visually represent the performance metrics, where larger numerical values correspond to longer time durations, indicating lower performance. In Figure 6.9, we present the code generation time for large visual source files. The time is measured in seconds, and tests are performed using a Victus Laptop [13th Gen Intel(R) Core(TM) i7-13700H, Windows 11, PWCT 1.9].

| File Name | LT1 | CGT1 | LT2 | CGT2 | LT3 | CGT3 | LT4 | CGT4 | LT5 | CGT5 | LT6 | CGT6 | LT7 | CGT7 | LT8 | CGT8 | LT9 | CGT9 | LT10 | CGT10 |
|---|---|---|---|---|---|---|---|---|---|---|---|---|---|---|---|---|---|---|---|---|
| ring | 0.141 | 0.208 | 0.126 | 0.204 | 0.126 | 0.213 | 0.13 | 0.201 | 0.13 | 0.213 | 0.119 | 0.205 | 0.12 | 0.206 | 0.13 | 0.188 | 0.11 | 0.192 | 0.11 | 0.207 |
| state | 0.518 | 1.307 | 0.505 | 1.288 | 0.519 | 1.228 | 0.5 | 1.318 | 0.51 | 1.321 | 0.486 | 1.255 | 0.49 | 1.241 | 0.49 | 1.286 | 0.505 | 1.227 | 0.534 | 1.259 |
| general | 0.22 | 0.315 | 0.202 | 0.283 | 0.22 | 0.299 | 0.2 | 0.298 | 0.21 | 0.299 | 0.204 | 0.313 | 0.22 | 0.298 | 0.22 | 0.299 | 0.204 | 0.289 | 0.204 | 0.314 |
| hashtable | 0.173 | 0.211 | 0.142 | 0.205 | 0.157 | 0.214 | 0.15 | 0.205 | 0.13 | 0.205 | 0.141 | 0.22 | 0.14 | 0.203 | 0.14 | 0.205 | 0.125 | 0.204 | 0.142 | 0.204 |
| item | 0.236 | 0.385 | 0.22 | 0.346 | 0.219 | 0.375 | 0.22 | 0.377 | 0.22 | 0.377 | 0.221 | 0.393 | 0.22 | 0.383 | 0.22 | 0.365 | 0.219 | 0.394 | 0.204 | 0.371 |
| items | 0.031 | 0.017 | 0.037 | 0.015 | 0.032 | 0.016 | 0.05 | 0.016 | 0.05 | 0.016 | 0.032 | 0.016 | 0.03 | 0.031 | 0.05 | 0.031 | 0.047 | 0.006 | 0.047 | 0.016 |
| list | 1.002 | 5.564 | 0.959 | 5.646 | 0.958 | 5.633 | 0.96 | 5.627 | 0.96 | 5.632 | 0.989 | 5.605 | 0.98 | 5.641 | 0.99 | 5.936 | 0.982 | 5.554 | 0.975 | 5.612 |
| string | 0.252 | 0.457 | 0.236 | 0.472 | 0.252 | 0.489 | 0.25 | 0.489 | 0.24 | 0.471 | 0.252 | 0.486 | 0.24 | 0.485 | 0.24 | 0.488 | 0.236 | 0.481 | 0.236 | 0.487 |
| hashlib | 0.063 | 0.015 | 0.062 | 0.016 | 0.041 | 0.015 | 0.05 | 0.016 | 0.05 | 0.016 | 0.047 | 0.016 | 0.05 | 0.015 | 0.05 | 0.016 | 0.047 | 0.016 | 0.047 | 0.015 |
| codegen | 0.377 | 0.865 | 0.346 | 0.91 | 0.345 | 0.834 | 0.35 | 0.896 | 0.35 | 0.928 | 0.346 | 0.871 | 0.35 | 0.881 | 0.35 | 0.906 | 0.346 | 0.853 | 0.363 | 0.878 |
| expr | 0.999 | 2.864 | 1.043 | 2.818 | 1.024 | 2.825 | 1.04 | 2.854 | 1.04 | 2.84 | 1.034 | 2.77 | 1.04 | 2.798 | 1.06 | 2.854 | 1.035 | 2.843 | 1.039 | 2.778 |
| objfile | 0.441 | 1.668 | 0.439 | 1.675 | 0.441 | 1.528 | 0.44 | 1.725 | 0.44 | 1.555 | 0.452 | 1.654 | 0.44 | 1.557 | 0.44 | 1.718 | 0.441 | 1.551 | 0.455 | 1.671 |
| parser | 0.204 | 0.394 | 0.204 | 0.393 | 0.189 | 0.393 | 0.19 | 0.401 | 0.2 | 0.401 | 0.188 | 0.408 | 0.19 | 0.393 | 0.19 | 0.382 | 0.189 | 0.377 | 0.188 | 0.377 |
| scanner | 0.846 | 2.975 | 0.737 | 3.011 | 0.726 | 3.028 | 0.75 | 3.026 | 0.72 | 2.981 | 0.731 | 2.986 | 0.74 | 3.006 | 0.74 | 2.982 | 0.74 | 3.01 | 0.736 | 3.025 |
| stmt | 0.956 | 4.468 | 0.942 | 4.402 | 0.95 | 4.44 | 0.86 | 4.5 | 0.93 | 4.453 | 0.943 | 4.44 | 0.99 | 4.505 | 0.88 | 4.439 | 0.897 | 4.661 | 0.972 | 4.437 |
| vm | 1.102 | 8.629 | 1.082 | 8.681 | 1.099 | 8.573 | 1.1 | 8.561 | 1.13 | 8.625 | 1.184 | 8.525 | 1.17 | 8.587 | 1.15 | 8.629 | 1.135 | 8.547 | 1.102 | 8.576 |
| vmapi | 0.471 | 1.377 | 0.472 | 1.321 | 0.471 | 1.401 | 0.5 | 1.34 | 0.49 | 1.371 | 0.47 | 1.385 | 0.47 | 1.358 | 0.5 | 1.403 | 0.47 | 1.395 | 0.488 | 1.338 |
| vmduprange | 0.063 | 0.041 | 0.047 | 0.032 | 0.047 | 0.032 | 0.05 | 0.04 | 0.06 | 0.045 | 0.063 | 0.039 | 0.06 | 0.045 | 0.06 | 0.048 | 0.063 | 0.047 | 0.47 | 0.48 |
| vmerror | 0.094 | 0.16 | 0.11 | 0.141 | 0.11 | 0.14 | 0.11 | 0.14 | 0.11 | 0.141 | 0.11 | 0.155 | 0.11 | 0.142 | 0.11 | 0.141 | 0.11 | 0.141 | 0.11 | 0.141 |
| vmeval | 0.157 | 0.327 | 0.15 | 0.362 | 0.137 | 0.357 | 0.16 | 0.366 | 0.16 | 0.363 | 0.157 | 0.362 | 0.16 | 0.345 | 0.14 | 0.347 | 0.157 | 0.362 | 0.157 | 0.361 |
| vmexit | 0.079 | 0.063 | 0.078 | 0.062 | 0.094 | 0.048 | 0.08 | 0.063 | 0.08 | 0.069 | 0.079 | 0.063 | 0.08 | 0.048 | 0.08 | 0.062 | 0.079 | 0.063 | 0.087 | 0.058 |
| vmexpr | 1.098 | 5.743 | 1.161 | 5.725 | 1.144 | 5.842 | 1.2 | 5.805 | 1.14 | 5.7 | 1.163 | 5.772 | 1.18 | 5.722 | 1.16 | 5.754 | 1.165 | 5.728 | 1.161 | 1.754 |
| vmext | 0.106 | 0.026 | 0.11 | 0.016 | 0.094 | 0.015 | 0.09 | 0.031 | 0.1 | 0.015 | 0.094 | 0.016 | 0.1 | 0.016 | 0.11 | 0.016 | 0.094 | 0.015 | 0.091 | 0.016 |
| vmfuncs | 0.565 | 1.84 | 0.549 | 1.818 | 0.565 | 1.748 | 0.55 | 1.868 | 0.57 | 1.811 | 0.582 | 1.815 | 0.57 | 1.83 | 0.58 | 1.85 | 0.549 | 1.826 | 0.55 | 1.838 |
| vmgc | 0.977 | 5.235 | 1.084 | 5.252 | 1.034 | 5.243 | 1.04 | 5.266 | 1.05 | 5.237 | 1.067 | 5.269 | 1.05 | 5.413 | 1.06 | 5.305 | 1.067 | 5.213 | 1.086 | 5.255 |
| vmjump | 0.126 | 0.112 | 0.11 | 0.111 | 0.11 | 0.109 | 0.11 | 0.103 | 0.11 | 0.127 | 0.11 | 0.11 | 0.09 | 0.11 | 0.11 | 0.11 | 0.11 | 0.109 | 0.126 | 0.122 |
| vmlists | 0.442 | 0.815 | 0.471 | 0.784 | 0.455 | 0.849 | 0.44 | 0.866 | 0.45 | 0.879 | 0.458 | 0.843 | 0.44 | 0.879 | 0.45 | 0.882 | 0.449 | 0.878 | 0.465 | 0.817 |
| vmoop | 0.898 | 3.888 | 0.925 | 3.923 | 0.879 | 3.937 | 0.93 | 3.922 | 0.97 | 3.862 | 0.926 | 3.873 | 0.93 | 3.93 | 0.86 | 3.95 | 0.911 | 3.899 | 0.879 | 3.968 |
| vmperformance | 0.162 | 0.215 | 0.16 | 0.22 | 0.173 | 0.226 | 0.16 | 0.196 | 0.17 | 0.22 | 0.161 | 0.229 | 0.17 | 0.236 | 0.17 | 0.206 | 0.172 | 0.204 | 0.173 | 0.22 |
| vmstackvars | 0.551 | 1.804 | 0.534 | 1.777 | 0.533 | 1.819 | 0.55 | 1.851 | 0.53 | 1.772 | 0.534 | 1.819 | 0.55 | 1.806 | 0.54 | 1.8 | 0.534 | 1.757 | 0.549 | 1.775 |
| vmstate | 0.283 | 0.725 | 0.282 | 0.738 | 0.283 | 0.66 | 0.27 | 0.676 | 0.28 | 0.723 | 0.283 | 0.733 | 0.28 | 0.662 | 0.28 | 0.674 | 0.28 | 0.724 | 0.268 | 0.707 |
| vmstrindex | 0.047 | 0.007 | 0.047 | 0.016 | 0.032 | 0.015 | 0.05 | 0.016 | 0.05 | 0.016 | 0.031 | 0.016 | 0.05 | 0.016 | 0.05 | 0.007 | 0.047 | 0.015 | 0.047 | 0.022 |
| vmthread | 0.11 | 0.127 | 0.094 | 0.142 | 0.095 | 0.142 | 0.09 | 0.141 | 0.11 | 0.142 | 0.094 | 0.143 | 0.1 | 0.149 | 0.1 | 0.142 | 0.094 | 0.142 | 0.094 | 0.142 |
| vmtrycatch | 0.032 | 0.016 | 0.032 | 0.016 | 0.031 | 0.006 | 0.03 | 0.016 | 0.04 | 0.01 | 0.017 | 0.004 | 0.03 | 0.015 | 0.03 | 0.015 | 0.031 | 0.004 | 0.016 | 0.016 |
| vmvars | 0.487 | 0.911 | 0.503 | 0.894 | 0.526 | 0.863 | 0.49 | 0.867 | 0.5 | 0.898 | 0.511 | 0.92 | 0.5 | 0.891 | 0.49 | 0.839 | 0.488 | 0.879 | 0.502 | 0.895 |
| vminfo_ext | 0.284 | 0.371 | 0.283 | 0.378 | 0.283 | 0.362 | 0.3 | 0.36 | 0.28 | 0.366 | 0.277 | 0.362 | 0.28 | 0.363 | 0.3 | 0.348 | 0.282 | 0.388 | 0.267 | 0.376 |
| dll_ext | 0.205 | 0.042 | 0.188 | 0.047 | 0.171 | 0.048 | 0.19 | 0.047 | 0.19 | 0.047 | 0.172 | 0.047 | 0.19 | 0.039 | 0.19 | 0.043 | 0.178 | 0.047 | 0.174 | 0.047 |
| file_ext | 0.81 | 2.704 | 0.75 | 2.669 | 0.802 | 2.666 | 0.77 | 2.713 | 0.82 | 2.656 | 0.747 | 2.661 | 0.76 | 2.675 | 0.8 | 2.661 | 0.8 | 2.651 | 0.754 | 2.732 |
| genlib_ext | 2.545 | 14.66 | 2.429 | 14.78 | 2.431 | 14.751 | 2.43 | 14.66 | 2.47 | 14.78 | 2.426 | 17.04 | 2.43 | 15.113 | 2.44 | 14.77 | 2.339 | 14.81 | 2.415 | 14.84 |
| list_ext | 0.439 | 1.728 | 0.456 | 1.759 | 0.445 | 1.73 | 0.44 | 1.762 | 0.44 | 1.778 | 0.44 | 1.761 | 0.44 | 1.71 | 0.46 | 1.717 | 0.439 | 1.735 | 0.456 | 1.683 |
| math_ext | 0.314 | 0.79 | 0.283 | 0.816 | 0.299 | 0.829 | 0.31 | 0.816 | 0.29 | 0.769 | 0.289 | 0.756 | 0.28 | 0.815 | 0.3 | 0.8 | 0.283 | 0.786 | 0.314 | 0.774 |
| os_ext | 0.33 | 0.642 | 0.329 | 0.647 | 0.33 | 0.629 | 0.31 | 0.598 | 0.3 | 0.597 | 0.314 | 0.598 | 0.31 | 0.597 | 0.33 | 0.643 | 0.298 | 0.603 | 0.314 | 0.646 |
| refmeta_ext | 0.611 | 2.01 | 0.595 | 2.015 | 0.596 | 2.062 | 0.62 | 2.055 | 0.59 | 2.038 | 0.596 | 2.047 | 0.6 | 2.014 | 0.6 | 2.043 | 0.628 | 1.993 | 0.698 | 2.013 |

***Figure 6.8** The loading time (LT) and code generation time (CGT).*

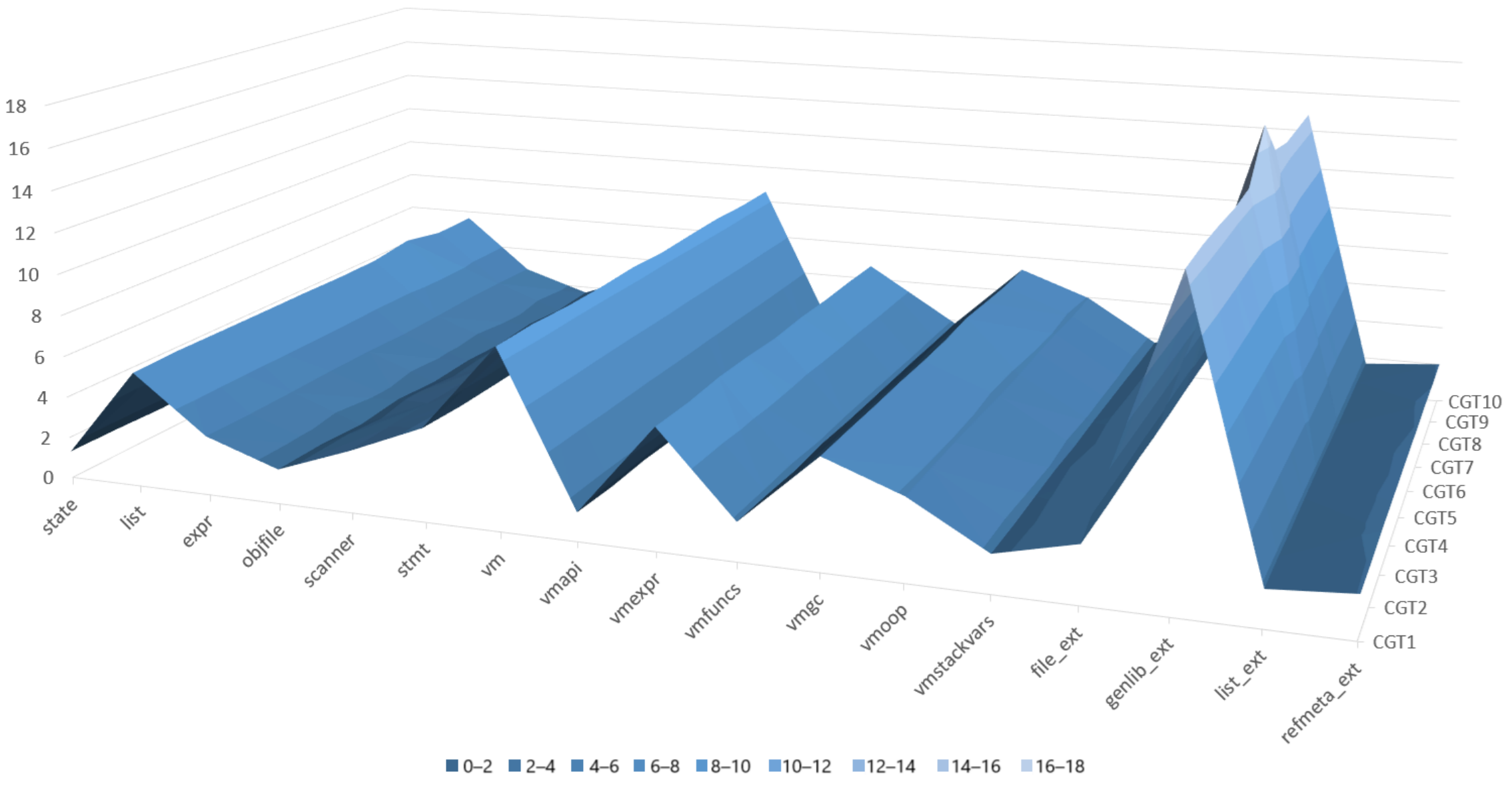

***Figure 6.9** Code generation time (CGT) for large visual source files.*

### 6.2.7 Lightweight Implementation

Developing a lightweight programming language is not just about providing a language with a small implementation. It is merely the beginning, and we must pay attention to the growth in the language size over time. In Table 6.6, we present the growth percentage in implementation size for the Ring programming language and other known lightweight programming languages. The table presents the LOC of the first release and the LOC of the latest release. The LOC includes the compiler, VM, and the built-in functions. The growth in code size can be attributed to several factors, including fixing bugs, adding new features, performance improvements, the expansion of libraries and built-in capabilities, and enhancements in compatibility and interoperability.

***Table 6.6*** *Growth in implementation size.*

| Language | Period | Implementation | LOC (FR) | LOC (LR) | Growth |
|---|---|---|---|---|---|
| Ring | 2016–2024 | C | 16,402 | 24,743 | 51% |
| mRuby | 2014–2024 | C | 18,134 | 23,742 | 31% |
| Squirrel | 2004–2022 | C++ | 9311 | 13,991 | 50% |
| Lua | 1993–2024 | C | 5603 | 20,081 | 258% |

Since Ring is designed to be a lightweight language, we have monitored the growth of the implementation size over the years. From 2016 to 2024, the implementation size has increased from 16 KLOC in Ring 1.0.0 to 24.7 KLOC in Ring 1.21.2, as demonstrated in Figure 6.10. The growth percentage in the implementation size is 51%. In Figure 6.11, we present the code size for the Lua Compiler/VM. The source code was written from 1993 to 2024, and the implementation size increased from 5.6 KLOC in Lua 1.0.0 to 20 KLOC in Lua 5.4.7. The growth percentage in implementation size is 258%.

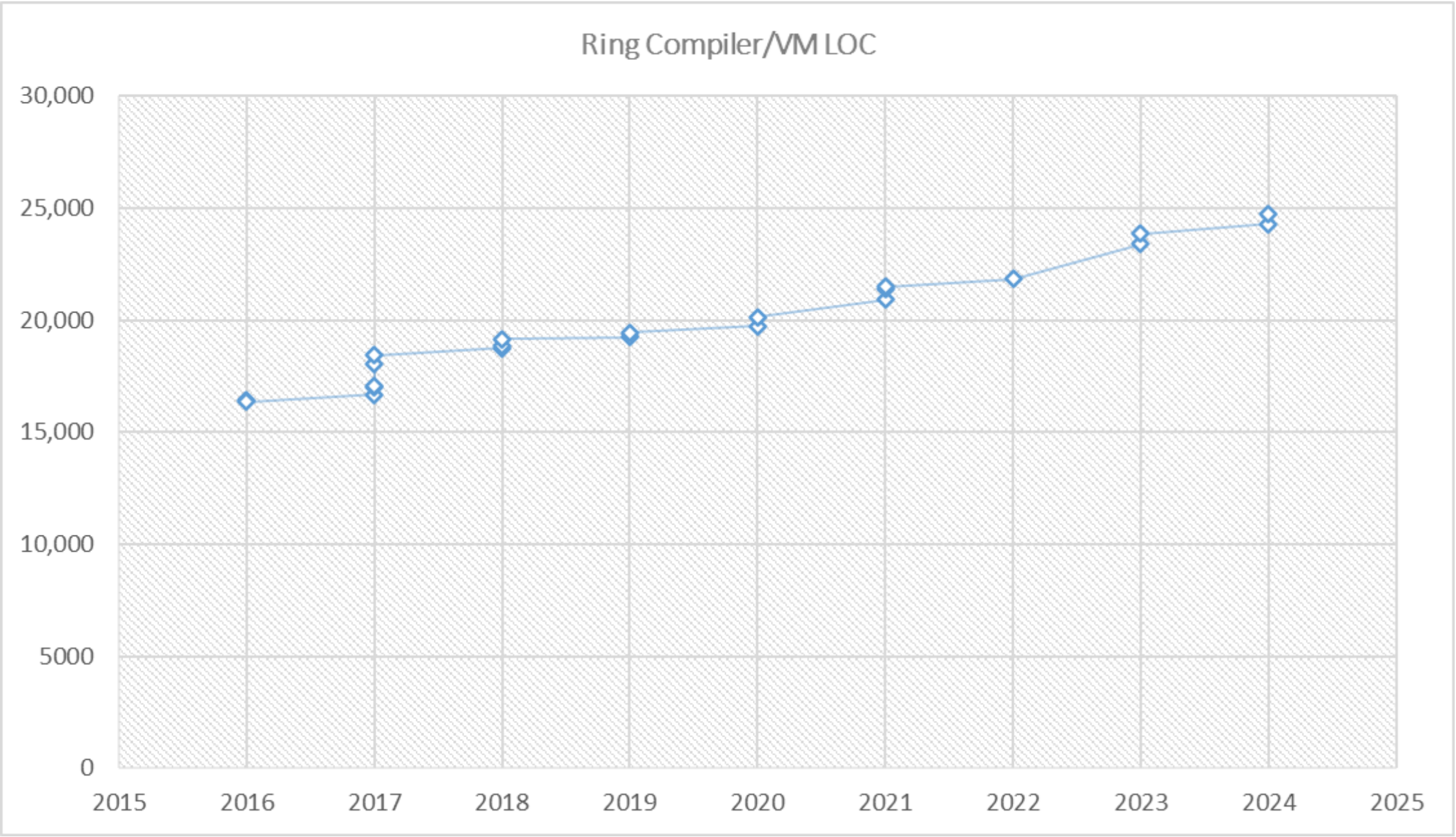

***Figure 6.10*** *Generated code size for Ring Compiler/VM from 2016 to 2024.*

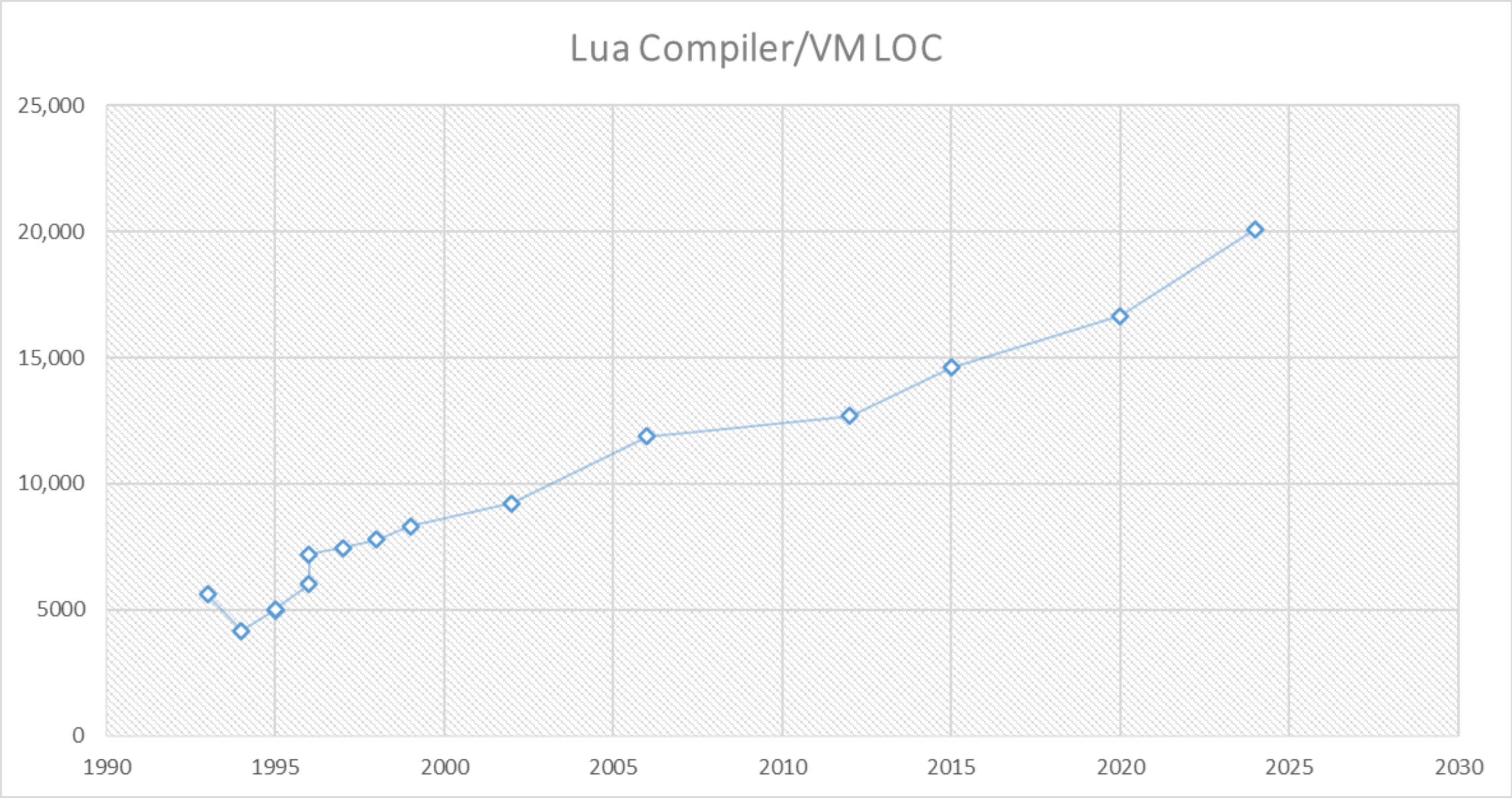

***Figure 6.11** Code size for Lua Compiler/VM from 1993 to 2024.*

In Figure 6.12, we present the generated code size for the Ring Compiler/VM for different Ring releases. The textual source code is generated in ANSI C and can be used by traditional programmers who may prefer text-based coding. This approach also enables adoption in settings where visual programming tools are less practical.

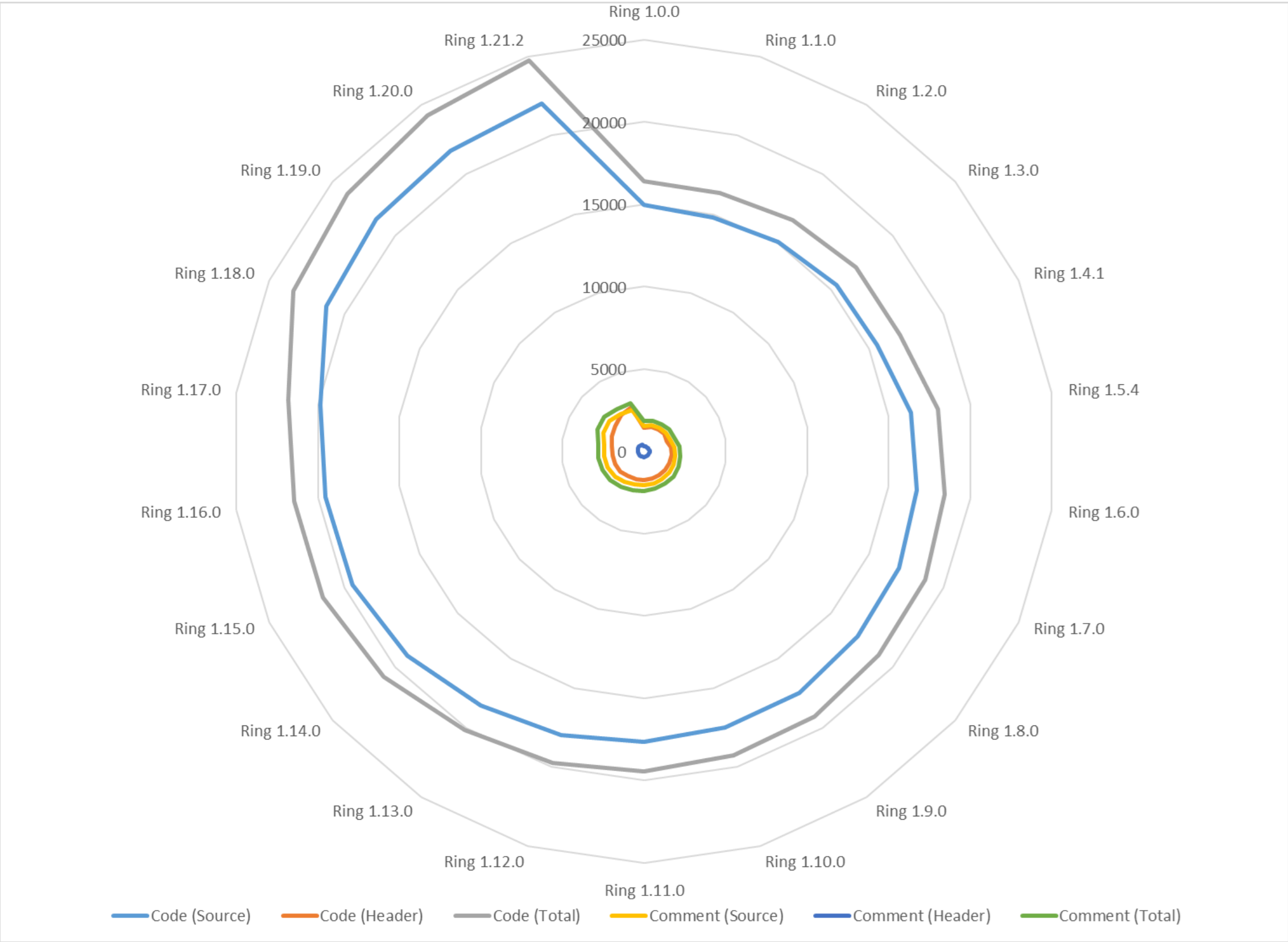

***Figure 6.12** Generated code size from Ring 1.0.0 to Ring 1.21.2.*

### 6.2.8 Performance Benchmarks

In Table 6.7, we provide a benchmark comparison of various versions of the Ring programming language (Ring 1.17, Ring 1.19, and Ring 1.21), including its WebAssembly implementations on Edge and Chrome browsers, against VFP 9.0 and Python 3.13. Tests are performed using a Victus Laptop [13th Gen Intel(R) Core(TM) i7-13700H, Windows 11]. The benchmarks cover a range of computational tasks, including looping (Loop), mathematical calculations (MathMax), function calls (FuncCall), dynamic programming Fibonacci calculations (FibDP), recursive Fibonacci calculations (FibRec), and list filling (ListFill), with varying input sizes. These benchmarks are designed to reflect features that are very common in many programs, ensuring their relevance and applicability across different use cases. VFP was selected for this comparison because it is a multiparadigm, dynamic language used in the development of PWCT. Python 3.13 was selected due to its popularity and versatility as a dynamic language that supports multiple programming paradigms. The performance, measured in milliseconds, indicates substantial improvements in the newer versions of Ring, particularly in Ring 1.21. For instance, the execution time for FuncCall (100 M) decreased dramatically from 113,142 ms in Ring 1.17 to 4058 ms in Ring 1.21. Figure 6.13 presents the performance of this benchmark.

WebAssembly implementations show a slight increase in time compared to native executions. Overall, the data highlights significant performance enhancements in newer versions of Ring and offers a comparison of the efficiency of different programming environments. With respect to Ring support for microcontrollers, which is relatively new (first support started with Ring 1.21, released in September 2024), the performance results for Ring 1.21 running on the Raspberry Pi Pico reveal some interesting insights when compared to Ring 1.21 on a desktop. The Loop (500 k) benchmark was completed in 3.35 s, while MathMax (100 k) and FuncCall (100 k) took 3.54 s and 3.32 s, respectively. The Fibonacci Recursive (FibRec) at 25 iterations took 5.81 s, and the Dynamic Programming Fibonacci (FibDP) at 500 iterations was notably fast at 0.89 s. However, the ListFill (100 k) benchmark resulted in an “OUT OF MEMORY” error, which is expected given that the Raspberry Pi Pico has only 264 KB of SRAM. A dynamically typed language like Ring may encounter challenges on resource-constrained devices such as the Raspberry Pi Pico due to its limited memory and processing power.

*Table 6.7 Performance benchmarks (Time in Milliseconds).*

| Benchmark | Ring 1.17 (2022) | Ring 1.19 (2023) | Ring 1.21 (2024) | Ring 1.21 WebAsm Edge | Ring 1.21 WebAsm Chrome | VFP 9.0 | Python 3.13 |
|---|---|---|---|---|---|---|---|
| Loop (500 K) | 9 | 5 | 4 | 7 | 7 | 8 | 7 |
| Loop (1 M) | 18 | 11 | 9 | 13 | 13 | 15 | 14 |
| Loop (10 M) | 185 | 113 | 91 | 133 | 132 | 47 | 149 |
| Loop (100 M) | 1896 | 1154 | 954 | 1362 | 1332 | 595 | 1534 |
| MathMax (100 K) | 136 | 25 | 7 | 11 | 12 | 16 | 7 |
| MathMax (1 M) | 1384 | 245 | 69 | 117 | 119 | 94 | 66 |
| MathMax (10 M) | 13,847 | 2474 | 708 | 1161 | 1204 | 909 | 776 |
| MathMax (100 M) | 139,373 | 24,868 | 7178 | 11,833 | 11,935 | 8968 | 7315 |
| FuncCall (100 K) | 111 | 19 | 4 | 10 | 9 | 16 | 3 |
| FuncCall (1 M) | 1134 | 194 | 39 | 97 | 94 | 110 | 32 |
| FuncCall (10 M) | 11,337 | 1943 | 398 | 1001 | 962 | 1102 | 444 |
| FuncCall (100 M) | 113,142 | 19,542 | 4058 | 10,164 | 9563 | 11,214 | 3297 |
| FibDP (500) | 6 | 5 | 3 | 3 | 4 | 6 | 0.1 |
| FibDP (700) | 11 | 10 | 5 | 5 | 6 | 13 | 0.3 |
| FibDP (1000) | 21 | 19 | 10 | 10 | 11 | 15 | 0.4 |
| FibDP (1200) | 29 | 27 | 15 | 15 | 17 | 16 | 0.4 |
| FibDP (500) | 6 | 5 | 3 | 3 | 4 | 6 | 0.1 |
| FibRec (20) | 22 | 6 | 1 | 2 | 3 | 16 | 0.3 |
| FibRec (25) | 244 | 65 | 13 | 22 | 23 | 31 | 3 |
| FibRec (30) | 2887 | 763 | 148 | 252 | 254 | 377 | 39 |
| FibRec (35) | 33,847 | 8660 | 1691 | 2795 | 2810 | 4357 | 431 |
| ListFill (100 K) | 10 | 10 | 6 | 11 | 12 | 14 | 5 |
| ListFill (500 K) | 54 | 50 | 31 | 56 | 56 | 30 | 19 |
| ListFill (1 M) | 108 | 99 | 62 | 112 | 113 | 63 | 35 |
| ListFill (10 M) | 1085 | 1007 | 643 | 1143 | 1145 | 565 | 376 |

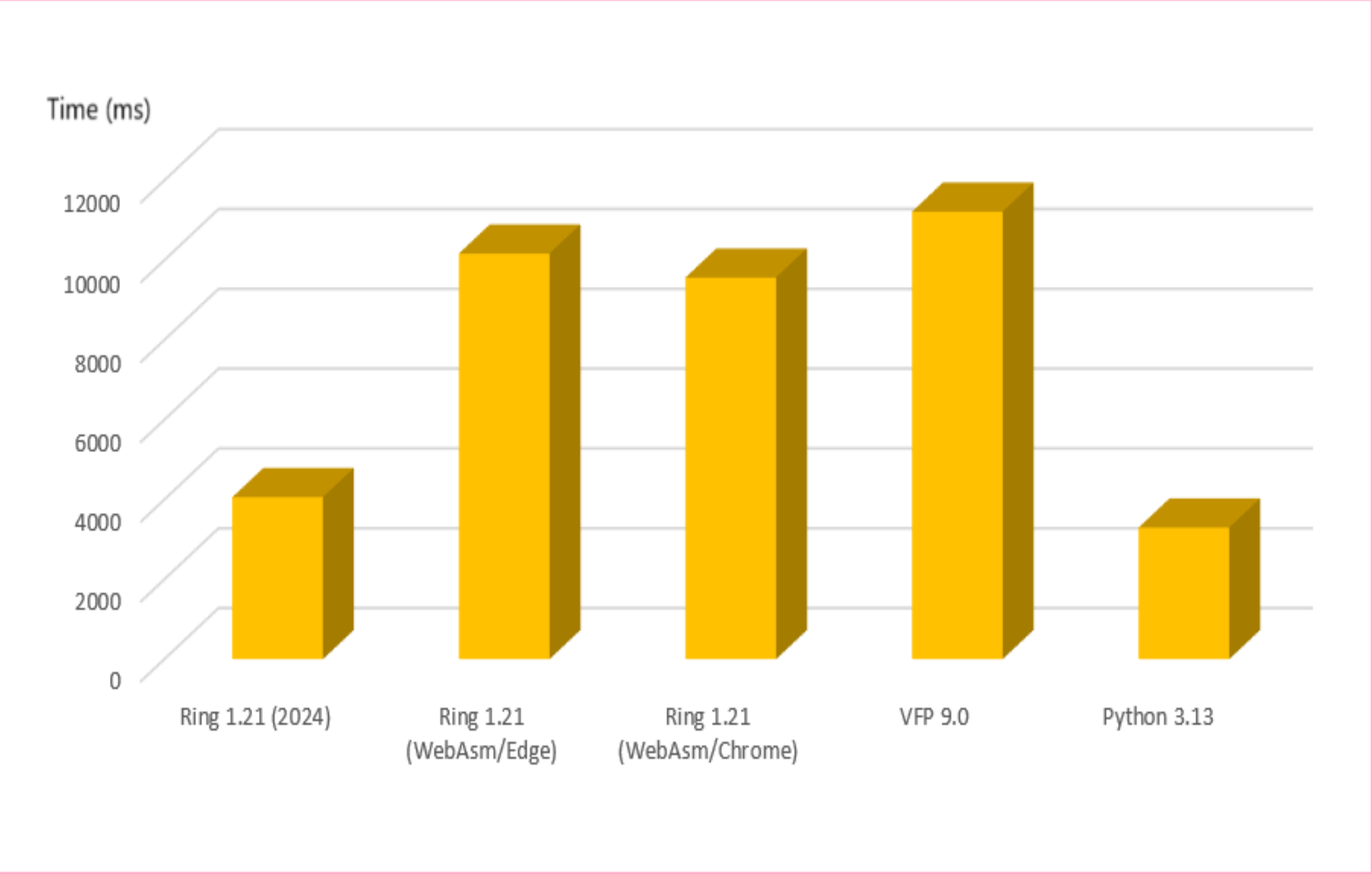

***Figure 6.13*** *Function call (100 M) benchmark for Ring, VFP, and Python.*

Ring is distributed with support for game programming libraries, which enable us to create benchmarks for graphics and animations. In Figure 6.14, we demonstrate different frames from the Waving Cubes sample provided by the RayLib open-source library.

This sample presents an animation of 3375 cubes by changing their position, color, and size. In Table 6.8, we present the performance results for this sample in Ring, C, and Python. C demonstrates the highest efficiency with 480 frames per second (FPS). Ring 1.21 significantly improves upon its predecessor, achieving 170 FPS compared to Ring 1.19's 40 FPS, showcasing notable advancements in performance. Python 3.13 provides 85 FPS.

***Table 6.8*** *The Waving Cubes benchmark.*

| Language | FPS (Min) | FPS (Max) |
|---|---|---|
| C | 470 | 480 |
| Ring 1.21 | 161 | 170 |
| Python 3.13 | 80 | 85 |
| Ring 1.20 | 33 | 40 |

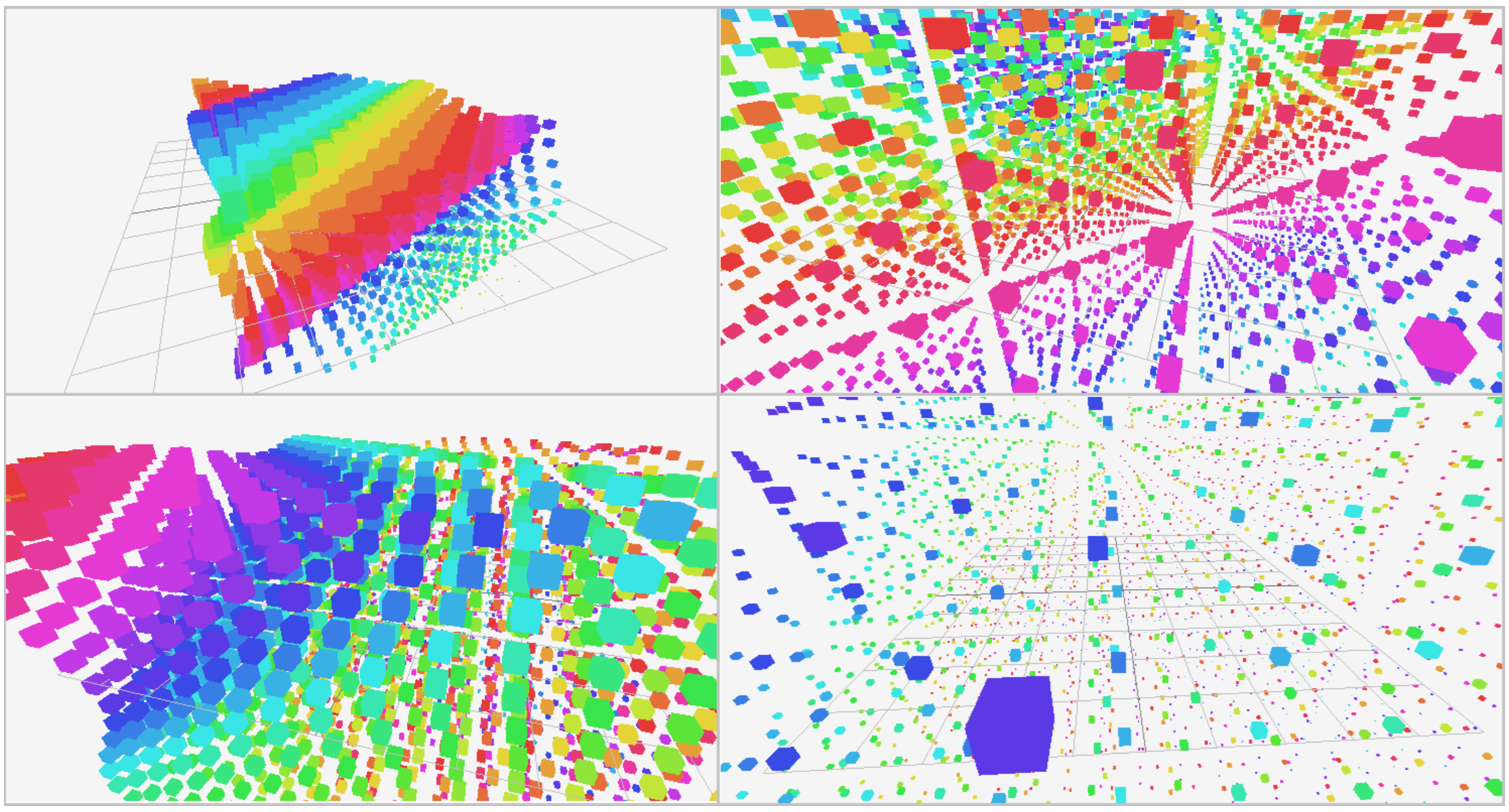

***Figure 6.14*** *Different frames from the waving cubes animation.*

The Binding Generator for Ring extensions is the first significant program written in the Ring language itself and serves as an example and benchmark for file and text processing. The generator comprises 1407 lines of Ring code. It operates very efficiently by processing configuration files and generating the C/C++ code required for the extensions. The largest extension we have with the Ring language is RingQt, which generates over 211,000 lines of C/C++ code in just 3.42 s, as demonstrated in Table 6.9. Other extensions are smaller, and their code-generation process is completed in less than a second.

***Table 6.9*** *Using the Code Generator to generate RingQt source code.*

| Variable | Value |
|---|---|
| Extension | RingQt |
| Configuration Files (Input) | 439 |
| Input Size | 478 KB |
| Generated Files | 197 |
| Generated Lines of Code | 211,174 |
| Output Size | 6.27 MB |
| Processing Time | 3420 ms |

The Ring IDE is designed as a project that includes a Code Editor, Form Designer, Web Browser, and a “Find in Files” application. Over the past eight years, it has proven to be a robust and reliable tool, supporting the development of all Ring samples and applications distributed with the language. The IDE's performance has consistently been impressive, with no notable issues encountered during regular use. To ensure its reliability, a stress test was conducted by opening all Ring applications and samples distributed with the language. Memory usage was observed to be 348 MB at startup, slightly increasing to 365 MB after opening 78 applications (253 files, totaling 76,924 lines of code) and further to 439 MB after opening each file in the samples (1302 files, totaling 65,563 lines of code). This increase is attributed to the autocomplete feature, which caches all the words in each opened file. Tests are performed using a Victus Laptop [13th Gen Intel(R) Core(TM) i7-13700H, Windows 11, and Ring 1.21.2].

Additionally, the performance of opening and displaying a source code file was less than 250 ms for most files, as shown in Table 6.10. This performance depends on the file size. This demonstrates the Ring IDE's capability to handle extensive development tasks without compromising performance. However, we continue to state in the Ring Group that the Ring IDE is just an example of Ring usage. Ring, as a language, can be used with different code editors based on programmer preference, which makes sense if the programmer is using multiple languages in a project. The Ring IDE only supports Ring source code files, which is a limitation in situations requiring a mix of programming languages. To load and display files, Ring Notepad uses functions written in C/C++, ensuring high performance.

***Table 6.10*** *Loading and displaying files in Ring IDE.*

| Application/Sample | Size (LOC) | Loading Time (ms) |
|---|---|---|
| Analog Clock | 256 | 36 |
| Image Pixel | 548 | 66 |
| Knight Tour | 646 | 67 |
| Othello | 780 | 78 |
| Visualize Sort | 817 | 81 |
| Game Of Life | 903 | 90 |
| Laser | 1051 | 94 |
| Checkers | 1354 | 124 |
| Get Quotes History | 3401 | 117 |
| Discrete Fourier Transform | 6417 | 203 |

## 6.3 Results related to PWCT2

PWCT2 was released on the Steam platform in March 2023, based on Ring 1.17, and has received continuous updates based on community feedback [172]. We also keep updating the Ring version included in the software after each Ring release. The current version of the software is PWCT 2.0 Rev. 2025.01.20, which is based on Ring 1.22. In this section, we present the various results related to our study. First, we explore a variety of use cases. Subsequently, we examine our observations regarding the implementation of PWCT2 using the Ring language. Furthermore, we discuss the performance metrics of PWCT2. Finally, we provide details about download numbers, usage time, and user feedback.

### 6.3.1 Use Cases

The primary use case for the PWCT2 software is as an alternative to the Ring Code Editor, allowing users to create new projects based on Ring or to import and update/execute Ring programs instead of using Ring Notepad. Ring2PWCT plays a significant role in this process, while the Time Machine feature enables users to read the program step by step and run it from a previous point in time. This functionality facilitates the continued development and maintenance of numerous Ring samples and applications.

In Figure 6.15, we introduce the SuperMan game, which is distributed with the Ring language. Using PWCT2, it was possible to import the game implementation and check it step by step. The game is based on the Ring game engine, which is designed for developing simple 2D games. PWCT2 comes with many samples related to game programming libraries like Allegro, RayLib, and Tilengine. These samples are converted from the Ring programming language project to PWCT2 through the Ring2PWCT tool, which converts Ring textual code to a PWCT2 visual source file.

Since PWCT2 comes with a Form Designer and the required visual components to build GUI applications, it was possible to use it in developing multiple GUI applications, including an application to predict citation counts in the Otology field using the research paper title, author, and abstract [166]. In the Citations Prediction application, the user interface is designed using the PWCT2 Form Designer (the same designer provided by the Ring language), and the application logic is designed using PWCT2 before generating

Ring code. The application receives input from the user and submits it to the machine learning model through the internet using the LibCurl library. The models are developed using Microsoft Azure Machine Learning. The 394 visual components provided by RingPWCT and introduced in Table 3 offer sufficient features to develop such applications.

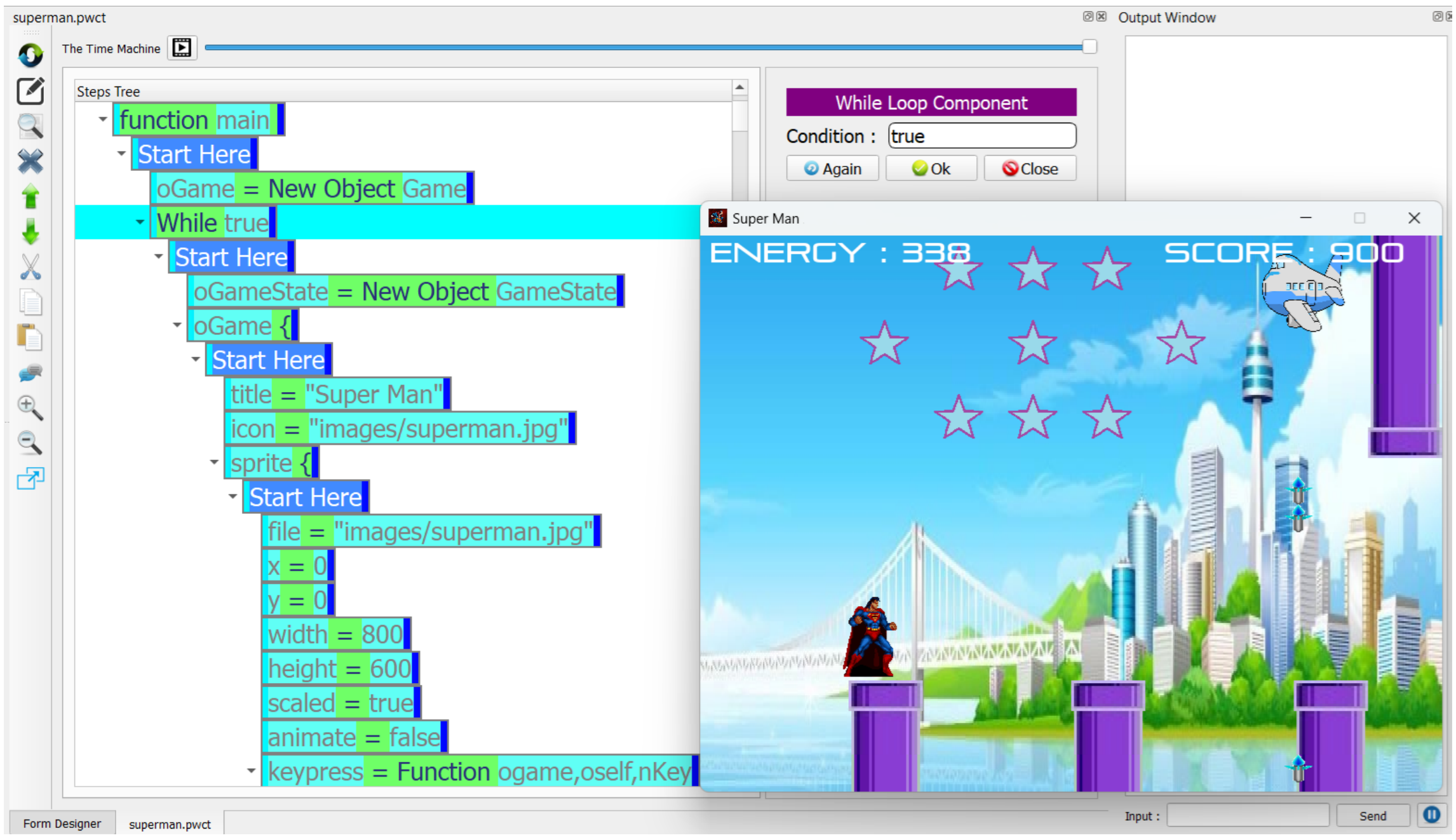

***Figure 6.15** Using PWCT2 to visualize and execute Ring language projects.*

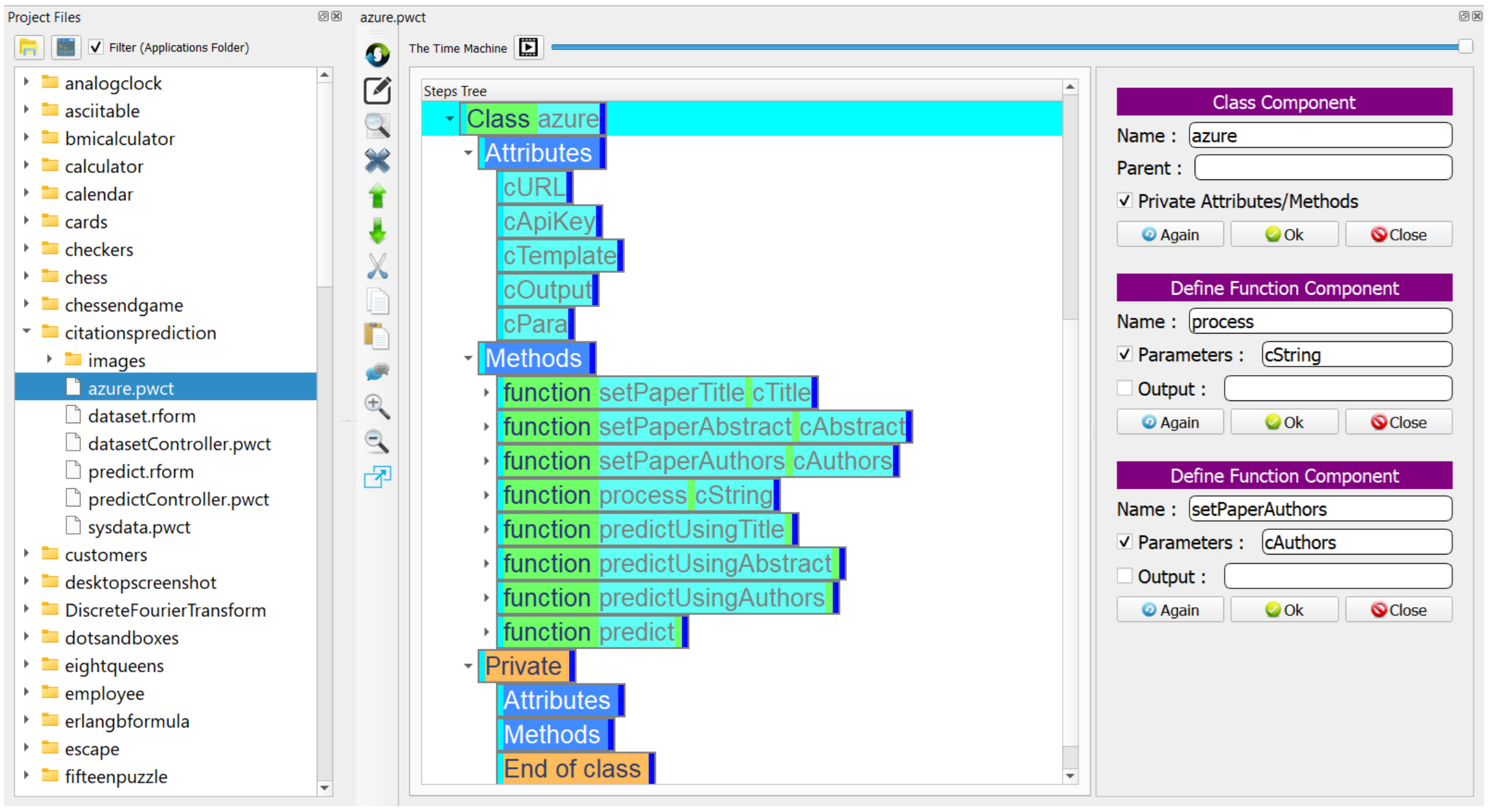

***Figure 6.16** Using PWCT2 to develop the Citations Prediction application.*

In Figure 6.16, we demonstrate the Azure class used by the Citations Prediction application to connect to the Machine Learning model. This class could submit the research paper title, authors, or abstract as input to the model and receives the predicted citation count as output. Different machine learning models are utilized based on the selected input. The class contains the web service URL and the API key as attributes. Within the same project, in addition to the azure.pwct file, there are other files, such as sysdata.pwct, which includes some research paper samples. There are two form designer files: dataset.rform and predict.rform. The first file provides a table to select a research paper from the available samples. The second form is the main form of our application and allows users to enter the paper information to be used as input for the model and includes a button to open the other form in the dataset.rform file. These forms respond to GUI events that are handled by the files datasetController.pwct and predictController.pwct. When we open a form file using the project files window, PWCT2 will automatically open the corresponding visual source file for the controller class.

Another known application in the Ring language community that is developed using PWCT2 is the Find in Files application, which is distributed with Ring Notepad [145]. The application is shown in Figure 6.17. The application's user interface is developed using the form designer and utilizes Qt layouts to automatically resize the controls.

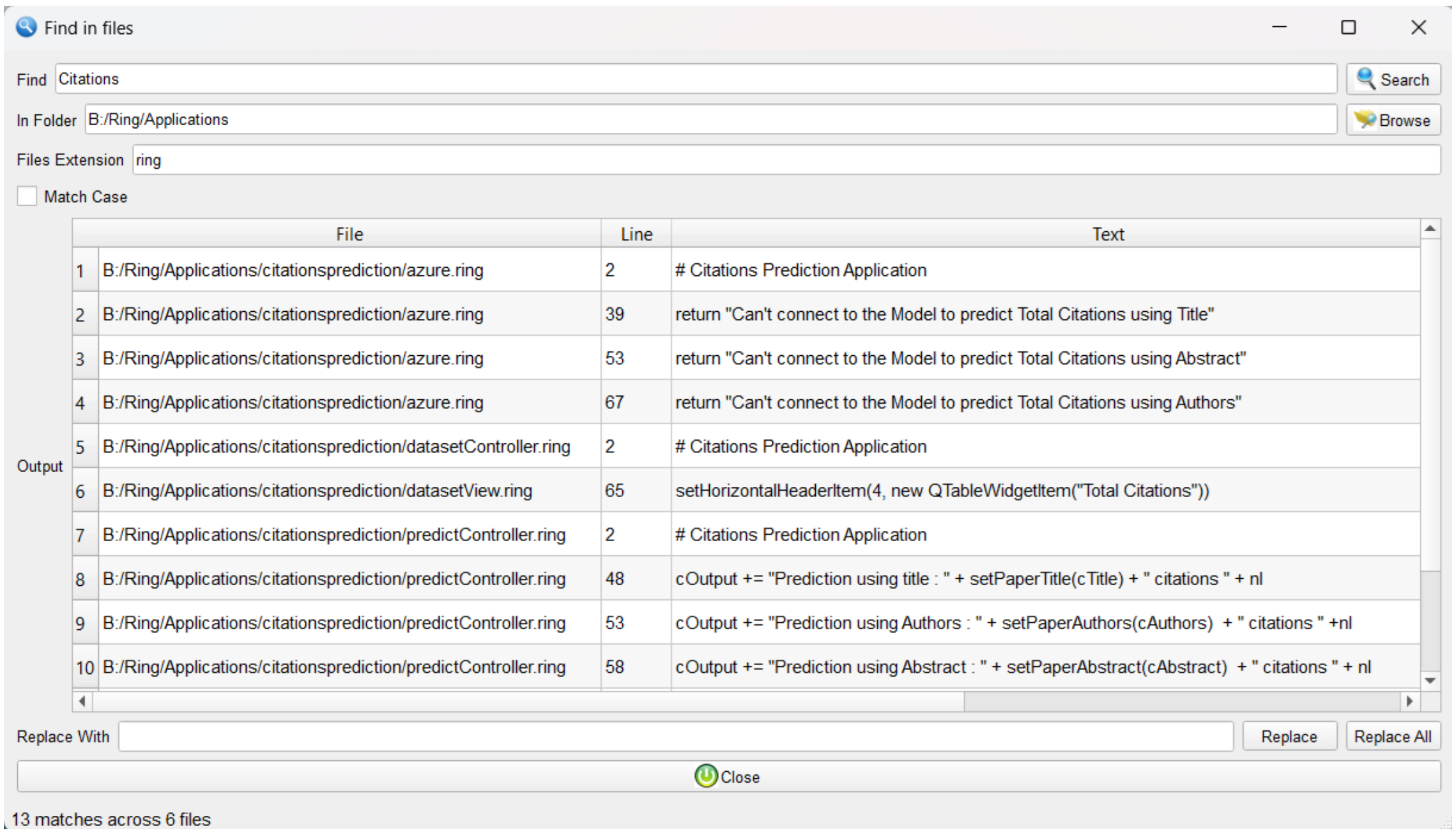

***Figure 6.17** Find in Files application developed using PWCT2.*

The Find in Files application supports search, replace, and replace all operations on one or more files. During a search operation, the user can enable the (Match Case) checkbox to perform a case-sensitive search. Using the (Browse) button, the user can select a specific folder to be used when searching for files. The application utilizes visual components related to file and text processing, as shown in Figure 6.18.

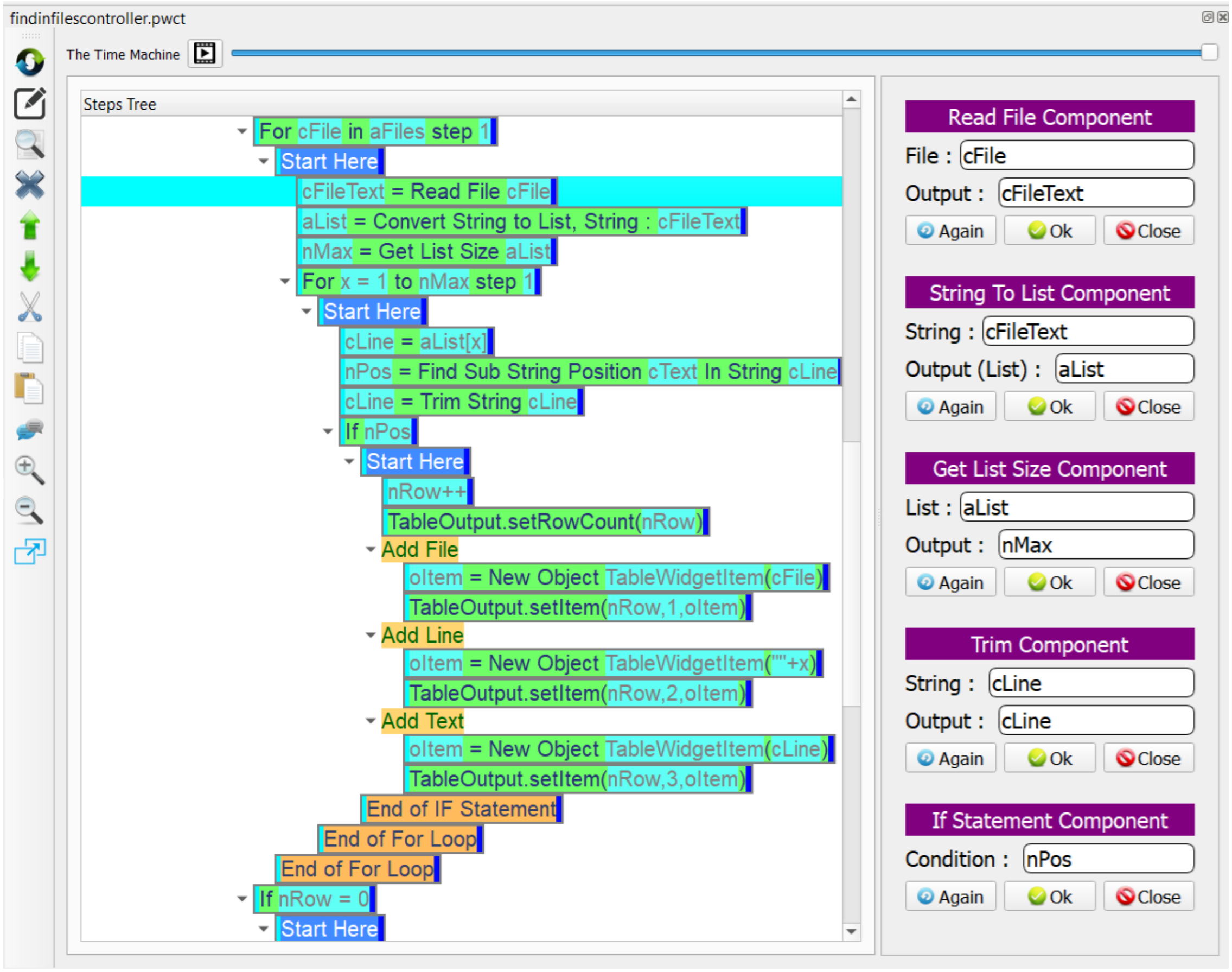

***Figure 6.18*** *Using PWCT2 to develop the Find in Files application.*

The (Read File) component receives two parameters: the (File) variable, which contains the file name to read, and the (Output) variable, which contains the file content as a string. The (String to List) component converts a Ring string to a Ring list that can be processed using list functions. Each list item will contain one line of text and can be processed using string processing components. For more efficient and high-performance implementation, it is recommended to avoid converting the string to a list since the string can be processed directly.

### 6.3.2 Implementation of PWCT2 with the Ring Language

The implementation of PWCT2 using Ring involves using the language compiler, virtual machine, libraries, and tools. Ring 1.17 was sufficient to release the PWCT2 software on the Steam platform in 2023, but we found the language features and stability more satisfactory starting from Ring 1.19, where Ring 1.18 improved support for references and Ring 1.19 improved performance. Later versions released in 2024, such as Ring 1.20, Ring 1.21, and Ring 1.22, provided additional features, stability, and performance improvements. In Table 6.11, we list information about the project size, including dependencies such as the standard library and the GUI library from Ring 1.22.

***Table 6.11*** *PWCT2 project size including dependencies.*

| Attribute | Value |
|---|---|
| Source code files | 1354 |
| Lines of Code | 92 KLOC |
| Dependencies | 27 KLOC |
| Total Lines of Code | 119 KLOC |

In Table 6.12, we present the compile-time using different versions of the Ring language, the size of the generated Ring VM byte-code instructions, and the size of the generated Ring object file (default size and compressed size). The tests are done using a Victus Laptop [13th Gen Intel(R) Core (TM) i7-13700H, Windows 11].

***Table 6.12*** *Using Ring to build the PWCT2 project.*

| Attribute | Ring 1.17 (2022) | Ring 1.19 (2023) | Ring 1.22 (2024) |
|---|---|---|---|
| Compile-time (ms) | 1480 | 1152 | 871 |
| Byte-code Instructions | 900,113 | 899,984 | 724,382 |
| Ring Object File Size (KB) | 22,184 | 22,825 | 18,952 |
| Object File Compressed (KB) | 2463 | 2483 | 2322 |

The data show a substantial reduction in compile-time from 1480 milliseconds in Ring 1.17 to 871 milliseconds in Ring 1.22 (see Figure 6.19), highlighting significant enhancements in the compilation process. There is a notable decrease in the number of byte-code instructions from 900,113 in Ring 1.17 to 724,382 in Ring 1.22 (see Figure 6.20), indicating more efficient bytecode generation.

We can compile the PWCT2 project to a Ring object file, which can be run directly using the Ring Virtual Machine without the need to compile the project again using the Ring compiler. The size of the Ring object file initially increased slightly but then decreased significantly (see Figure 6.21) to 18,952 KB in Ring 1.22, reflecting optimization improvements.

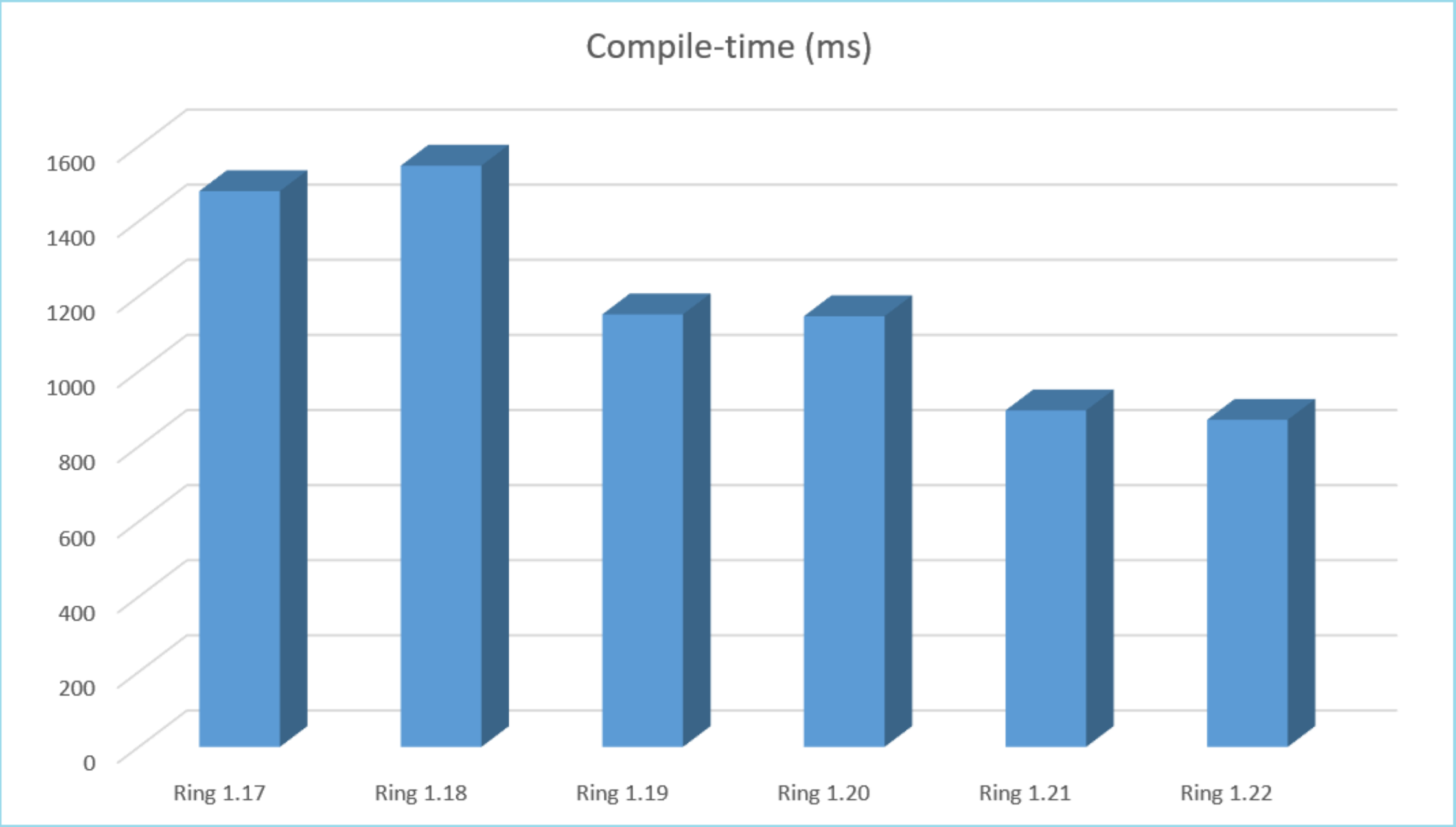

***Figure 6.19*** *Ring compile-time for PWCT2 from Ring 1.17 to Ring 1.22.*

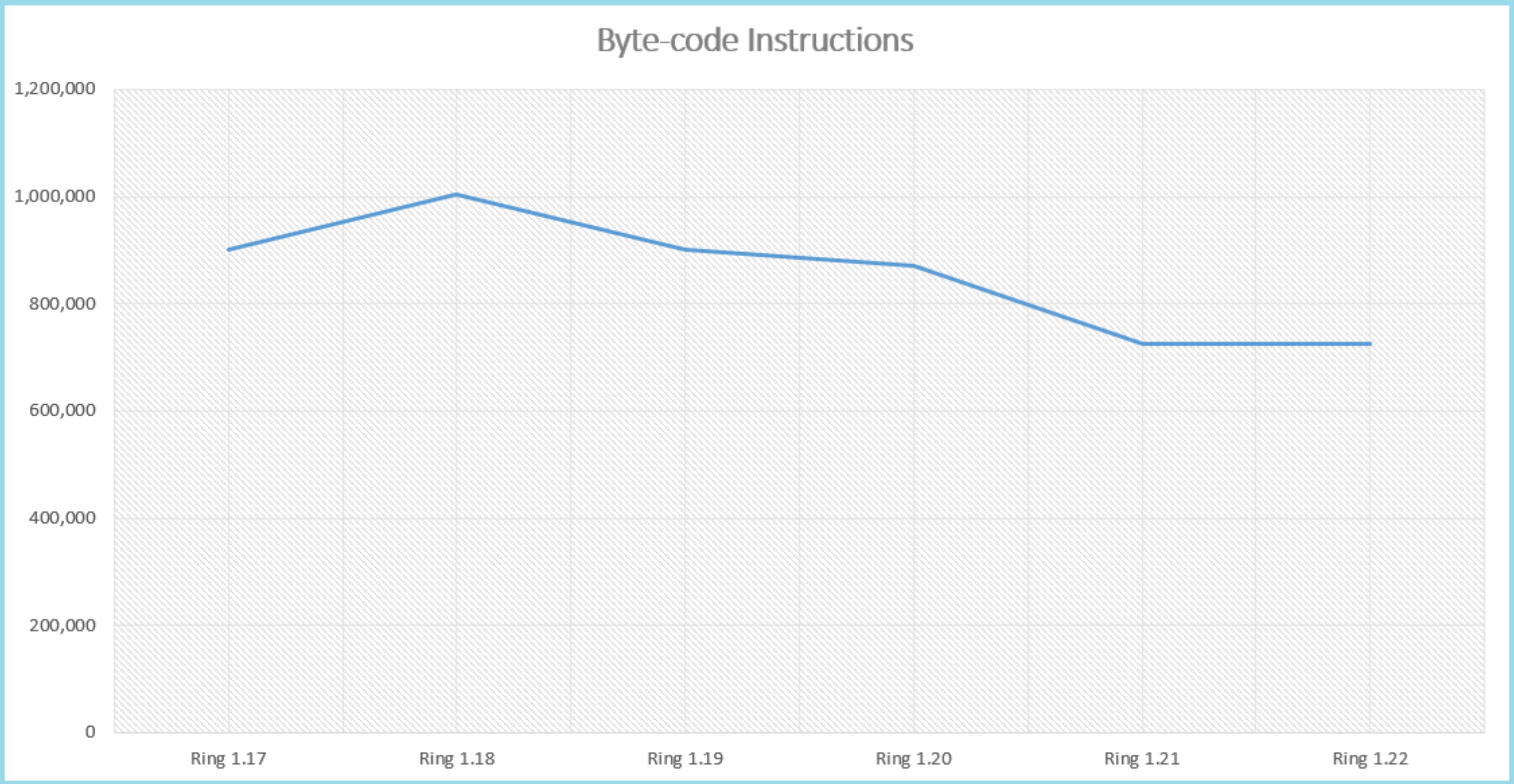

***Figure 6.20*** *Generated bytecode instructions for PWCT2 from Ring 1.17 to Ring 1.22.*

The compressed object file size remained relatively stable. Compressing the object file and reducing the required storage could be useful when distributing the project over the internet. However, the real size of the object file (without compression) matters because the Ring Virtual Machine processes and loads this file into memory when running the project. These results collectively illustrate the progress achieved with the subsequent versions of the Ring language.

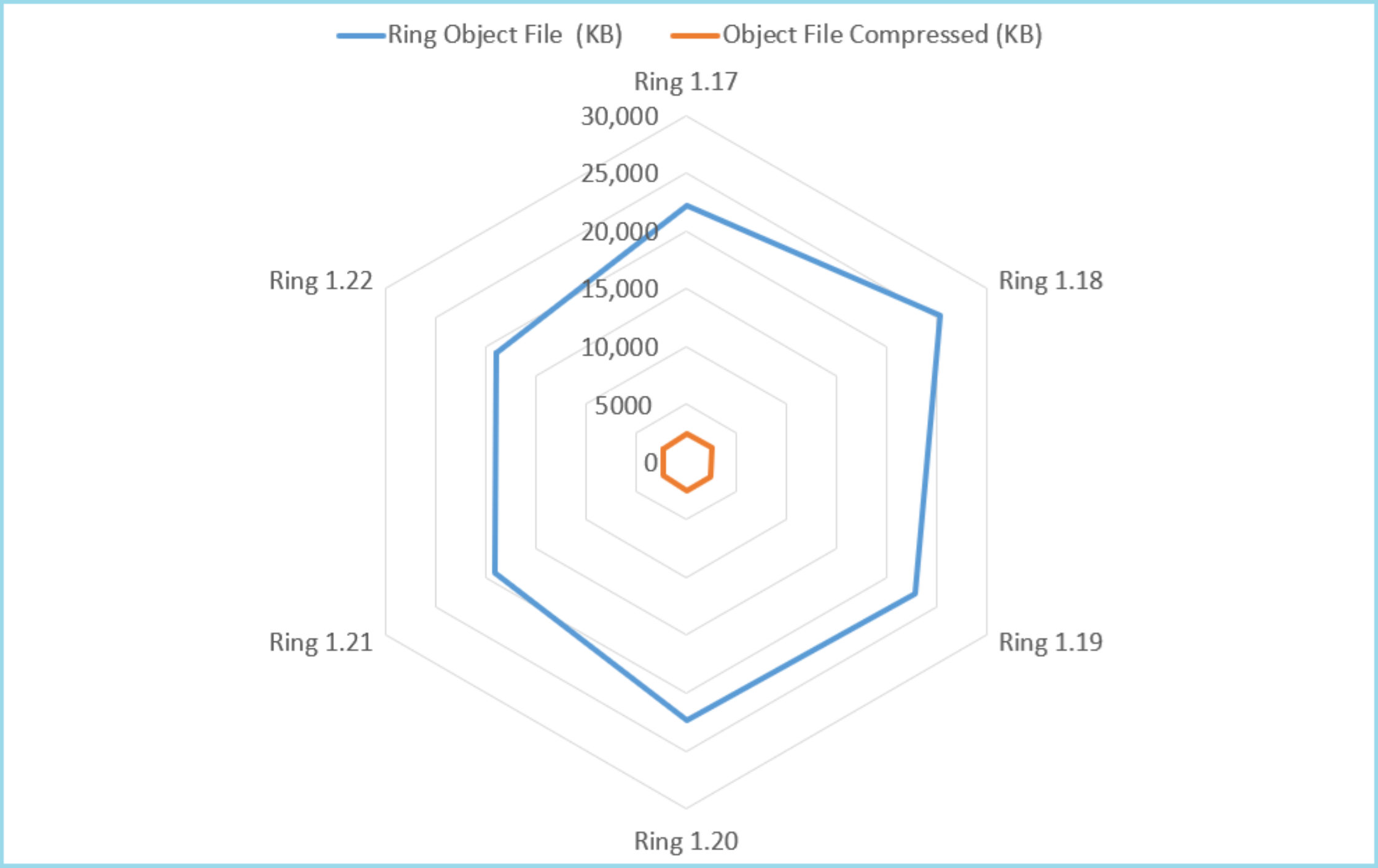

***Figure 6.21*** *Ring Object File Size for PWCT2 from Ring 1.17 to Ring 1.22.*

### 6.3.3 PWCT2 Performance

Table 6.13 presents performance results related to the PWCT2 environment, showcasing various visual source files and their corresponding metrics. These files represent games and utilities such as StopWatch, Snake, Matching, Sokoban, Escape, Tessera, FlappyBird3000, and others.

Each sample is evaluated based on several criteria: file storage in kilobytes (KB), the number of visual components, steps involved, generated lines of code (LOC), loading time (LT) in milliseconds required to display the visual representation, and code generation time (CGT) in milliseconds. The data demonstrates a range of sample complexities, with file storage sizes varying from 41 KB to over 1000 KB, and visual components ranging from 40 to 1269.

The number of steps and lines of code also vary significantly, highlighting the diverse nature of these visual source files. Loading and code generation times provide insights into the performance efficiency of PWCT2 across different applications. The measurements are done using a Victus Laptop [13th Gen Intel(R) Core(TM) i7-13700H, Windows 11, PWCT 2.0].

***Table 6.13*** *Measuring performance of PWCT2 (Time in Milliseconds).*

| ID | File | Storage (KB) | Components | Steps | LOC | LT (ms) | CGT (ms) |
|---|---|---|---|---|---|---|---|
| 1 | StopWatch | 41 | 40 | 74 | 54 | 73 | 3 |
| 2 | Snake | 138 | 149 | 234 | 185 | 199 | 9 |
| 3 | Matching | 218 | 215 | 330 | 275 | 287 | 13 |
| 4 | Sokoban | 217 | 235 | 330 | 269 | 269 | 13 |
| 5 | Escape | 250 | 277 | 401 | 329 | 330 | 14 |
| 6 | Tessera | 268 | 269 | 418 | 337 | 370 | 17 |
| 7 | FlappyBird3000 | 260 | 263 | 429 | 341 | 358 | 17 |
| 8 | Pairs | 283 | 287 | 432 | 362 | 374 | 17 |
| 9 | Cards | 298 | 305 | 457 | 399 | 363 | 15 |
| 10 | SquaresPuzzle | 299 | 327 | 513 | 378 | 429 | 20 |
| 11 | StarsFighter | 321 | 329 | 530 | 423 | 449 | 21 |
| 12 | KnightTour | 347 | 386 | 549 | 465 | 448 | 21 |
| 13 | Tetris | 351 | 402 | 586 | 483 | 459 | 20 |
| 14 | Game2048 | 371 | 356 | 588 | 461 | 499 | 22 |
| 15 | Othello | 427 | 422 | 653 | 536 | 562 | 25 |
| 16 | DFT | 787 | 538 | 677 | 603 | 784 | 22 |
| 17 | MagicBalls | 438 | 510 | 700 | 602 | 567 | 24 |
| 18 | Minesweeper | 449 | 509 | 701 | 611 | 559 | 24 |
| 19 | SuperMan | 465 | 480 | 757 | 607 | 622 | 30 |
| 20 | Laser | 707 | 825 | 1139 | 955 | 918 | 39 |
| 21 | GameOfLife | 773 | 870 | 1253 | 1066 | 1068 | 43 |
| 22 | Checkers | 845 | 968 | 1307 | 1128 | 1077 | 43 |
| 23 | GoGame | 946 | 1089 | 1453 | 1271 | 1166 | 47 |
| 24 | GetQuotesHistory | 1021 | 1269 | 1555 | 1407 | 1228 | 51 |
| 25 | Chess | 1012 | 1166 | 1560 | 1340 | 1270 | 52 |

In Figure 6.22, we present the relationship between the number of steps and the required storage. The figure is a stacked bar chart comparing the storage (in KB) and steps for various applications or games developed using PWCT2. The x-axis lists the names of the applications/games, while the y-axis represents the values for storage and steps. The blue segments of the bars represent storage, and the orange segments represent the number of steps in the visual source file.

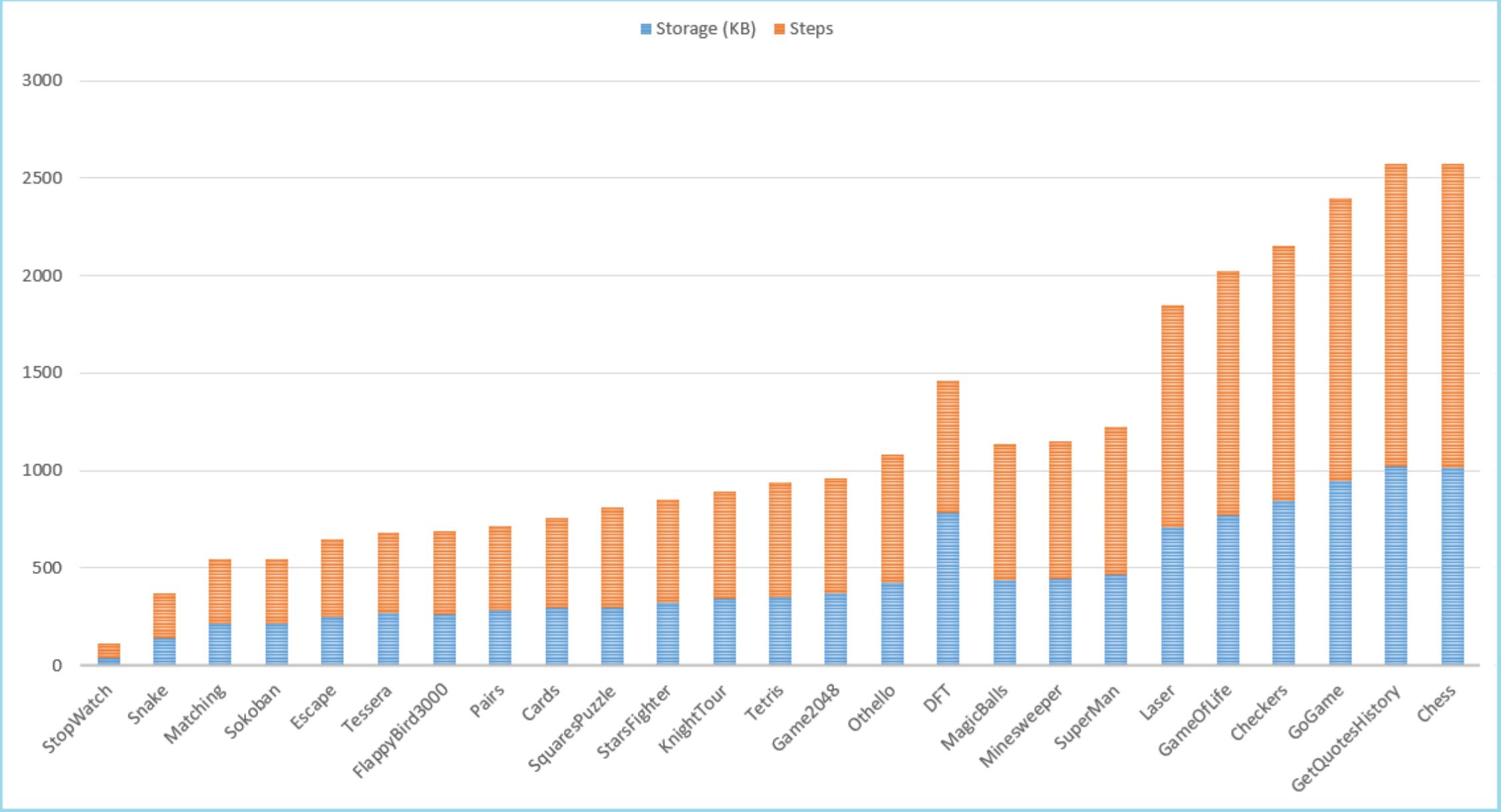

***Figure 6.22** The relationship between the number of steps and the required storage.*

In Figure 6.23, we present the relationship between the number of visual components and the generated source code, showing that using visual components increases the abstraction level. In Figure 6.24, we demonstrate the relationship between both the loading time/code generation time and the step count in the visual source file.

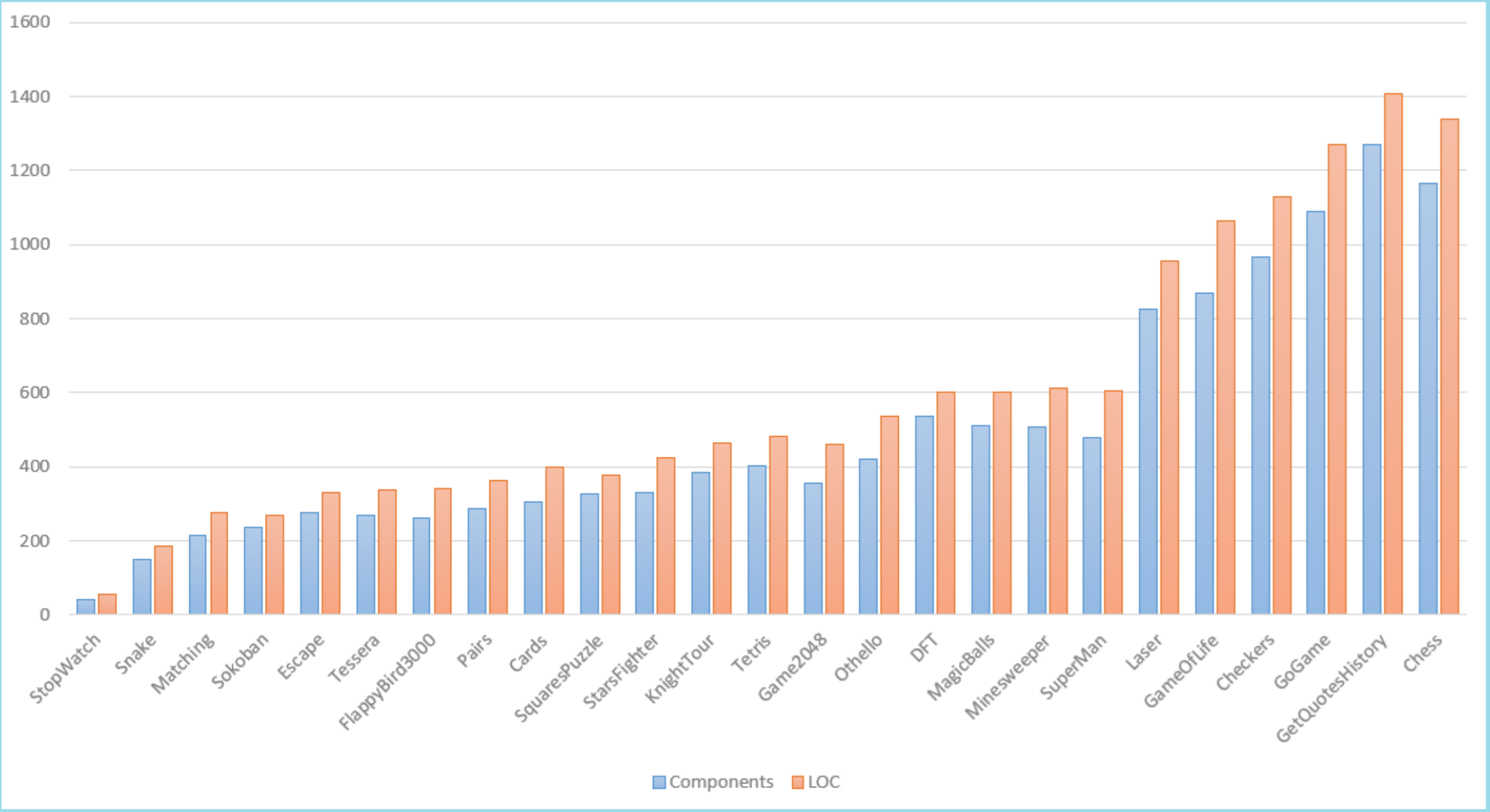

***Figure 6.23** Using visual components increases the abstraction level.*

We notice from Figure 6.24 that the loading time increases significantly as the file size and complexity grow, with a noticeable spike around the 600 ms mark, indicating that more complex files require more time to load. The orange line represents the code generation time, which remains relatively low throughout the different samples, with only minor fluctuations. The code generation time is more critical for developers because they may need to run the program multiple times during development and updating the visual code, making it essential to keep this time low for a smoother and more efficient development cycle. On the other hand, the loading time, while important, is typically incurred only once per file. By maintaining consistently low code generation time, PWCT2 enhances the development experience, making it more efficient and less time-consuming for developers.

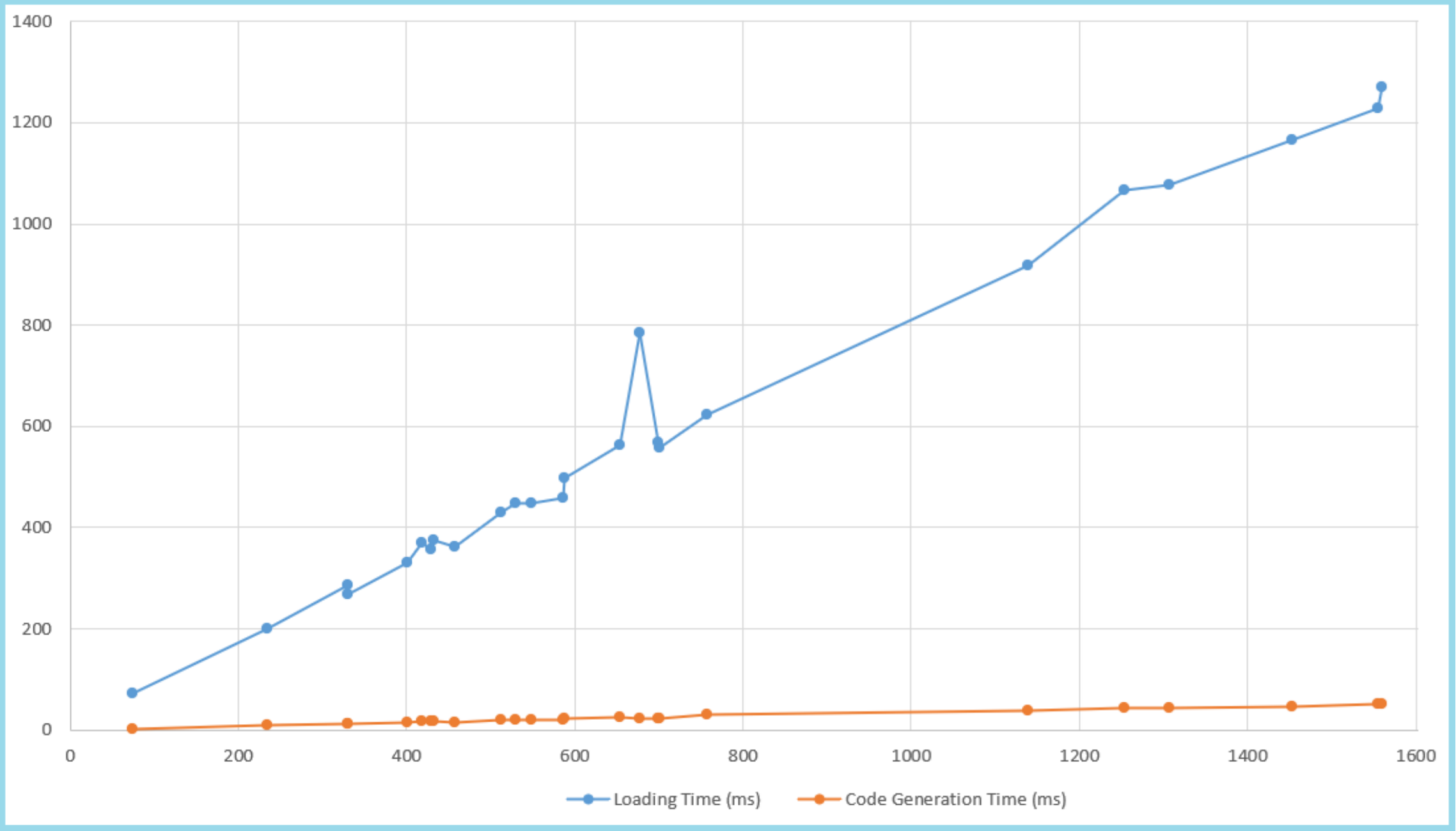

***Figure 6.24*** *The relationship between the LT/CGT and the step count.*

Table 6.14 demonstrates the performance and storage requirements of PWCT2 and PWCT [145] when using visual source files that contain at least 1000 steps by evaluating various metrics: file storage in kilobytes (KB), the number of steps, loading time (LT) in milliseconds (ms), and code generation time (CGT) in milliseconds (ms). These visual source files pertain to different visual programming languages (CPWCT vs. RingPWCT) and various programs. However, the step counts fall within the same range, representing the size of the visual programs, which influences the code generation performance and storage requirements.

PWCT2 shows significant improvements in file storage efficiency and code generation time, with much smaller file sizes (773 KB to 1012 KB) and faster code generation (43 ms to 52 ms) compared to PWCT, which has larger file sizes (7966 KB to 11,397 KB) and much longer code generation times (1748 ms to 3862 ms). Although PWCT2 has higher loading times for the visual source files (1068 ms to 1270 ms) than PWCT (549 ms to 860 ms), this difference is not substantial and is due to new optional features in PWCT2, such as rich colors where the same steps can use more than one color. The faster code generation time in PWCT2 is crucial for developers who need to run the program multiple times during development, making it more efficient and less time-consuming.

***Table 6.14*** *Some large visual source files.*

| Generation | VPL | File | Storage (KB) | Steps | LT (ms) | CGT (ms) |
|---|---|---|---|---|---|---|
| PWCT2 | RingPWCT | GameOfLife | 773 | 1253 | 1068 | 43 |
| PWCT2 | RingPWCT | Checkers | 845 | 1307 | 1077 | 43 |
| PWCT2 | RingPWCT | GoGame | 946 | 1453 | 1166 | 47 |
| PWCT2 | RingPWCT | Chess | 1012 | 1560 | 1270 | 52 |
| PWCT | CPWCT | Vmfuncs | 7966 | 1000 | 549 | 1748 |
| PWCT | CPWCT | Refmeta_ext | 8243 | 1075 | 593 | 1993 |
| PWCT | CPWCT | File_ext | 9799 | 1235 | 747 | 2651 |
| PWCT | CPWCT | Vm_oop | 11397 | 1497 | 862 | 3862 |

In Table 6.15, we present a summary of the key statistical metrics derived from analyzing the relationship between the number of steps, storage requirements, and code generation time (CGT). The analysis for PWCT2 is done for 25 visual source files presented in Table 11 which used RingPWCT. The analysis for PWCT is done for 43 visual source files related to CPWCT and introduced in the literature with storage size and code generation time [145]. We used the same hardware for all experiments. Also, to ensure the reliability and validity of our findings, we calculated the *p*-values for both Pearson and Spearman correlation coefficients across storage vs. steps and code generation time vs. steps. This statistical analysis allowed us to determine the significance of the observed relationships. By verifying that the p-values were well below the accepted threshold of 0.05.

***Table 6.15*** *Statistical Analysis of RingPWCT and CPWCT samples.*

| Attribute | PWCT | PWCT2 |
|---|---|---|
| Visual Programming Language | CPWCT | RingPWCT |
| Visual Source Files Count (Sample Size) | 43 | 25 |
| Pearson Correlation (Storage vs. Steps) | 0.8693 | 0.9662 |
| Pearson Correlation (CGT vs. Steps) | 0.9105 | 0.9947 |
| Spearman Correlation (Storage vs. Steps) | 0.8198 | 0.9867 |
| Spearman Correlation (CGT vs. Steps) | 0.9914 | 0.9855 |
| Average Storage per Step (KB/step) | 13.6751 | 0.6543 |
| Average CGT per Step (ms/step) | 1.2956 | 0.0353 |
| RMSE for Storage | 23.4063 | 0.1082 |
| RMSE for CGT | 1.0539 | 0.0032 |

For PWCT2, our analysis revealed a very strong positive correlation between the number of steps and both storage and code generation time. This indicates that as the number of steps increases, both storage and CGT also increase in a nearly linear fashion. The average storage required per step was found to be 0.6543 KB, while the average CGT per step was 0.0353 milliseconds. Furthermore, the Root Mean Square Error (RMSE) values for storage per step (0.1082) and CGT per step (0.0032) were low, suggesting that the average values are reliable and well represented by the data. These findings demonstrate the efficiency and scalability of code generation in PWCT2, particularly in handling visual source files with a high number of steps.

For PWCT, the average storage required per step was found to be 13.6751 KB, while the average CGT per step was 1.2956 milliseconds. Furthermore, the Root Mean Square Error (RMSE) values for storage per step (23.4063) and CGT per step (1.0539) were calculated, indicating some variability around the averages. Despite this variability, the correlations remain strong and statistically significant, underscoring the reliability of the data and the performance characteristics of PWCT. These findings highlight the substantial resource demands of code generation in PWCT, particularly as the number of steps increases. PWCT2 improves upon PWCT by storing the Steps Tree directly in visual source files according to the actual control flow, either appending or inserting steps as needed, whereas PWCT appends all steps to the end and reorders them only during code generation.

The statistically significant data and strong performance metrics allow us to generalize the findings when discussing the performance improvements of PWCT2 over PWCT. On average, PWCT2 is approximately 36.7 times faster in code generation time (CGT) per step and uses approximately 20.9 times less storage per step compared to PWCT. This significant improvement is evidenced by the lower average CGT and storage per step for PWCT2, as well as the low RMSE values, indicating high efficiency and consistency. These results highlight the advancements and effectiveness of PWCT2, making it a more efficient tool for visual programming tasks.

### 6.3.4 User Feedback

Once we started distributing the software and describing it as a visual programming tool and a replacement for the Ring Code Editor, users began joining the community group, which now has over 750 members, more than 100 discussion topics, and over 50 announcements about software updates. In early 2023, most discussion topics were questions about using the software, its capabilities, and a few bug reports. We fixed the reported bugs and provided educational videos (see Table 6.16) to help users learn about the software and how to use it [173]. Then, we guided them to the PWCT2 samples and Ring language documentation to learn more about the software. Many of the recent topics in 2024 are questions about future directions and requests to support other textual programming languages like Lua, Python, and C#.

***Table 6.16*** *Statistics about educational videos introducing PWCT2.*

| Attribute | Value |
| --- | --- |
| Total Videos Count | 39 |
| Average Duration (M:S) | 8:47 |
| Total Duration (H:M:S) | 5:42:27 |

Based on statistics from the Steam platform regarding PWCT2 web page visits from March 2023 to December 2024, the platform displayed the project title, logo, and short description over 1.73 million times to users through various lists. The software web page received over 159 K visits (a 9.2% clickthrough rate), including over 72 K visits from the United States, over 33 K visits from the Russian Federation, and over 11 K visits from Saudi Arabia. Over 20,000 users own the software and have added it to their Steam library, enabling them to download and use it at any time.

In Table 6.17, we present the minimum usage time and the percentage of users as reported by the Steam platform.

***Table 6.17*** *Usage time as reported by the Steam platform.*

| Minimum Usage Time | Percentage of Users |
|---|---|
| 30 min | 27% |
| 1 h 0 min | 19% |
| 2 h 0 min | 13% |
| 5 h 0 min | 8% |
| 10 h 0 min | 5% |

Steam reported that 1772 users have downloaded and launched the software through Steam, with an average usage time of 9 h and 40 min. This means the software has been used for over 17,000 h as shown in Table 6.18.

***Table 6.18*** *Statistics about the PWCT2 from 1 March 2023 to 21 December 2024.*

| Attribute | Value |
|---|---|
| Impressions | 1.72 M |
| Web page visits | 159 K |
| Software owners | 20,623 |
| Users launched the software | 1772 |
| Average usage time | 9 h and 40 min |
| Total usage time | Over 17,000 h |

In Figure 6.25, we present a horizontal bar chart that shows the download statistics of the PWCT2 software across different regions, listing only the top regions, such as North America and Western Europe.

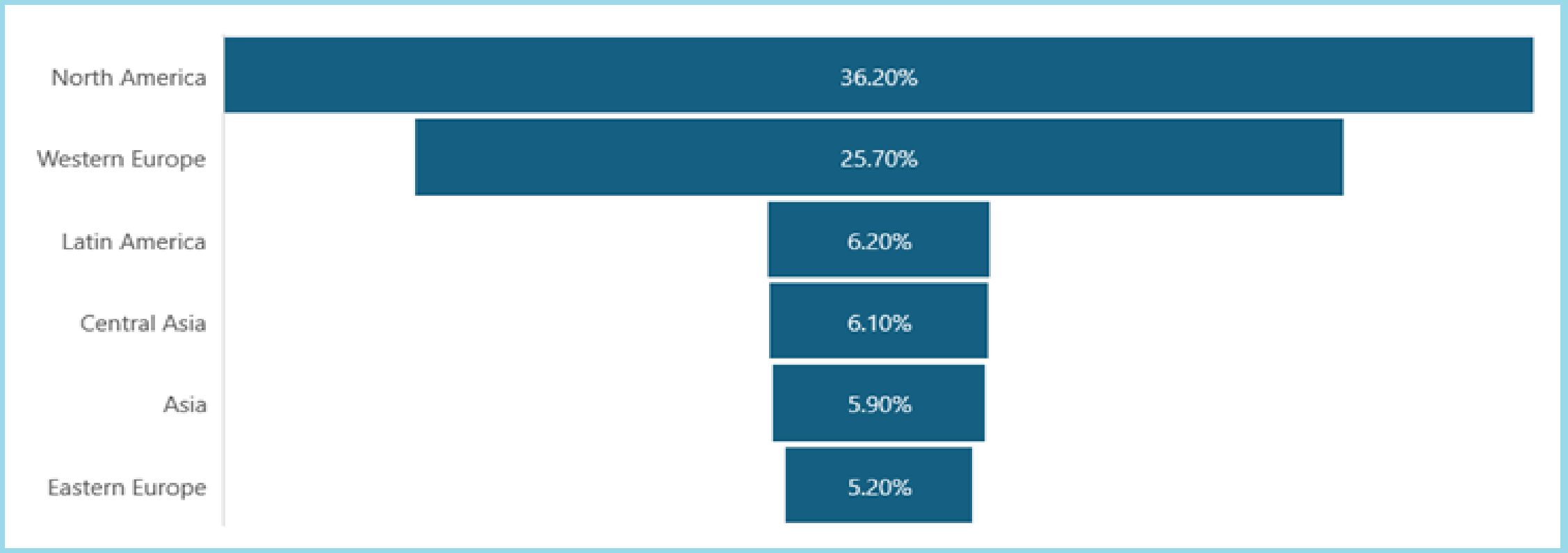

***Figure 6.25*** *Download statistics of the PWCT2 software across top regions.*

The regions and their corresponding percentages are as follows: North America at 36.20%, Western Europe at 25.70%, Latin America at 6.20%, Central Asia at 6.10%, Asia at 5.90%, and Eastern Europe at 5.20%. The chart highlights the significant differences in the distribution of downloads across these regions, with North America having the highest percentage and Eastern Europe having the lowest. In Figure 6.26, we present the downloads across the top countries, showing that the United States has the highest number of downloads (31%), followed by Germany (10%) and Canada (5%).

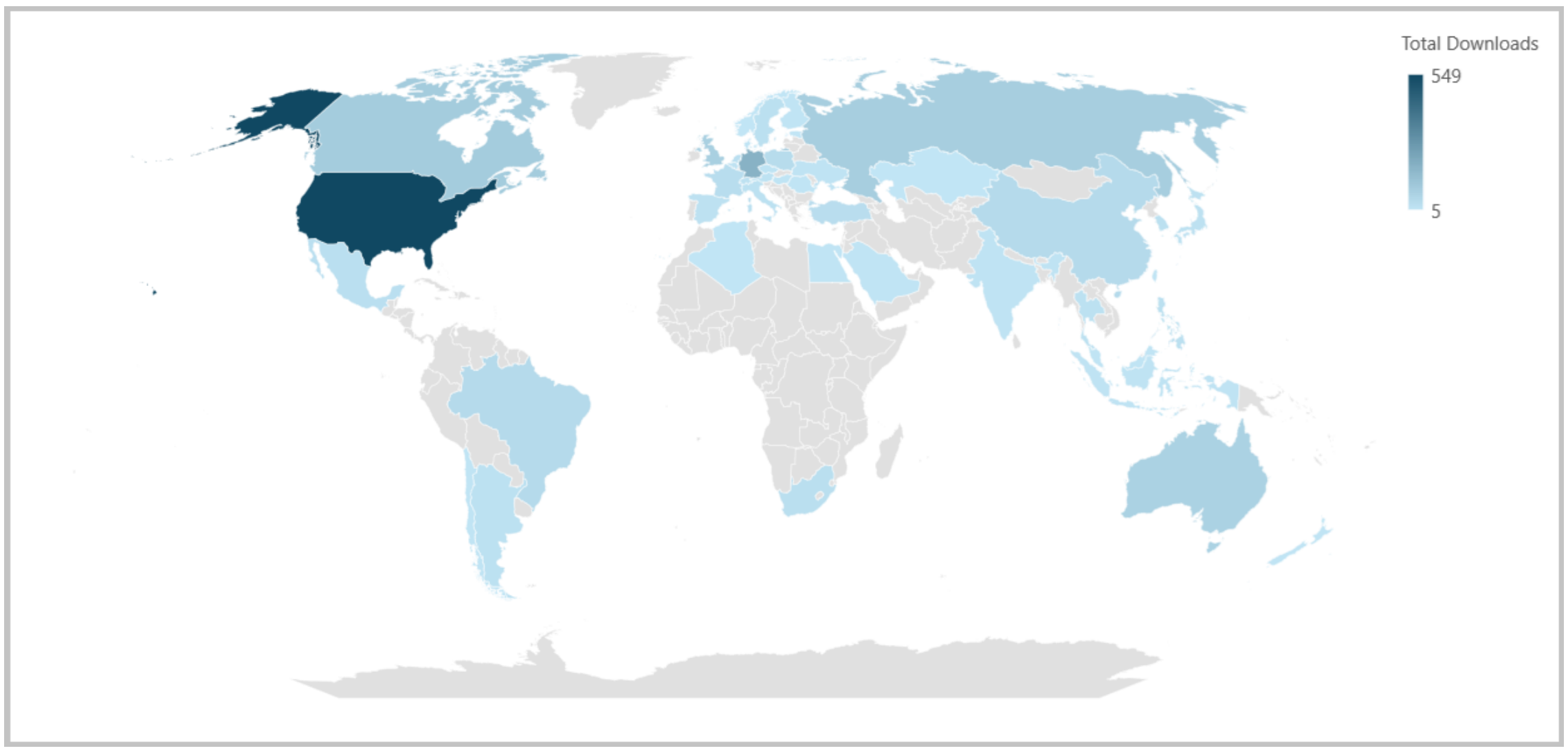

***Figure 6.26*** *PWCT2 Software downloads across top countries.*

We conducted an analysis for the user reviews for the PWCT2 software on the Steam platform, while noticing important details such as review type, language used to write the review, and usage time in hours. The reviews are written in various languages, indicating a diverse user base. Out of the thirty-one reviews, twenty-eight users recommended the software, while three users did not.

The languages used in the reviews include Arabic, Chinese, English, Italian, Polish, Portuguese, Russian, Thai, and Turkish. Usage varies significantly, from as little as 0.1 h to an extensive 560.1 h. Notably, a user who used the software for 560.1 h recommended it, showing a high level of engagement. Other notable usage times (in their respective review languages) include 93.4 h (Thai), 61.4 h (English), and 56 h (Portuguese), all with positive recommendations. Figure 6.27 illustrates the proportion of positive and negative reviews for the PWCT2 software on the Steam platform.

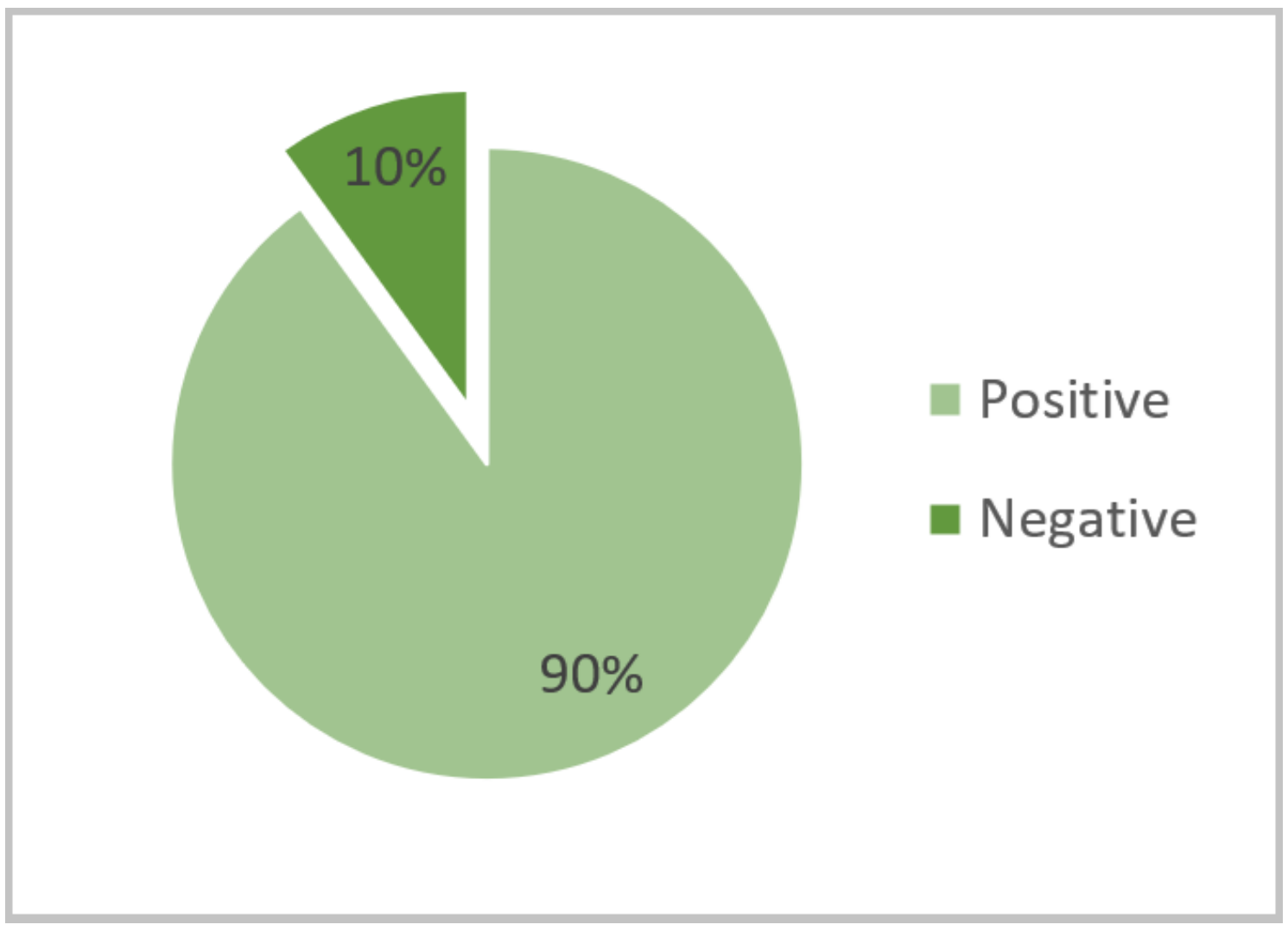

***Figure 6.27*** *User satisfaction according to steam statistics.*

The user reviews provide a mix of positive and critical feedback. Users appreciate the software's educational value, especially for beginners. The visual and organized approach to programming is praised, with some comparisons to Scratch. Users find it helpful for those with little coding knowledge and believe it has great potential for learning programming.

With respect to negative feedback, we noticed that it was based on various reasons, including encountering bugs, requiring more educational resources, and seeking additional features to enhance usability. During 2023 and 2024, we worked on resolving all reported bugs, introduced more educational resources (videos and documentation), and added requested features such as Drag-and-Drop support in the Goal Designer, an expression builder in the interaction pages, as well as preventing composition errors in the Steps Tree. However, we believe there is room for further improvement in the educational resources and the features provided.

The software’s interface is described as smooth, with many features to learn. Some users highlight the convenience of having the entire programming interface visible and the ability to keep necessary interaction pages open. Some users also express satisfaction with features like Ring2PWCT and the Time Machine, which enhance their programming experience.

There are positive remarks about the potential for serious projects. Many users express satisfaction with the software’s development and updates. On the other hand, a few users feel the software is not beginner-friendly and recommend it only for those with prior programming knowledge.

Continued development and support are encouraged, with some users acknowledging our responsiveness in fixing issues. Overall, the reviews highlight both the strengths and areas for improvement in PWCT2, reflecting diverse user experiences and perspectives.

## 6.4 Chapter Summary

In this chapter, we presented the results and experiments that evaluated the effectiveness and performance of both the Ring dynamic programming language and the PWCT2 visual programming language. Additionally, we examined storage usage for visual source files in PWCT2, highlighting its impact on resource efficiency. Furthermore, we discussed the portability of Ring, which supports deployment across Desktop, WebAssembly, and Microcontroller environments, showcasing its versatility. Ring also features automatic memory management, equipped with a built-in garbage collector, ensuring efficient handling of memory allocation and deallocation. the results also incorporated user feedback, providing valuable insights into the usability and practical experiences with both languages.

In the next chapter, we will engage in a discussion of our findings, delving into the implications and limitations discovered during our research.

# Chapter 7: Discussion

## 7.1 Introduction

In this chapter, we will provide a comprehensive summary of the experiments and results presented in the previous chapter. We will analyse the findings in detail, discussing their implications and how they support the objectives of this thesis. Additionally, we will address the limitations encountered during our research, providing a critical evaluation of the potential challenges and areas for improvement. By examining these aspects, we aim to offer a thorough understanding of the strengths and weaknesses of the Ring dynamic programming language and the PWCT2 visual programming language.

## 7.2 Ring dynamic programming language

From Table 6.2, we notice that dozens of users learned the language through the available resources (documentation, samples, applications, and videos), and their feedback helped us grow the educational resources distributed with the language. For example, chapters such as FAQ, Scope Rules, Performance Tips, and General Information were added and enhanced based on this feedback. Also, many graphics programming samples (OpenGL Camera and background, Collision detection, Chess 3D, etc.) are developed by one of those users. Another important lesson learned from being close to users and responding to their feedback is that this not only encourages more people to get involved and report issues but also motivates users to become active contributors to the open-source project. However, we acknowledge the limitations and potential biases, especially regarding the demographic homogeneity (predominantly male participants) and the regional limitations of early feedback, which could affect the generalizability of the results.

From the statistics in Figures 6.3 and 6.4, we can conclude that Ring, as a programming language and research prototype, has been tried by thousands of users [83,84]. Based on the use cases demonstrated in Table 6.3, the Ring programming language has proven to be versatile. It has been effectively utilized in a variety of domains, including desktop development (Ring IDE and Chess End Game application), game development (Shooter Game and Gold Magic 800 puzzle game), and data analysis (Arabic poetry analysis application). The language's lightweight and embeddable nature, combined with its support for many programming paradigms, allows for rapid development [91,152,166–171]. However, this versatility still needs to benefit more from the language's ability to create domain-specific languages and localization packages.

The remainder of this section addresses the second research question (RQ2). Abstraction is a known dimension in the Cognitive Dimensions Framework (CDF), which is utilized in many research studies for the usability analysis of visual programming languages [10]. Abstraction involves grouping elements or entities into a single entity to either reduce viscosity (making it less difficult to modify) or align the notation with the user's conceptual structure. Abstractions are useful for modification and transcription tasks (copying content from one structure to another). They play a crucial role in visual implementation, as they significantly influence ease of use and can also increase protection against errors [174,175].

From the results in Table 6.5, we observe that Ring visual implementation comprises 18,945 visual components, which in turn generated 24,743 lines of code. This finding highlights the significant advantage of visual implementation: it increases the abstraction level by 23.5% while concealing unnecessary details. Specifically, this advantage becomes apparent when specifying component data through interaction pages (data-entry forms). These interaction pages generate and update the steps tree based on the provided data. However, a different scenario emerges when considering the generated steps tree. The steps tree aims to provide additional information about the program structure and related details, resulting in a total number of steps that exceed the lines of code, as demonstrated in the "Steps" column. On the other hand, PWCT offers a "Read Mode" that allows users to hide many of these implementation details. In this mode, the "Visible Steps" column shows a count slightly less than the total lines of code. Despite this difference, the Steps Tree has a clear advantage: it facilitates easy interaction with groups of steps. Its tree structure directly provides two dimensions of interaction—siblings and children—which enhances usability and navigation within the visual implementation. This higher level of abstraction translates into a more productive development process by allowing developers to focus on the overall structure and functionality of the program rather than getting bogged down in the minutiae of code syntax.

The visual components provide a more intuitive and accessible interface, making it easier for developers to understand and modify the codebase. Additionally, this approach improves usability by reducing the likelihood of syntax errors and simplifying the debugging process. [34,66,120].

However, we have observed certain disadvantages while using PWCT:

- Large Storage Size: Visual implementations tend to occupy more storage space compared to their textual counterparts. This is an important consideration, especially when dealing with large projects.
- Memory Requirements for Multiple Instances: The PWCT environment is designed to open one visual source file at a time. To work with multiple files simultaneously, you need to run multiple instances of PWCT. Unfortunately, this approach comes with a memory cost. Opening all the visual source files related to the Ring compiler and virtual machine implementation (43 files) requires approximately 1.3 GB of memory. This could become an issue for larger projects that contain more visual source files, where opening these files simultaneously for quick navigation could be problematic. This might be required when searching in multiple files. We also noticed that PWCT supports search/replace in a single visual source file, which is not practical for large projects. To work around this issue, we used external tools to search in the generated textual source code.
- Limitations of the Steps Tree Editor: The Steps Tree editor lacks support for drag-and-drop functionality. Moving steps within the tree is possible only through cut-and-paste operations. While this may not be a critical issue, it is worth noting for usability purposes because it forces us to use the keyboard to move steps faster from one location to another.
- Performance Challenges with Large Visual Source Files: When dealing with visual source files containing thousands of components, performance can become an issue. Loading such files or generating source code may exhibit slower behavior. For example, our largest file (genlib_ext) contains 1732 components and 2965 steps. Loading the file and displaying the visual representation takes over two seconds, while generating the source code takes over 14 s, as demonstrated in Figures 6.8 and 6.9. This results in slow development and iteration when tasks involve updating many visual source files.

- No support for importing textual source code. This missing feature introduces many limitations. If a programmer submits a GitHub pull request by modifying the textual source code of the Ring Compiler/VM, we cannot simply import these changes. Additionally, this is a barrier to integrating with AI tools that generate textual code. While in visual implementation, we could use external libraries provided through textual code (without visual implementation), the problem occurs when the generated source code (from visual implementation) is modified directly by other contributors in the project without making the change through visual programming first and then generating the textual source code. This leads to effort duplication. This demonstrates a limitation on collaborative development efforts. For users who might wish to integrate Ring with existing workflows or tools and are looking to modify the language implementation, we recommend making a choice and sticking to it: either use the visual implementation and make any changes through it first or use the generated textual source code and continue development based on it.
- PWCT is designed to work only under Microsoft Windows. The support for other operating systems is not native and requires extra tools (Like Wine for Linux).

Despite these challenges, our successful use of PWCT to develop and maintain the Ring programming language compiler and virtual machine demonstrates its value. However, addressing these scalability issues will be crucial if PWCT is to be adopted in larger projects in the future.

Suggestions to mitigate these challenges:

1. Separate the visual source into many files with clear names and purposes.
2. Keep each visual source file to fewer than a few thousand steps.
3. Open related visual source files according to the current task while closing unrelated visual source files (or not opening them) to provide easy navigation between PWCT instances through the operating system features.
4. External tools are needed when searching multiple generated source code files.

From Table 6.6 and Figures 6.10 and 6.12, we demonstrated the growth of the Ring language over eight years; while being a lightweight language, we noticed a growth in the implementation size from 16 KLOC in 2016 to 24 KLOC in 2024. This percentage of growth (51%) requires attention, and we could focus in the next years on reducing the implementation size since the core features have been implemented and the implementation is stable and usable. With respect to adding new features, we will try to keep most of these new features in the libraries and the new domain-specific languages. From Tables 6.7, 6.8 & 6.9, we notice that the performance of the Ring programming language has improved over time, and it is now fast enough for many use cases as a scripting language. However, improving Ring's performance remains a challenge, and we aim to provide optimizations and enhancements with each new release.

## 7.3 PWCT2 visual programming language

The diverse use cases of PWCT2 demonstrate its adaptability and utility in various programming scenarios. The primary use case for PWCT2 is as an alternative to the Ring Code Editor, which allows users to create new projects based on Ring, as well as visualize and execute Ring programs. This flexibility is further enhanced by the Time Machine feature, which enables users to read programs step by step and run them from a previous point in time, facilitating the continued development and maintenance of numerous Ring samples and applications. Additionally, PWCT2 supports the development of GUI applications through its Form Designer, which has been utilized to create applications for predicting citation counts in the Otology field and for finding files distributed with Ring Notepad. These applications showcase PWCT2's capability to handle different tasks, such as integrating machine learning models developed using Microsoft Azure Machine Learning and providing user-friendly interfaces. Moreover, PWCT2's ability to convert Ring textual code into visual code and vice versa using the Ring2PWCT tool underscores its potential as a flexible tool in both educational and professional settings. The practical use cases of PWCT2 highlight its contributions to simplifying programming and enhancing the overall development process based on the Ring programming language.

The results presented in Tables 6.11 and 6.12 highlight the excellent compilation performance achieved during the development of PWCT2 using various versions of the Ring language. Furthermore, the number of byte-code instructions decreased in Ring 1.22, indicating more efficient byte-code generation. However, despite these improvements, the size of the generated Ring object file remains relatively large, even though there was a reduction from 22,825 KB in Ring 1.19 to 18,952 KB in Ring 1.22. Addressing the large object file size could reduce the storage and memory requirements of the PWCT2 project. Tables 6.14 and 6.15 present an insightful comparison of PWCT2's performance and its improvements over PWCT [145]. PWCT2 demonstrates significant enhancements in file storage efficiency and code generation time. The results show that PWCT2 has much smaller file sizes, and a notably faster code generation compared to PWCT, which is crucial during multiple programs runs and updates.

User feedback has played a crucial role in the development and refinement of PWCT2. Since its distribution as a visual programming tool and a replacement for the Ring Code Editor, the community group has grown significantly, now boasting over 750 members. Notably, over 20,000 users have added the software to their Steam libraries, enabling them to download and use it anytime. Steam reported that 1772 users have launched the software, with an average usage time of 9 h and 40 min, cumulatively amounting to over 17,000 h of usage (Table 6.18). Regional and country-specific download data highlight that North America and the United States lead in downloads (Figures 6.25 and 6.26). User reviews on Steam (Figure 6.27) also present a mix of positive and critical feedback, with notable appreciation for the educational value, smooth interface, and unique features like Ring2PWCT and the Time Machine.

Since PWCT2 currently supports the Ring programming language through RingPWCT visual components, these visual components follow the Ring language approach for representing different data structures, where Ring lists are used instead of arrays, linked lists, trees, hash tables, etc. However, by using classes and operator overloading, we can create custom types. The Ring Standard Library comes with specific classes for lists, trees, hash tables, etc., and the RingPWCT includes visual components that enable the use of these classes. With respect to scalability and creating large projects, the PWCT2 approach is based on organizing large projects into folders, subfolders, and different visual source files, where a visual source file can load other files and use the functionality provided by them, such as functions and classes. However, more improvements are required in this area to provide visualizations that highlight information from different visual source files.

The development and evaluation of PWCT2, the proposed research prototype and successor to PWCT, bring several noteworthy limitations to light. First and foremost, PWCT2 supports only the Ring programming language, whereas PWCT provided visual components for various textual programming languages, including Harbour, C, Supernova, Python, and C# [66]. This limitation restricts the versatility of PWCT2 and may reduce its appeal to users who require support for multiple programming languages. Secondly, PWCT2 is not compatible with PWCT in terms of visual component design or visual source file formats. This incompatibility necessitates the development of tools to convert projects from PWCT to PWCT2, which could pose challenges for users looking to transition their existing projects. Additionally, PWCT2 is currently distributed as a desktop tool rather than a web-based application. The lack of a web-based version limits accessibility and convenience, highlighting a potential area for future development to enhance user experience and broaden the tool's reach.

Since PWCT2 is based on the Qt framework, which supports WebAssembly, and the Ring language also supports the WebAssembly platform, we plan to develop PWCT2 for WebAssembly in the future. At this stage, we have an online version of the Form Designer, and other components will be ported as well. While PWCT2 includes the Ring2PWCT feature, which allows it to accept Ring code generated by large language models, the process currently requires a copy-and-paste operation (e.g., from Copilot to PWCT2) [160]. A fully integrated solution, where writing a natural language prompt directly generates a visual representation without needing manual transfer, would be more efficient and user-friendly. This enhancement could significantly streamline the workflow and improve usability.

Moreover, despite providing 39 instructional videos to explain PWCT2 features, using PWCT2 still demands a general understanding of programming and specific knowledge of the Ring language [173]. To achieve broader adoption, it may be necessary to expand educational resources, including tutorials and documentation, to support users with varying levels of expertise.

In summary, while PWCT2 advances the capabilities of its predecessor and offers a robust platform for developing Ring-based applications, addressing the limitations identified could enhance its functionality, user experience, and appeal to a broader audience. Future work should focus on expanding language support, ensuring compatibility with PWCT, developing a web-based version, improving integration with natural language processing tools, and enhancing educational resources to support a diverse user base.

## 7.4 Chapter Summary

In this chapter, we have thoroughly discussed the experiments and results, providing a comprehensive analysis of the strengths and limitations of the Ring dynamic programming language and the PWCT2 visual programming language.

In the next chapter, we will present our overall conclusions and outline potential future work. We will summarize the key findings of this thesis and propose directions for further research and development, aiming to build upon the foundations laid by the Ring and PWCT2 languages.

# Chapter 8: Conclusion and Future Work

In this chapter, we present the conclusion of this thesis, along with the planned future work.

## 8.1 Conclusion

In this thesis we introduced the design, implementation and evaluation of the Ring textual programming language and the PWCT2 self-hosting visual programming language. Ring is a dynamically typed language developed and maintained for over eight years using visual programming through the PWCT visual programming language, where the generated code is based on ANSI C.

The visual implementation is composed of 18,945 components that generate 24,743 lines of ANSI C code, which increases the abstraction level and hides unnecessary details. Using PWCT to develop Ring allowed us to realize several issues in PWCT like large storage size and performance challenges with large visual source files. We addressed these issues through the development of the PWCT2 visual programming language using the Ring textual programming language.

Ring combines a lightweight implementation with several advantages, such as a rich and versatile standard library, along with direct support for classes and object-oriented programming. Ring is adaptable across diverse platforms. Rather than creating separate language implementations for specific contexts, the same Ring implementation serves a wide range of environments. From desktop systems to WebAssembly and even 32-bit microcontrollers like the Raspberry Pi Pico, Ring addresses the problem of missing language features that exist in other implementations.

To achieve this, we applied specific design decisions such as using a single-pass compiler, grouping built-in functions in optional modules through preprocessor directives, opting for Ring Lists over C structures, selectively using C structures for critical features, implementing flexible lists using various data structures and optimization techniques, storing bytecode in a single continuous memory block, using a writable long-byte code format for performance improvements, and avoiding the use of a global interpreter lock for better thread performance.

Customization is a key feature of Ring, allowing developers to easily modify the language syntax multiple times. Moreover, Ring empowers the creation of domain-specific languages through novel features that extend object-oriented programming. Beyond its language design, the underlying idea relies on using braces to access objects, granting us the ability to utilize the attributes and methods provided by those objects. Ring does not require semicolons or new lines between statements. We can type different statements on the same line without any fuss. Additionally, in Ring, every expression is an acceptable statement, giving us the freedom to write various values, all of which will be accepted by the compiler. Ring classes also support properties. Typing a property name can invoke the getter method and execute the associated code. Moreover, Ring goes a step further by allowing us to define methods like braceStart() and braceEnd(). These methods are automatically called when we access an object using braces. Furthermore, the language automatically invokes a method called braceExprEval() when we write an expression inside braces. With these features, coupled with the ability to customize language keywords and operators, we can construct domain-specific languages that resemble external DSLs such as CSS, QML, SQL, and Supernova. Also, Ring provides a practical development environment and facilitates rapid GUI application development.

In summary, Ring emerges as a lightweight, versatile, and customizable dynamic language developed using visual programming, adapting seamlessly to the ever-evolving landscape of software development. The implementation based on visual programming increases the abstraction level, hides unnecessary details and provides a more user-friendly implementation through visual programming advantages, such as avoiding syntax errors.

Visual programming languages are helpful in making programming easier and faster to learn. This study introduces some useful improvements. We have developed the second generation of the PWCT visual programming language that includes enhanced features, works on different systems, performs code generation more efficiently than before, and reduces storage requirements. PWCT2 is a self-hosting visual programming language developed and maintained for over eight years using the Ring programming language. PWCT2 consists of approximately 92,000 lines of Ring code.

We also created the first visual programming language (RingPWCT) that generates code in the Ring programming language. RingPWCT contains 394 visual components that enable the development of a wide range of applications and tools, including PWCT2 itself. This helps combine the ease of visual programming with the flexibility of Ring. Additionally, we tested how well the Ring programming language Compiler/VM works for developing an advanced project like PWCT2. We also designed a tool called Ring2PWCT to convert textual Ring code into visual code in PWCT2, making it easier to use.

PWCT2 has been widely distributed to users via the Steam platform, receiving positive feedback. On Steam, the software has been launched by 1,772 users, with a total recorded usage time exceeding 17,000 hours. These improvements show that there is potential for making visual programming languages more accessible and effective for Ring developers. Our work provides a foundation for further development in this area.

## 8.2 Future Work

In the future, we plan to build multiple projects on top of the Ring programming language, such as a localization package for many human languages, various domain-specific languages for different fields, and a modern framework that includes many templates for database applications. Our priority is to provide a complete translation of all language syntax and libraries into Arabic. Following this, we aim to develop a domain-specific language for GUI development, like the Supernova language, but based on Ring's features that extend object-oriented programming to support the creation of internal domain-specific languages. After that, we will focus on creating the framework and templates for database applications.

In the future, we aim to enhance the PWCT2 visual programming language by supporting additional textual programming languages such as C, Java, C# and Python. We also plan to improve the environment by offering translations in various human languages, to make it more accessible to a global audience. Moreover, we intend to add more components that provide better support for Ring libraries, further enriching the functionality and usability of PWCT2. These enhancements will continue to build on the progress made and open new possibilities for users and developers.